\newcommand\apjcls{1}
\newcommand\aastexcls{2}
\newcommand\othercls{3}

\newcommand\papercls{\apjcls}
\documentclass[iop]{emulateapj}

\if\papercls \apjcls
\usepackage{apjfonts}
\else\if\papercls \othercls
\usepackage{epsfig}
\usepackage{margin}
\usepackage{times}
\fi\fi
\usepackage{ifthen}
\usepackage{natbib}
\usepackage{amssymb, amsmath}
\usepackage{appendix}
\usepackage{etoolbox}
\usepackage[T1]{fontenc}
\usepackage{paralist}

\usepackage{graphicx}
\usepackage{latexsym,amsmath,amssymb}
\usepackage{subfigure}
\usepackage{rotating}
\usepackage{enumerate}
\usepackage{color}
\usepackage{multirow}

\if\papercls \apjcls
\newcommand\aas{\ref@jnl{AAS Meeting Abstracts}}
\newcommand\dps{\ref@jnl{AAS/DPS Meeting Abstracts}}
\newcommand\maps{\ref@jnl{MAPS}}
\else\if\papercls \othercls
\usepackage{astjnlabbrev-jh}
\fi\fi

\if\papercls \aastexcls
\hypersetup{citecolor=blue, 
            linkcolor=blue, 
            menucolor=blue, 
            urlcolor=blue}  
\else
\usepackage[
bookmarks=true,           
bookmarksnumbered=true,   
colorlinks=true,          
citecolor=blue,           
linkcolor=blue,           
menucolor=blue,           
urlcolor=blue,            
linkbordercolor={0 0 1},  
pdfborder={0 0 1},
frenchlinks=true]{hyperref}
\fi

\if\papercls \othercls

\else

\fi

\providecommand{\adsurl}[1]{\href{#1}{ADS}}

\makeatletter
\patchcmd{\NAT@citex}
  {\@citea\NAT@hyper@{%
     \NAT@nmfmt{\NAT@nm}%
     \hyper@natlinkbreak{\NAT@aysep\NAT@spacechar}{\@citeb\@extra@b@citeb}%
     \NAT@date}}
  {\@citea\NAT@nmfmt{\NAT@nm}%
   \NAT@aysep\NAT@spacechar\NAT@hyper@{\NAT@date}}{}{}

\patchcmd{\NAT@citex}
  {\@citea\NAT@hyper@{%
     \NAT@nmfmt{\NAT@nm}%
     \hyper@natlinkbreak{\NAT@spacechar\NAT@@open\if*#1*\else#1\NAT@spacechar\fi}%
       {\@citeb\@extra@b@citeb}%
     \NAT@date}}
  {\@citea\NAT@nmfmt{\NAT@nm}%
   \NAT@spacechar\NAT@@open\if*#1*\else#1\NAT@spacechar\fi\NAT@hyper@{\NAT@date}}
  {}{}
\makeatother

\makeatletter
\DeclareRobustCommand{\lowcase}[1]{\@lowcase#1\@nil}
\def\@lowcase#1\@nil{\if\relax#1\relax\else\MakeLowercase{#1}\fi}
\makeatother

\DeclareSymbolFont{UPM}{U}{eur}{m}{n}
\DeclareMathSymbol{\umu}{0}{UPM}{"16}
\let\oldumu=\umu
\renewcommand\umu{\ifmmode\oldumu\else\math{\oldumu}\fi}

\if\papercls \othercls

\else

\fi

\let\oldsim=\sim
\renewcommand\sim{\ifmmode\oldsim\else\math{\oldsim}\fi}
\let\oldpm=\pm
\renewcommand\pm{\ifmmode\oldpm\else\math{\oldpm}\fi}
\newcommand\by{\ifmmode\times\else\math{\times}\fi}

\newbox{\wdbox}
\renewcommand\c{\setbox\wdbox=\hbox{,}\hspace{\wd\wdbox}}
\renewcommand\i{\setbox\wdbox=\hbox{i}\hspace{\wd\wdbox}}

\newcount\timect
\newcount\hourct
\newcount\minct
\newcommand\now{\timect=\time \divide\timect by 60
         \hourct=\timect \multiply\hourct by 60
         \minct=\time \advance\minct by -\hourct
         \number\timect:\ifnum \minct < 10 0\fi\number\minct}

\catcode`@=11

\newcommand\comment[1]{}

\newcommand\commenton{\catcode`\%=14}

\renewcommand\math[1]{$#1$}
\newcommand\mathshifton{\catcode`\$=3}

\let\atab=&
\newcommand\atabon{\catcode`\&=4}

\let\oldmsp=\sp
\let\oldmsb=\sb
\def\sp#1{\ifmmode
           \oldmsp{#1}%
         \else\strut\raise.85ex\hbox{\scriptsize #1}\fi}
\def\sb#1{\ifmmode
           \oldmsb{#1}%
         \else\strut\raise-.54ex\hbox{\scriptsize #1}\fi}
\newbox\@sp
\newbox\@sb
\def\sbp#1#2{\ifmmode%
           \oldmsb{#1}\oldmsp{#2}%
         \else
           \setbox\@sb=\hbox{\sb{#1}}%
           \setbox\@sp=\hbox{\sp{#2}}%
           \rlap{\copy\@sb}\copy\@sp
           \ifdim \wd\@sb >\wd\@sp
             \hskip -\wd\@sp \hskip \wd\@sb
           \fi
        \fi}
\def\msp#1{\ifmmode
           \oldmsp{#1}
         \else \math{\oldmsp{#1}}\fi}
\def\msb#1{\ifmmode
           \oldmsb{#1}
         \else \math{\oldmsb{#1}}\fi}

\def\supon{\catcode`\^=7}

\def\subon{\catcode`\_=8}

\def\supsubon{\supon \subon}

\newcommand\actcharon{\catcode`\~=13}

\newcommand\paramon{\catcode`\#=6}

\comment{And now to turn us totally on and off...}

\newcommand\reservedcharson{ \commenton  \mathshifton  \atabon  \supsubon 
                             \actcharon  \paramon}

\catcode`@=12
\reservedcharson

\if\papercls \apjcls

\else

\fi

\newcommand\chisq{\ifmmode{\chi\sp{2}}\else\math{\chi\sp{2}}\fi}
\newcommand\redchisq{\ifmmode{ \chi\sp{2}\sb{\rm red}}
                    \else\math{\chi\sp{2}\sb{\rm red}}\fi}
\newcommand\Teq{\ifmmode{T\sb{\rm eq}}\else$T$\sb{eq}\fi}
\newcommand\mjup{\ifmmode{M\sb{\rm Jup}}\else$M$\sb{Jup}\fi}
\newcommand\rjup{\ifmmode{R\sb{\rm Jup}}\else$R$\sb{Jup}\fi}
\newcommand\msun{\ifmmode{M\sb{\odot}}\else$M\sb{\odot}$\fi}
\newcommand\rsun{\ifmmode{R\sb{\odot}}\else$R\sb{\odot}$\fi}
\newcommand\mearth{\ifmmode{M\sb{\oplus}}\else$M\sb{\oplus}$\fi}
\newcommand\rearth{\ifmmode{R\sb{\oplus}}\else$R\sb{\oplus}$\fi}

\usepackage{rotating}

\shorttitle{Ultra-Fast Outflows in the Quasar PDS 456}
 \shortauthors{Rozenn Boissay-Malaquin {\em et al.}}

\def\apjl{ApJ}
\def\apjs{ApJS}
\def\ao{Appl.~Opt.}

\def\aap{A\&A}

\newcommand{\oviii}{O\,{\sc viii}}

\newcommand{\neix}{Ne\,{\sc ix}}

\newcommand{\civ}{C\,{\sc iv}}

\newcommand{\sixiv}{Si\,{\sc xiv}}
\newcommand{\sxvi}{S\,{\sc xvi}}
\newcommand{\fexxv}{Fe\,{\sc xxv}}
\newcommand{\fexxvi}{Fe\,{\sc xxvi}}
\newcommand{\nixxviii}{Ni\,{\sc xxviii}}
\newcommand{\ovii}{O\,{\sc vii}}
\newcommand{\nex}{Ne\,{\sc x}}
\newcommand{\fexvii}{Fe\,{\sc xvii}}
\newcommand{\fexviii}{Fe\,{\sc xviii}}
\newcommand{\cvi}{C\,{\sc vi}}
\newcommand{\mgxii}{Mg\,{\sc xii}}
\newcommand{\mgxi}{Mg\,{\sc xi}}
\newcommand{\sixiii}{Si\,{\sc xiii}}
\newcommand{\fexxiv}{Fe\,{\sc xxiv}}
\newcommand{\nevi}{Ne\,{\sc vi}}

\begin{document}
   \title{Relativistic components of the Ultra-Fast Outflow in the Quasar PDS 456 from Chandra/HETGS, NuSTAR and XMM-Newton observations}


   \author{Rozenn Boissay-Malaquin
          \altaffilmark{1},
          Ashkbiz Danehkar
	\altaffilmark{2},
          Herman L. Marshall
	\altaffilmark{1},
	Michael A. Nowak
	\altaffilmark{1,3}
          }
\affil{
\sp{1} Massachusetts Institute of Technology, Kavli Institute for Astrophysics, Cambridge, MA 02139, USA\\
\sp{2} Harvard-Smithsonian Center for Astrophysics, 60 Garden Street, Cambridge, MA 02138, USA\\
\sp{3} Department of Physics, Washington University, One Brookings Drive, St. Louis, MO 63130-4899, USA
}

   \email{rboissay@mit.edu}
\accepted{January 19, 2019}

\begin{abstract}
We present the spectral analysis of \textit{Chandra}/HETGS and \textit{NuSTAR} observations of the quasar PDS 456 from 2015, and \textit{XMM-Newton} and \textit{NuSTAR} archival data from 2013-2014, together with \textit{Chandra}/HETGS data from 2003. We analyzed these three different epochs in a consistent way, looking for absorption features corresponding to highly ionized blueshifted absorption lines from H-like and He-like ions of iron (and nickel), as well as of other elements (O, Ne, Si, and S) in the soft band. We confirm the presence of a persistent ultra-fast outflow (UFO) with a velocity of $v_{out}=-0.24$-$-0.29c$, previously detected. We also report the detection of an additional faster component of the UFO with a relativistic velocity of $v_{out}=-0.48c$. We implemented photoionization modeling, using XSTAR analytic model \texttt{warmabs}, to characterize the physical properties of the different kinematic components of the ultra-fast outflow and of the partial covering absorber detected in PDS 456. These two relativistic components of the ultra-fast outflow observed in the three epochs analyzed in this paper are powerful enough to impact the host galaxy of PDS 456 through AGN feedback.
\end{abstract}


   
\section{Introduction}
\label{intro}

Outflows are commonly detected in Active Galactic Nuclei (AGN), through absorption lines visible in X-rays and in UV \citep{Crenshaw2003}, with moderate velocity of hundreds to several thousands km/s. The so-called warm absorbers are found in more than 50\% of AGN \citep{Blustin2005,Piconcelli2005,
Mckernan2007}. They are thought to result from efficient accretion onto supermassive black holes, if the radiative energy exceeds the binding energy of the gas \citep{King2003,King2010}. Higher velocity ($v_{out}=-0.1$-$-0.4c$) and dense outflows ($N_{H}\sim10^{23}\text{ cm}^{-2}$) have been also observed through blueshifted iron absorption lines above 7 keV \citep{Chartas2002,Reeves2009,Tombesi2010} in about 40\% of AGN \citep{Tombesi2010,Gofford2013}, and in particular in bright and distant quasars \citep{Chartas2003, Reeves2003,Pounds2003,Lanzuisi2012,Chartas2014,Vignali2015,Dadina2018}. These are often called Ultra-Fast Outflows (UFOs). Their frequent detection suggests that they are characterized by a wide angle (e.g. \citealt{Nardini2015, Reeves2018b}). Their outflowing rates can reach several solar masses per year (up to $\sim 10^{3} M_{\odot} \text{yr}^{-1}$), for a kinetic power of $10^{45}-10^{46}$ erg s$^{-1}$ \citep{Reeves2009,Tombesi2011,Tombesi2013}.  
The link between galaxy parameters and the growth of the central supermassive black hole (SMBH), such as the black-hole mass -- velocity dispersion $M-\sigma$ relation \citep{Ferrarese2000,Gebhardt2000}, is thought to be regulated by AGN feedback processes, via these powerful winds. 
The kinetic power of UFOs, launched from the accretion disk at a few gravitational radii ($R_{g}$) from the SMBH, can be 0.5-5\% of the bolometric luminosity, and affect the AGN host galaxy, by sweeping away the galaxy's reservoir of gas and hence quench the star formation  \citep{Silk1998,King2003,King2010,DiMatteo2005,Hopkins2010,Alexander2012,Fabian2012b,
Kormendy2013,Tombesi2015,Fiore2017}. Several mechanisms have been suggested to play a role in the acceleration of these fast winds: UFOs can be radiatively driven winds (e.g. \citealt{Proga2004,Sim2010,King2010,King2015,Reeves2014,Hagino2015,Nomura2017}), thermal winds (e.g. \citealt{Begelman1983}), or magneto-hydrodynamic
(MHD) flows (e.g. \citealt{Blandford1982,Fukumura2010,Chakravorty2016,Fukumura2015,Fukumura2017,
Kraemer2017,Fukumura2018}).

PDS 456, which was first identified by \cite{Torres1997} in the Pico dos Dias survey (PDS), is the most luminous radio-quiet quasar in the local universe (z=0.184), 
with a bolometric luminosity of $L_{Bol}=10^{47}$ erg s$^{-1}$ \citep{Reeves2000} and a black-hole mass of
$M_{BH}=10^{9}M_{\odot}$ \citep{Reeves2009}.
Its initial observations taken with \textit{RXTE}, \textit{ASCA}, \textit{BeppoSAX} and then \textit{XMM-Newton} \citep{Reeves2000,Vignali2000,Reeves2002} have shown a steep absorption feature around 9 keV, as well as a rapid variability, suggesting the presence of an absorbing highly ionized outflow in the line-of-sight. The deep absorption trough detected above 7 keV has been associated with the blueshifted K-shell transition of highly ionized iron (\fexxvi\ ). Broad absorption features in \textit{XMM-Newton/RGS} data near 1 keV are additional signatures of a dense and highly ionized UFO with an extreme velocity of about $v_{out}=-0.1$-$-0.2c$ \citep{Reeves2003,Behar2009}.

This persistent UFO was also identified in \textit{Suzaku} observations, via an absorption feature near 9 keV corresponding to a Compton-thick and clumpy wind of velocity $v_{out}=-0.25$-$-0.30c$ \citep{Reeves2009}. Rapid variability of the high velocity iron K-shell absorption lines may result from wind clumpiness \citep{Gofford2014,Reeves2016}. 
The spectral variability of PDS 456 can be explained by the variations in the partial covering absorber observed in most of the data (e.g. \citealt{Reeves2014,Nardini2015,Matzeu2016a,Reeves2016}; see also the Principal Component Analysis, PCA, performed by \citealt{Parker2017} on \textit{XMM-Newton} and \textit{Suzaku} data). In this model, the clouds responsible for the partial covering have a size of 20 $R_{g}$ and  would be the denser clumps of the inhomogeneous accretion disk wind \citep{Matzeu2016b,Reeves2018b}. \cite{Luminari2018} recently determined that the $0.23c$-velocity wind may be a wide-angle outflow, according to the covering factor of 0.7 and the opening angle of 71 deg resulting from the application of an AGN wind emission model (WINE) to \textit{XMM-Newton} and \textit{NuSTAR} data.

\cite{Nardini2015} analyzed five simultaneous \textit{XMM-Newton} and \textit{NuSTAR} observations from 2013-2014 and found in all these observations a broadened Fe K emission line in addition to an absorption trough that they first fitted with individual lines and then with a P Cygni profile characteristic of a spherically symmetric expanding gas. Their analysis revealed for the first time several blueshifted absorption lines (\fexxvi\ Ly$\alpha$, Ly$\beta$ and K-edge) with the same outflow velocity $v_{out}=-0.25c$. 
\cite{Nardini2015} performed photoionization modeling of PDS 456, using the photoionization program XSTAR \citep{Kallman2001}, constrained with its Spectral Energy Distribution (SED)  based on data from UV to hard X-rays (from 2 eV to 30 keV) collected by \textit{NuSTAR} and \textit{XMM-Newton} EPIC-\textit{pn} and \textit{OM} observations from 2013-2014, approximated by a three-segment broken powerlaw \citep{Nardini2015}. The hard X-ray powerlaw has a photon index of $\Gamma=2.4$ according to \textit{XMM-Newton}/EPIC and \textit{NuSTAR} data. The \textit{OM} photometric data are well described by a slope of $\Gamma=-0.7$, the connecting slope between 10 and 500 eV being $\Gamma=3.3$. A more complex description of this SED with a multi-temperature Comptonized accretion disk model was applied by \cite{Matzeu2016b} to \textit{Suzaku} data.
 Analysis of the \textit{RGS} data was performed by \cite{Reeves2016}, who found Broad Absorption Line (BAL) profiles around 1 keV, identified as He- and H-like Neon and L-shell Iron lines blueshifted by a velocity of $v_{out}=-0.1$-$-0.2c$, which could be the signature of a lower ionization and clumpy phase of the accretion disk wind responsible for the absorption trough around 9 keV. The emission lines detected by \cite{Nardini2015}, \cite{Reeves2016} and \cite{Matzeu2017b} in \textit{Suzaku} data are likely associated with the re-emission from the outflow in PDS 456. Reflection on ionized material has also been considered in several studies, but the scenario of an ultra-fast outflow absorbing the hot corona emission was often preferred \citep{Reeves2009,Behar2009,Reeves2014,Nardini2015,Chiang2017}. 

BAL profiles are expected in the UV spectrum of PDS 456 as signatures of the fast outflowing gas detected in X-rays. \cite{OBrien2005} indeed detected a Ly$\alpha$ BAL in the UV spectrum of the \textit{Hubble Space Telescope} (\textit{HST}), blueshifted with a velocity of $v_{out}=-0.05$-$-0.08c$, that could be the signature of a decelerating cooling outflow. However, a recent analysis of the same data by \cite{Hamann2018} identified this UV BAL as a \civ\ line blueshifted with a velocity of $v_{out}=-0.3c$, similar to the one measured in X-ray data. Such BAL could come from dense and low-ionization clumps embedded in the X-ray UFO.

Considering the fact that the supermassive black hole in PDS 456 is accreting at about the Eddington limit, its detected UFO should be radiatively driven \citep{Matzeu2017a,Nardini2015}, supported by a correlation between the velocity of the outflow and the ionizing luminosity, with UV line driving strongly contributing to the wind acceleration \citep{Hagino2015,Hamann2018,Reeves2018b}. However, a MHD-wind model has also been successfully applied recently to \textit{XMM-Newton} and \textit{NuSTAR} data of PDS 456 by \cite{Fukumura2018}. 

Despite the fact that PDS 456 is a radio-quiet quasar, it has been observed recently by the European VLBI Network (EVN) at 5 GHz \citep{Yang2019}. This observation revealed a radio structure made of two components, faint, diffuse and separated by about 20 pc, that could either be the signature of a recent jet, or the radio-emission of an outflow launched in the vicinity of the central supermassive black-hole. In the latest hypothesis, the radio-emission could originate from shocks produced by the interaction of the known powerful mildly-relativistic X-ray outflow and the surrounding material.

Recent observations with \textit{XMM-Newton} and \textit{NuSTAR} in 2017 have shown the presence of an additional faster UFO with a velocity of $v_{out}=-0.46c$, identified from a deep absorption line around 11 keV, that may be part of the multiple velocity components of the ultra-fast outflow around the accretion disk, launched from a radius of about 10$R_{g}$, and may be visible only during the lowest states \citep{Reeves2018b,Reeves2018}. In addition to the two fast components of the UFO, a less ionized soft absorber with a variable covering factor was detected, probably originating from denser clumps further out along the stratified outflow \citep{Reeves2018b}. Simultaneous UV data from \textit{HST} did not show significant absorption signature as the X-ray absorbers might be too ionized to be visible in this energy band.

In the present paper, we aim to analyze multiple epochs of PDS 456, namely \textit{Chandra}/HETGS data (from 2003 and 2015), \textit{NuSTAR} data (from 2015), and \textit{XMM-Newton} and \textit{NuSTAR} data from 2013-2014, to look for signatures of the persistent UFO previously detected and of the faster UFO recently identified, and constrain their physical properties. We describe the data, their selection and their reduction in section \ref{red}. We then present our dual-approach spectral analysis. We first explain the combination and binning of the data in section \ref{bin} and the continuum fitting in section \ref{cont}. We next present a model-independent analysis, involving the modeling of the absorption features at high energy with gaussian lines (section \ref{Fe}) and with P Cygni profiles (section \ref{pcyg}). 
We finally present in section \ref{warmabs} the model-dependent approach that consists in photoionization modeling, before discussing the results in section \ref{discussion} and conclude in section \ref{conclusion}.

\section{Data}
\label{red}
\subsection{Data selection}

\begin{table*}[!h]
\begin{center}
 
\caption{Observation log} 
\label{Obs}
\begin{tabular}{|c|lcccccccc|}
\hline
Label & Instrument & Obs. ID & Obs. Start & Obs. End & T$_{\text{tot}}$ &  T$_{\text{net}}$ & Channel &  Count rate & Flux \\
\hline
\multirow{6}{*}{CN} & \multirow{2}{*}{Chandra/HETGS} & \multirow{2}{*}{17452} & \multirow{2}{*}{2015-07-21, 12:52} & \multirow{2}{*}{2015-07-23, 03:48} & \multirow{2}{*}{138} & \multirow{2}{*}{136} & HEG  &  0.098 & 5.16 $\pm$ 0.07\\
 & & & & & & & MEG &  0.100& 6.00 $\pm$ 0.12\\
 \cline{2-10}
 & \multirow{4}{*}{NuSTAR} & \multirow{2}{*}{90101008002} & \multirow{2}{*}{2015-07-21, 11:01} & \multirow{2}{*}{2015-07-23, 07:46} & \multirow{2}{*}{160} &  \multirow{2}{*}{74} & FPMA &  0.048& 4.63 $\pm$ 0.14\\
  &  & & & & && FPMB  &0.044& 4.51 $\pm$ 0.16\\
   &  & \multirow{2}{*}{90101008004} & \multirow{2}{*}{2015-07-24, 11:36} & \multirow{2}{*}{2015-07-25, 10:51}  & \multirow{2}{*}{83} & \multirow{2}{*}{38} &FPMA &  0.044& 4.75 $\pm$ 0.81\\
  &  & &  & &  & & FPMB    & 0.042& 4.47 $\pm$ 0.20\\
\hline
\multirow{2}{*}{C} & \multirow{2}{*}{Chandra/HETGS} & \multirow{2}{*}{4063} & \multirow{2}{*}{2003-05-07, 03:29} & \multirow{2}{*}{2003-05-08, 20:08} & \multirow{2}{*}{145} & \multirow{2}{*}{143} & HEG  & 0.070&  3.85 $\pm$ 0.06\\
 & & & & & & & MEG   & 0.073& 4.23 $\pm$ 0.07 \\
 \hline
 
 \multirow{3}{*}{XN1} & \textit{XMM-Newton} & 0721010201 & 2013-08-27, 04:41 & 2013-08-28, 11:13 & 110 & 85.5 & EPIC/pn & 3.175 &10.87 $\pm$ 0.04  \\
 \cline{2-10}
 & \multirow{2}{*}{NuSTAR} & \multirow{2}{*}{60002032002} & \multirow{2}{*}{2013-08-27, 03:41} & \multirow{2}{*}{2013-08-28, 11:41} & \multirow{2}{*}{114} & \multirow{2}{*}{44} & FPMA &  0.134& 9.70 $\pm$ 0.22 \\
 & & & & & & & FPMB &  0.128& 9.50 $\pm$ 0.23\\
 \hline
 \multirow{3}{*}{XN2} & \textit{XMM-Newton} & 0721010301 & 2013-09-06, 03:24 & 2013-09-07, 10:36 & 112 & 92.1 & EPIC/pn &   2.098 & 5.63 $\pm$ 1.39 \\
 \cline{2-10}
 & \multirow{2}{*}{NuSTAR} & \multirow{2}{*}{60002032004} & \multirow{2}{*}{2013-09-06, 02:56} & \multirow{2}{*}{2013-09-07, 10:51} & \multirow{2}{*}{114} & \multirow{2}{*}{43} & FPMA &  0.051& 3.56 $\pm$ 0.38\\
 & & & & & & & FPMB & 0.049& 3.61 $\pm$ 0.17\\
 \hline
 \multirow{3}{*}{XN3} &XMM-Newton & 0721010401 & 2013-09-15, 18:47 & 2013-09-17, 03:57 & 119 & 102.0 & EPIC/pn & 1.974 &  6.08 $\pm$ 0.02  \\
 \cline{2-10}
 & \multirow{2}{*}{NuSTAR} & \multirow{2}{*}{60002032006} & \multirow{2}{*}{2013-09-15, 17:56} & \multirow{2}{*}{2013-09-17, 04:01} & \multirow{2}{*}{119} & \multirow{2}{*}{44} & FPMA & 0.072& 5.20 $\pm$ 0.18\\
 & & & & & & & FPMB &  0.068& 4.59 $\pm$ 0.18\\
  \hline
 \multirow{3}{*}{XN4} & \textit{XMM-Newton} & 0721010501 & 2013-09-20, 02:47 & 2013-09-21, 09:37 & 111 & 92.9 & EPIC/pn & 1.925  & 6.35 $\pm$ 0.03 \\
 \cline{2-10}
 & \multirow{2}{*}{NuSTAR} & \multirow{2}{*}{60002032008} & \multirow{2}{*}{2013-09-20, 03:06} & \multirow{2}{*}{2013-09-21, 11:11} & \multirow{2}{*}{119} & \multirow{2}{*}{44} & FPMA &0.075& 5.18 $\pm$ 0.16 \\
 & & & & & & & FPMB & 0.073& 5.38 $\pm$ 0.17\\
  \hline
 \multirow{3}{*}{XN5} & \textit{XMM-Newton} & 0721010601 & 2014-02-26, 08:03 & 2014-02-27, 22:51 & 140 & 103.9 & EPIC/pn & 1.383&  4.47 $\pm$ 1.38 \\
 \cline{2-10}
 & \multirow{2}{*}{NuSTAR} & \multirow{2}{*}{60002032010} & \multirow{2}{*}{2014-02-26, 08:16} & \multirow{2}{*}{2014-02-28, 22:56} & \multirow{2}{*}{224} & \multirow{2}{*}{110} & FPMA &  0.045& 3.01 $\pm$ 0.53\\
 & & & & & & & FPMB &  0.043& 3.00 $\pm$ 0.58\\
 \hline
 
\end{tabular}
\end{center}

 Summary of PDS 456 \textit{Chandra}, \textit{XMM-Newton} and \textit{NuSTAR} observations used in this study. The observation ``CN'' considers the contemporaneous observations with \textit{Chandra}/HETGS and \textit{NuSTAR} from 2015, while observation ``C'' refers to the archival \textit{Chandra}/HETGS observation from 2003. Observations ``XN1'' to ``XN5'' refer to \textit{XMM-Newton} and \textit{NuSTAR} observations performed in 2013-2014 and previously analyzed by \cite{Nardini2015}. We refer to the combination of observations ``XN1'' to ``XN5'' as observation ``XN''. Observation start and end times are in UT. T$_{\text{tot}}$ (in ks) is the total elapsed time, while T$_{\text{net}}$ (in ks) is the net exposure after screening and deadtime correction. The count rates (in $s^{-1}$) are the background subtracted count rates. The fluxes (in $10^{-12}$ erg cm$^{-2}$ s$^{-1}$) refer to the 0.8-7 keV (HEG), 0.4-7 keV (MEG), 0.4-10 keV (EPIC/pn) and 3-30 keV (FPMA/FPMB) bands, obtained using the ISIS function \texttt{data\_flux()} that calculates the absorbed X-ray flux solely from the spectral data, and estimates errors from the data and information from the instrumental responses.

\end{table*}

\begin{figure} [!h]
\resizebox{\hsize}{!}{\includegraphics[trim = 15mm 10mm 0mm 0mm, clip, angle=0]{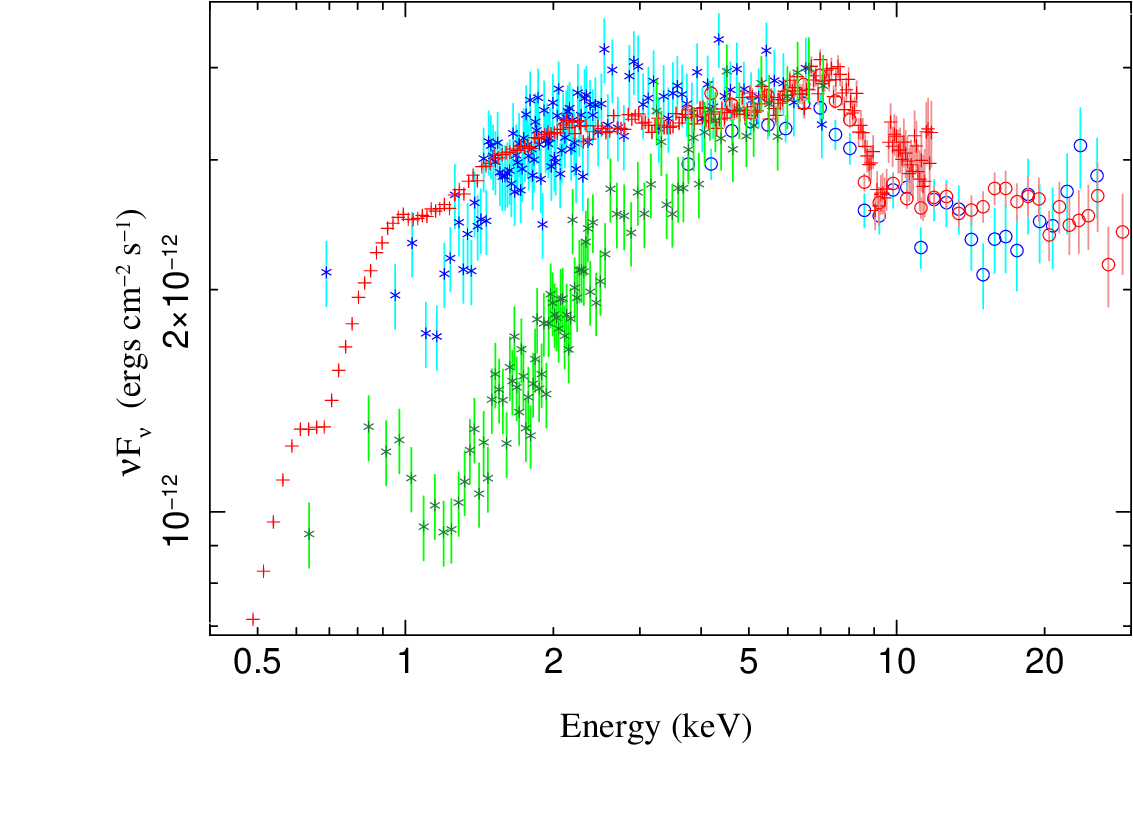}} 
\caption{Unfolded spectra of the three observations studied in this paper (see Table \ref{Obs}). Blue: observation ``CN'', Red: observation ``XN'', Green: observation ``C''. Symbols: ``x'' for HETGS data, ``o'' for \textit{NuSTAR} data, and ``+'' for EPIC/pn data. Note that the spectra have been strongly rebinned for plotting purpose.}
\label{AllObs}
\end{figure}

The aim of this work is to look for signatures of UFOs in PDS 456. For this purpose, we chose to focus on three different datasets, as presented on Table \ref{Obs}. Strongest signatures of UFOs are found at high energy, above 7 keV, as found in particular in PDS 456 according to previous studies. We thus selected observations with \textit{NuSTAR} data, therefore \textit{XMM-Newton} and \textit{NuSTAR} data from 2013-2014 (observations ``XN''), and \textit{Chandra} and \textit{NuSTAR} data from 2015 (observation ``CN''). Other signatures of the UFOs are expected at lower energy, so \textit{Chandra}/HETGS is an excellent instrument to look for such features. We also considered \textit{Chandra}/HETGS data from 2003 (observation ``C''), data that, together with ``CN'' data, have never been analyzed in detail. Despite the lack of high energy data for this ``C'' observation, our analysis performed consistently along the different datasets allowed us to constrain the winds parameters in this observation as well. For observations ``XN'', we focused on EPIC/\textit{pn} data, mostly to look for high energy signatures of the UFOs. We did not re-analyze \textit{RGS} data, already studied in details by \cite{Reeves2016}, as we focus on high velocity and high ionization outflows, whose signatures are expected to be outside the energy range of \textit{RGS}. 
The scope of the present paper is to look for signatures of UFOs at both high and low energies, so we did not examine other archival data.

\subsection{Data reduction}
PDS 456 was observed with the High Energy Transmission Grating Spectrometer (HETGS; \citealt{Markert1994,Canizares2005}) using the \textit{Chandra} 
Advanced CCD Imaging Spectrometer (ACIS; \citealt{Garmire2003}), from 2015 July 21 to July 23 (observation ``CN'', 138~ks of exposure time, see Table \ref{Obs}) and from 2003 May 07 to May 08 (observation ``C'', 145~ks of exposure time, see Table \ref{Obs}). The HETGS spectrometer is composed of two grating types: the medium energy gratings (MEGs), covering the 0.4-7~keV energy band, with a full-width at half-maximum (FWHM) resolution of 0.023 {\AA}, and the high energy gratings (HEGs), having a FWHM resolution of 0.012 {\AA}  in the 0.8-10~keV band. 
Data processing was performed with the TGCat software \citep{Huenemoerder2011}, 
which employs \textit{Chandra} Interactive Analysis of Observations tools (CIAO v.4.8; \citealt{Fruscione2006}) and Calibration Data base (CALDB v.4.8.0). \textit{Chandra} HETGS data for both periods were reduced in a standard way, using a narrow mask to avoid mask confusion above 6 keV. Plus and minus first-order ($m=\pm1$) MEG and HEG data were extracted from the -1 and the +1 arms of the MEG and HEG gratings,  for the source and the background for both observations, using the CIAO tool \texttt{tgextract}. Spectral redistribution matrix files (RMF) and effective area files (ARF) were generated with \texttt{mkgrmf} and \texttt{mkgarf}. 

PDS 456 was simultaneously observed by \textit{NuSTAR} during the observation ``CN'' of \textit{Chandra}, from 2015 July 21 to July 23 (74~ks exposure) and from July 24 to July 25 (38~ks exposure). \textit{NuSTAR} data were processed using the NuSTARDAS v1.7.1. The source spectra were extracted using a 45'' circular region centered on the source, and the background from a 45'' circular region clear of stray light, in the same detector, for both focal plane modules A and B (FPMA and FPMB). 
Observation reports depict a notable solar activity, which may impact the background event rate,  during both \textit{NuSTAR} observations. Passage of Solar Coronal Mass Ejections (CMEs) over the Earth induces a temporary increase in the low Earth orbit radiation environment (which can persist for many orbits) and can significantly increase the background event level in the detectors of \textit{NuSTAR} when the observatory is close to the South Atlantic Anomaly (SAA). We thus optimized the screening out of high-background periods near SAA passages using the module \texttt{nucalcsaa}. 

PDS 456 was observed in 2013-2014 by \textit{NuSTAR} and by \textit{XMM-Newton}, in five observations labeled XN1 to XN5 in Table \ref{Obs} (see \citealt{Nardini2015}). We reduced the \textit{NuSTAR} observations as described above, filtering the SAA passages when required, in order to compare our results to previous observations. We also reduced the \textit{XMM-Newton} original data files using the \textit{XMM-Newton} Standard Analysis Software (SAS v16.0.0 - \citealt{Gabriel2004}) considering the EPIC/pn \citep{Struder2001} spectrum for each observation of PDS 456. Events corresponding to flaring particle background were filtered using the SAS standard procedure. Single and double events were selected for extracting spectra. The data were screened for any increased flux of background particles. Spectra were extracted from a circular region of 30'' centered on the source. We checked for pile-up in all observations. The background was extracted from a nearby source-free region of 40'' in the same CCD as the source. Response matrices were generated for each source spectrum using the SAS \texttt{arfgen} and \texttt{rmfgen} tasks.

\section{Spectral analysis}
\label{anal}
We performed our dual-approach spectral analysis of the data listed in Table \ref{Obs} using the Interactive Spectral Interpretation System (ISIS; version 1.6.2-40, \citealt{Houck2002}). We analyzed the simultaneous data from \textit{Chandra}/HETGS and \textit{NuSTAR} from 2015 (observation ``CN''), the simultaneous time-averaged \textit{XMM-Newton} and \textit{NuSTAR} observation (combined data ``XN1'' to ``XN5'') and the archival \textit{Chandra}/HETGS observation from 2003 (observation ``C'') in a similar way, adapting the procedure to the available datasets, as described in the following. After describing the combination and binning of our data in section \ref{bin}, and the continuum modeling in section \ref{cont}, we performed a model-independent analysis of the selected observations, looking at individual features at high energy (sections \ref{Fe} and \ref{pcyg}) and in the broadband spectra (blind line search described in appendix). We then performed a model-dependent analysis using photoionization modeling (section \ref{warmabs}). This dual-method involving both model-dependent and model-independent analyses is an accurate way to check that the results found via different procedures are consistent and thus reliable. In this paper, all errors are quoted at 1$\sigma$ confidence level. All figures are presented in rest frame energy. Figure \ref{AllObs} shows the spectra of the three observations (coarsely rebinned for presentation purpose), and illustrates the changes in fluxes and shapes between the different epochs, as already noticed in previous studies (e.g. \citealt{Matzeu2017a}). 
\newpage
\subsection{Combination and binning}
\label{bin}

\begin{figure} [!h]
\resizebox{\hsize}{!}{\includegraphics[trim = 10mm 10mm 0mm 0mm, clip, angle=0]{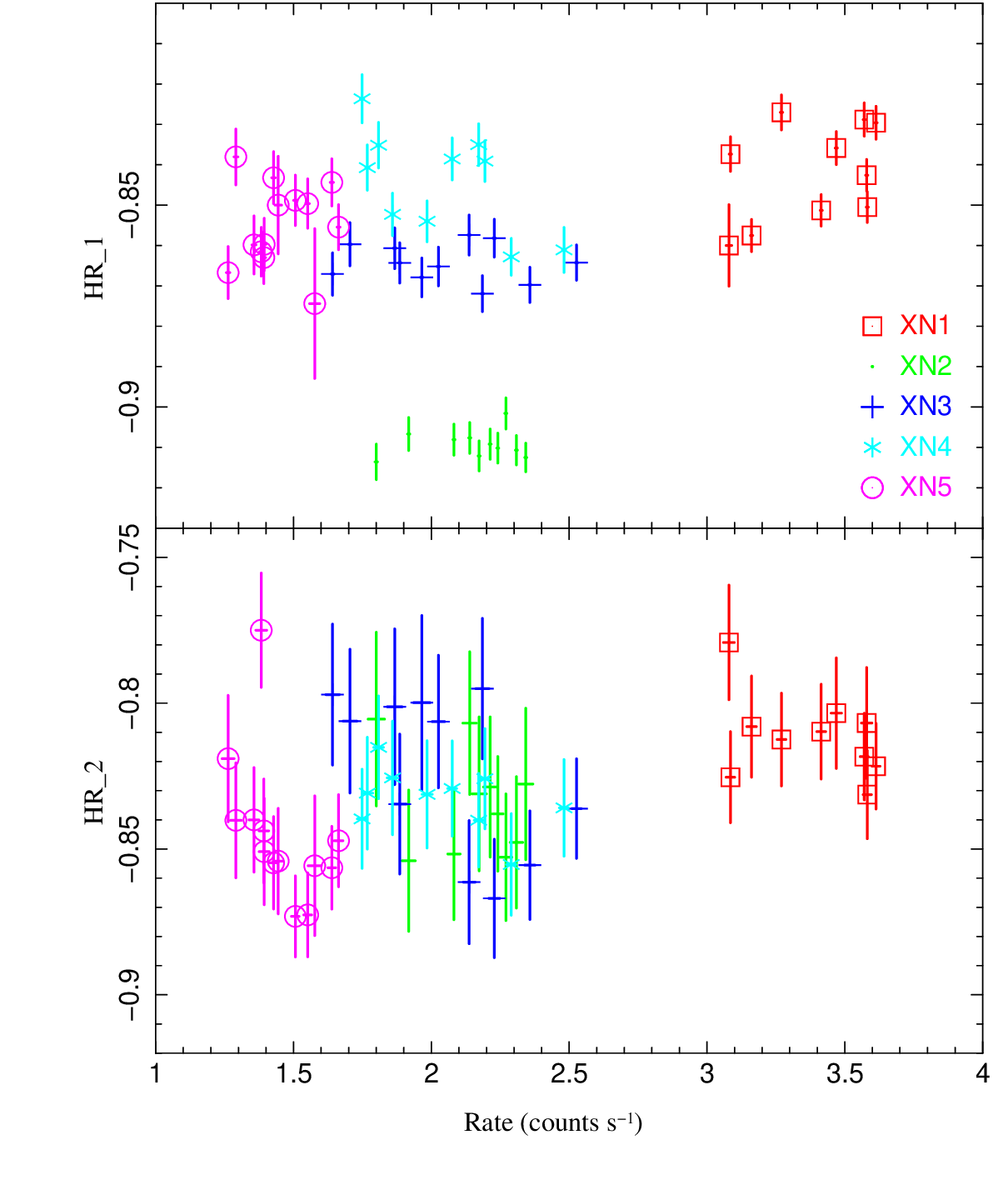}} 
\caption{The hardness ratios, defined in equations (1) and (2), for the five ``XN'' observations, plotted against the 0.5-10 keV EPIC/pn light curve. $HR_{1}$ is sensitive to the overall continuum shape; the top panel shows that this shape is consistent for all ``XN'' observations but the second, which is slightly softer.  $HR_{2}$ is sensitive to the slope of the spectrum around the 8-15 keV region and is generally consistent between these ``XN'' observations.}
\label{HR}
\end{figure}

In order to increase the signal-to-noise ratio and because of the complex shape of the broadband spectrum around 9 keV (as described in Section \ref{intro}), we combined the data to perform the spectral analysis, using the \texttt{combine\_datasets} function in ISIS. For the observation ``CN'' (see Table \ref{Obs}), we put the HEG data on the same grid as MEG and combined all HETGS spectra (HEG-1, HEG+1, MEG-1, MEG+1), and we also combined \textit{NuSTAR} FPMA and FPMB data of both observations 90101008002 and 90101008004. We used a combination of binning by signal-to-noise and channel binning, in order to get an adequate signal-to-noise even below 1 keV (as discussed by \citealt{Ash2017}). We have thus binned \textit{Chandra} and \textit{NuSTAR} data with a minimum of 1 channel per bin (binned to Half Width Half Maximum of MEG resolution) and a minimum signal-to-noise ratio of 2. We took into account the background. The same binning of \textit{Chandra} data has been used for observation ``C''. As our ``C'' and ``CN'' observations have a Poisson distribution, we used Cash statistics \citep{Cash1979} to analyze these data. For the estimation of the error bars and the evaluation of the significance of our results, we used a $\Delta$C value appropriate for a likelihood estimate with Poisson statistics (corresponding to a given confidence level and a given number of degrees of freedom) that is analogous to $\Delta\chi^{2}$ for the case of Gaussian statistics (see for example the statistical textbook from \citealt{Breiman1973}, works from \citealt{Cash1979} and \citealt{Wilks1938}, and the pedagogical discussions in \citealt{Arnaud2011}).

For the simultaneous \textit{XMM-Newton} and \textit{NuSTAR} dataset, we combined the \textit{pn} spectra of the 5 observations ``XN1'' to ``XN5'', as well as the FPMA and FPMB spectra. We refer to this combined dataset as ``XN''. Despite a significant variability of the shape of the spectra described by \cite{Nardini2015}, this combination of data is possible because of compatible fluxes for each observation, and because the absorption through around 9 keV is detected in all observations with similar parameters. We calculated the hardness ratios of the five observations (see Fig. \ref{HR}), using count rates $CR$ in different energy bands: 
\begin{equation}
\text{HR}_1=\frac{CR^{\text{ XMM}}_{\text{ 5-10 keV}} - CR^{\text{ XMM}}_{\text{ 0.5-4 keV}}}{CR^{\text{ XMM}}_{\text{ 5-10 keV}} + CR^{\text{ XMM}}_{\text{ 0.5-4 keV}}}
\end{equation} 
and 
\begin{equation}
\text{HR}_2=\frac{CR^{\text{ NuSTAR}}_{\text{ 15-30 keV}} - CR^{\text{ XMM}}_{\text{ 3-8 keV}}}{CR^{\text{ NuSTAR}}_{\text{ 15-30 keV}} + CR^{\text{ XMM}}_{\text{ 3-8 keV}}}
\end{equation} 
in order to study the possible impact of spectral changes among the observations. By their definitions, $\text{HR}_1$ gives information about spectral variability regarding the whole EPIC/pn energy band, while $\text{HR}_2$ represents the variability of the continuum shape around the absorption features at 9 and 11 keV. 
In the top panel of Figure \ref{HR}, the hardness ratio of the second ``XN'' observation is slightly lower, due to a small change in the spectral shape of the EPIC/pn spectrum, as shown in Figure 1 from \cite{Nardini2015}. However, by comparing the spectra resulting from the sum of ``XN'' observations including and excluding this second observation, we didn't notice any significant change in the spectral shape. In the bottom panel of Figure \ref{HR}, we see that the hardness ratio representative of the variability around the absorption features is consistent among the five ``XN'' observations. The hardness ratios presented in Fig. \ref{HR} show that all ``XN'' observations are consistent spectrally and can be combined for our analysis. The stability of $\text{HR}_2$ shows that the absorption features around 9 and 11 keV are not due to changes of the continuum. We thus combined the data in order to increase the signal-to-noise, and to get an average shape of the P Cygni-like profile detected in individual observations, to be compared with the ``CN'' observation. We binned \textit{NuSTAR} data to a minimum of 2 channels per bin and a minimum signal-to-noise of 2, and \textit{XMM-Newton} data to a minimum of 1 channel per bin and a minimum signal-to-noise of 5 (during the analysis described in the following sections, we checked that re-binning the data to a minimum of 5 channels per bin does not change the results on the absorption lines). The binning of these data is sufficient to get enough counts per channel to allow us to use $\chi^{2}$ statistics for the analysis.


\subsection{Continuum modeling}
\label{cont}

\begin{figure} [!h]
\resizebox{\hsize}{!}{\includegraphics[trim = 15mm 10mm 0mm 0mm, clip, angle=0]{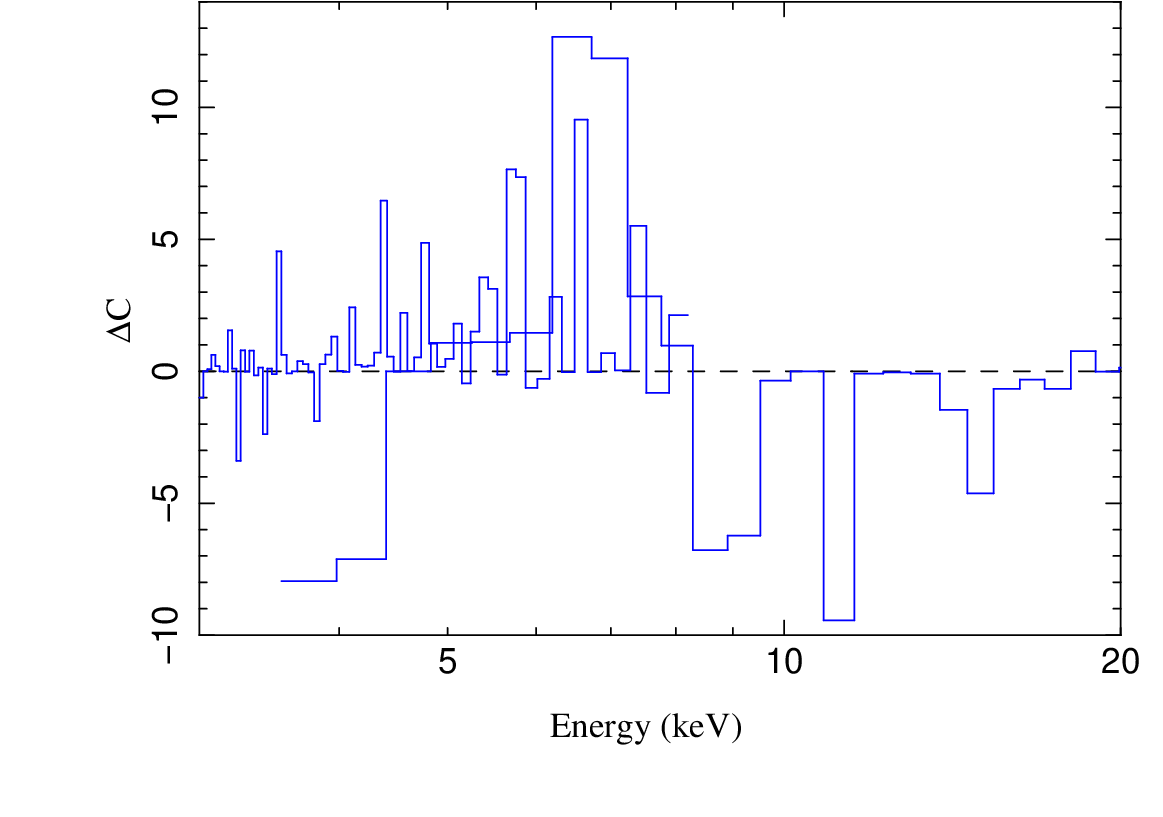}} 
\resizebox{\hsize}{!}{\includegraphics[trim = 16.5mm 10mm 0mm 0mm, clip, angle=0]{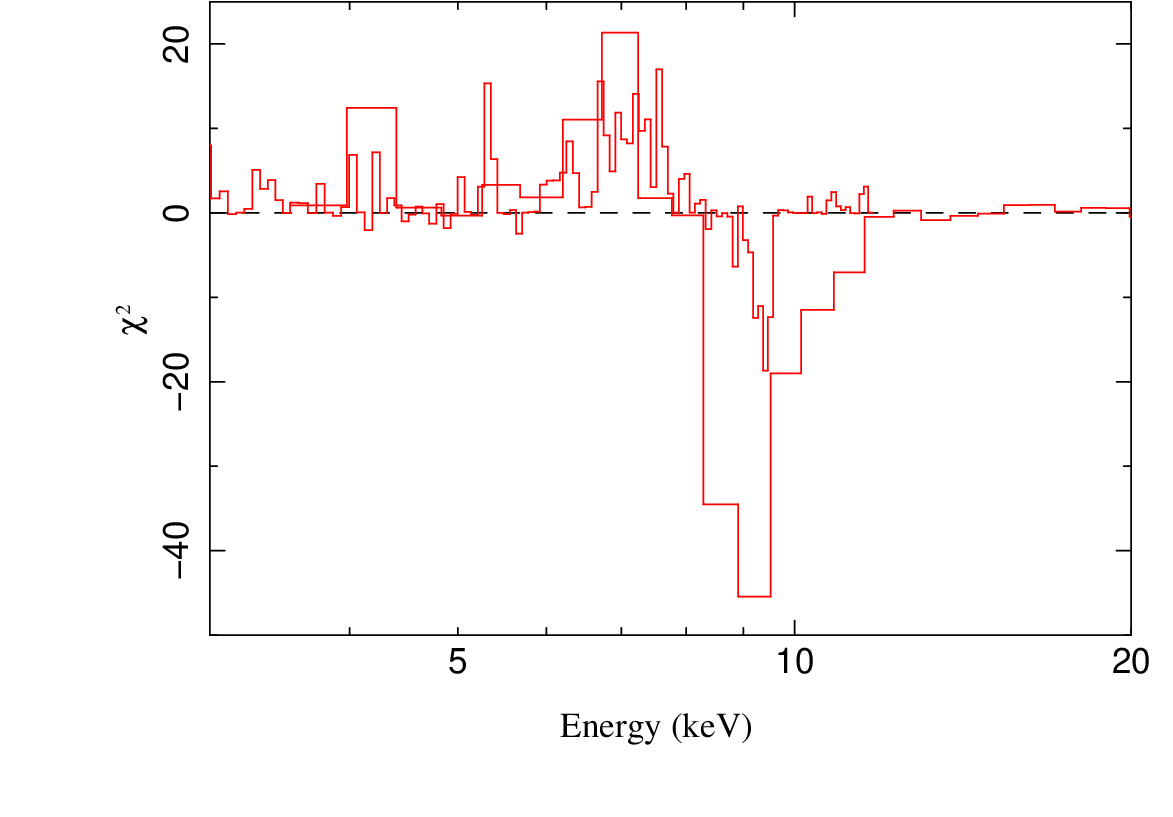}} 
\caption{Ratio between the data and their corresponding continuum model, between 3 and 20 keV. Top panel (blue): data-to-model ratio for observation ``CN'', expressed as $\Delta$C. Bottom panel (red): data-to-model ratio for observation ``XN'', expressed as $\chi^{2}$. Note that the spectra have been strongly rebinned for plotting purpose.}
\label{ResObs}
\end{figure}

To perform a precise analysis of the combined datasets described above, in particular the absorption features, we needed to determine the continuum carefully. Following the modeling of the continuum performed in previous studies, and in particular by \cite{Nardini2015}, on the same \textit{XMM-Newton} and \textit{NuSTAR} data, we first determined the continuum for observation ``CN''. We thus first fitted the hard energy band above 3 keV with a simple powerlaw, taking into account Galactic absorption (with $N_{H}^{\text{gal}}$ fixed to $2.4\times10^{21}\text{cm}^{-2}$ according to 21-cm measurements, \citealt{Dickey1990,Kalberla2005}), modeled with the \texttt{tbabs} absorption model \citep{Wilms2000}. We used a cross-calibration factor to allow to fit simultaneously data from different satellites, however the photon index of the powerlaw was the same. The fit resulted in a C statistic of 472.8 for 364 dof.
In order to account for the spectral curvature as found in previous studies between 2 and 5 keV (e.g. \citealt{Behar2009,Reeves2009,Turner2009,Reeves2014,Reeves2018b}), we added a partial covering absorption to the powerlaw, using the \texttt{zpcfabs} model, which improved the fit significantly, with a $\Delta$C of 15.4 (for two parameters of interest) and a ftest F-value of 4.05 and p-value of 0.007. This partial covering by a moderately-ionized absorber could well describe the continuum spectral variability, due to  patchy obscuration e.g. in NGC 5548 \citep{Kaastra2014} or NGC 3516 \citep{Turner2011}.
We also checked for the presence of neutral distant reflection by replacing the simple powerlaw with a \texttt{pexmon} model. However, the improvement on the fits is less significant, with a $\Delta$C of 11.1 (for three free parameters), a F-value of 2.97 but a p-value of 0.03.
Furthermore, the spectra of our datasets do not show an excess around 30 keV or a fluorescence Fe K$\alpha$ line at 6.4 keV.
The spectral complexity of our datasets above 3 keV is well illustrated by the data-to-model ratios of observations ``CN'' and ``XN'' on Figure \ref{ResObs}. 

\begin{figure*} [!h]
\begin{tabular}{cc}
\includegraphics[trim = 18mm 10mm 0mm 0mm, clip, scale=0.45,angle=0]{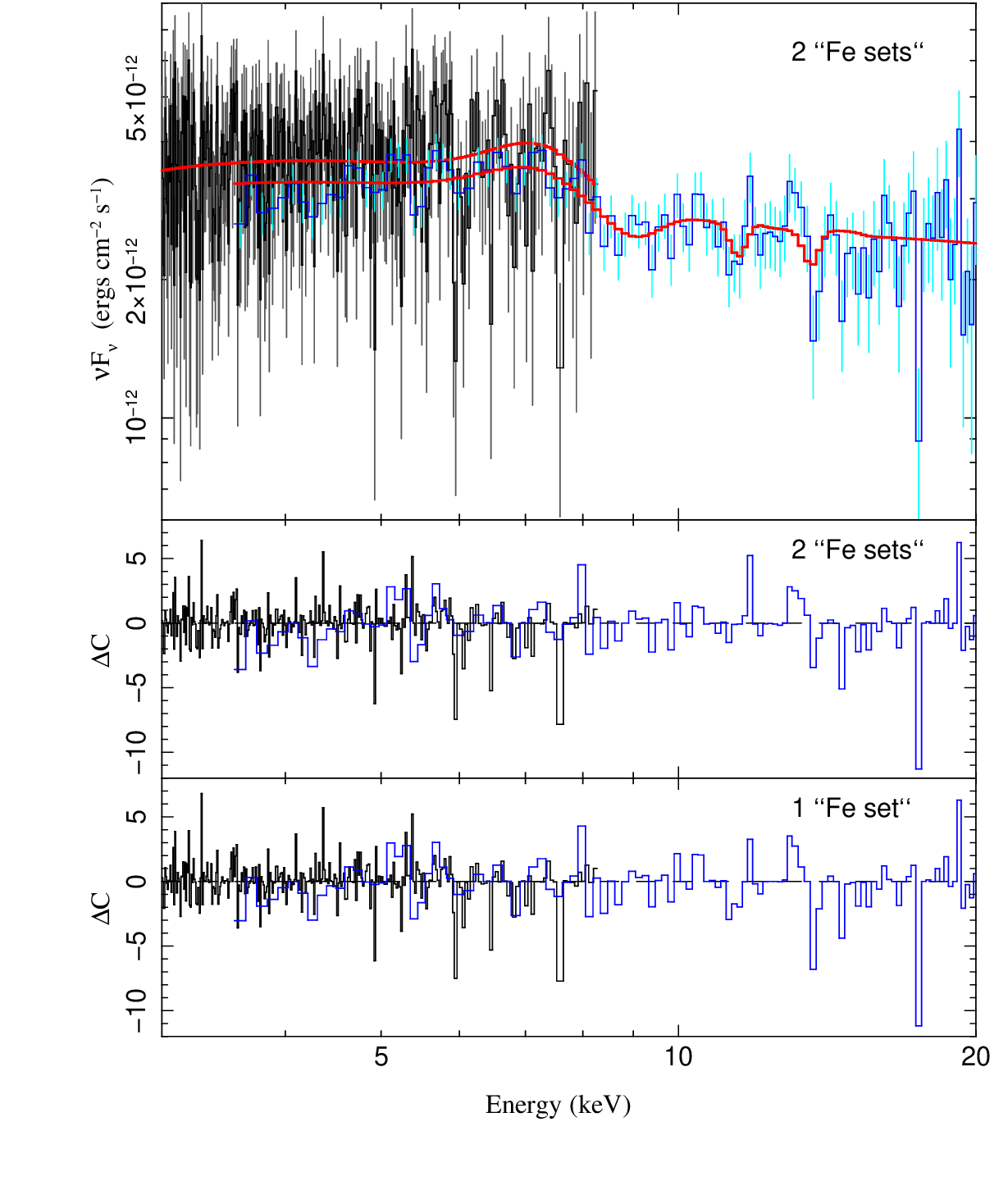} & \includegraphics[trim = 18mm 10mm 0mm 0mm, clip, scale=0.45, angle=0]{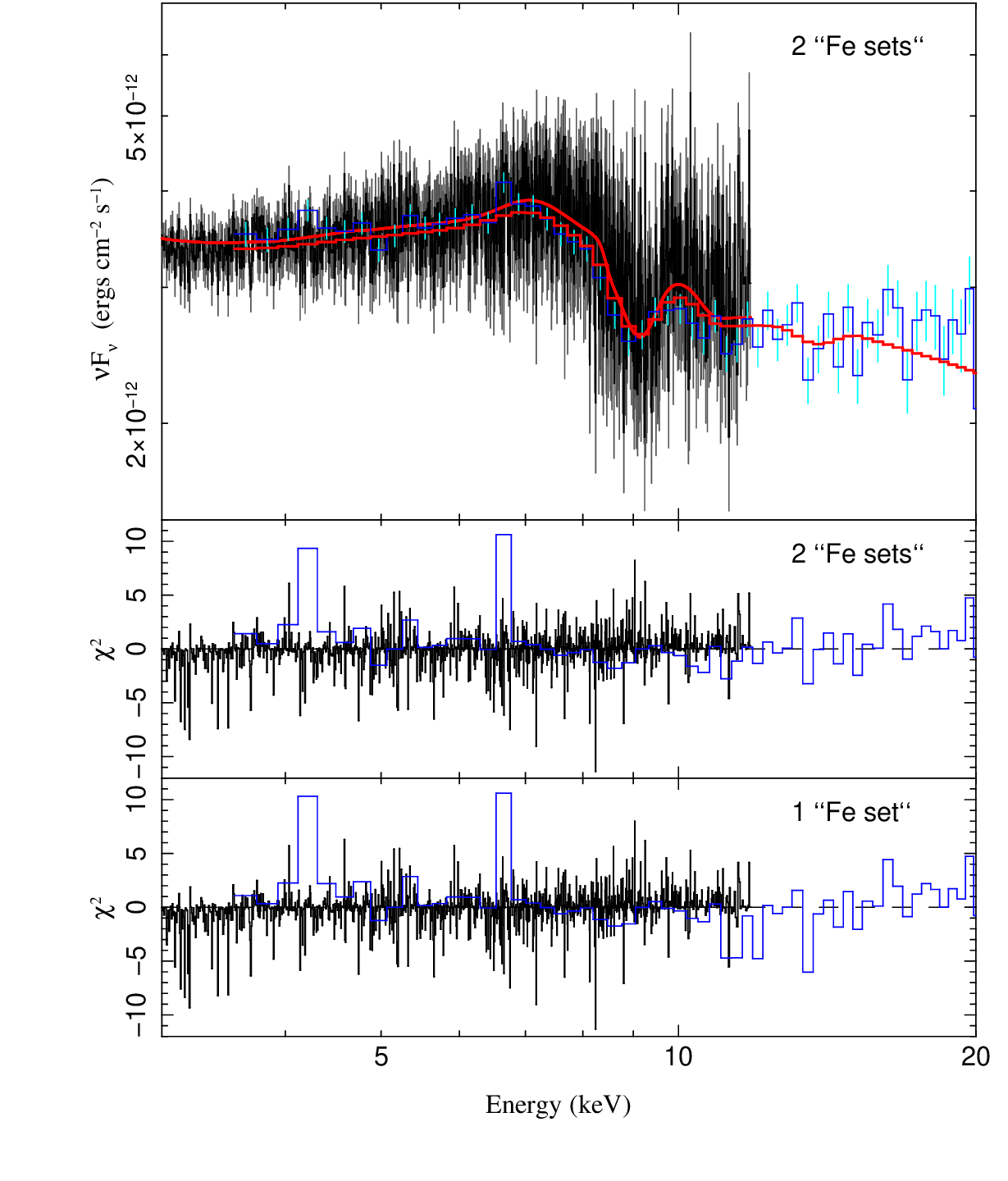} 
\end{tabular}
\caption{Unfolded hard spectra and residuals of observations ``CN'' (left) and ``XN'' (right) fitted above 3 keV with two sets (top and middle panels) or one (bottom panel) set of Fe lines (corresponding to two UFOs or one UFO, as described in Section \ref{Fe}). Especially note the improvement on the residuals above 10 keV, when taking into account the presence of the second UFO.}
\label{FeFig}
\end{figure*}

When adding the soft part of the spectra and extrapolating this continuum model fitted in hard X-rays, we clearly detect a soft excess below 1 keV, as well as strong emission and absorption features, mostly broadened, especially around 1 keV and above 7 keV (QSO frame energy). 
We modeled the soft-excess with a broad Gaussian emission line (\texttt{zgauss}), that is a phenomenological model. We chose to fit the soft-excess with a broad line to be consistent with the analysis from \cite{Nardini2015} (as we were re-analyzing the same data), and because using a more usual black-body component instead did not improve the fit. The resulting continuum model is:

\begin{center}
\texttt{constant*tbabs*zpcfabs*(powerlaw+zgauss)} 
\end{center}

As the continuum model of observation ``CN'' is totally consistent with the analysis of \textit{XMM-Newton} and \textit{NuSTAR} data from \cite{Nardini2015}, we applied it to our combined ``XN'' data. In order to analyze all our observations in a similar way, we also applied this model to our ``C'' observation, for which the continuum is determined on a smaller energy band.
The parameters of the broadband fitting with this continuum model are given in Table \ref{ContPars}. 

\begin{table}[!h]
\begin{center}
\caption{Continuum parameters}
\label{ContPars} 
\begin{tabular}{|l|ccc|}
\hline
Parameters & Obs. CN & Obs. XN & Obs. C\\
\hline
\underline{Powerlaw:} &&&\\
Photon Index & $2.30 _{-0.012}^{+0.0048}$ & $2.33_{-0.008}^{+0.009}$ & $2.47 _{-0.01}^{+0.01}$\\
$F_{7-30 keV}$ & $1.28_{-0.011}^{+0.011}$ & $1.24_{-0.001}^{+0.001}$ & $1.27_{-0.013}^{+0.013}$\\
$F_{0.4-30 keV}$ & $12.27_{-0.10}^{+0.10}$ & $13.31_{-0.0015}^{+0.0015}$ & $16.49_{-0.16}^{+0.16}$\\
\hline
\underline{Partial covering:} &&&\\
$N_{H}$ & $3.03  _{-0.53}^{+0.46}$ & $8.46_{-0.42}^{+0.49}$ & $3.42 _{-0.11}^{+0.14}$\\
Covering factor & $0.32  _{-0.006}^{+0.009}$ & $0.33_{-0.009}^{+0.009}$& $0.75 _{-0.004}^{+0.004}$\\
\hline
\underline{Soft-excess:}&&&\\
E & $0.060 _{-0.01}^{+0.008}$& $0.69_{-0.007}^{+0.007}$ & $0.14 _{-0.03}^{+0.05}$\\
$\sigma$ & $0.10 _{-0.0017}^{+0.0017}$ & $0.19_{-0.004}^{+0.004}$& $0.06 _{-0.006}^{+0.003}$\\
$F_{0.4-30 keV}$ & $65.59_{-15.45}^{+13.67}$& $0.97_{0.007}^{+0.007}$ & $9.43_{-2.87}^{+3.44}$\\
\hline
Cross-calibration & $0.90 _{-0}^{+0.005}$ & $0.97_{-0.007}^{+0.007}$& -\\
\hline
C or $\chi^{2}$ / dof & 1693.48 / 1609 & 2240.58 / 1704  & 1409.60 / 1331 \\
\hline
\end{tabular}

\end{center}
Parameters of the fit of the continuum, performed above 0.4 keV, for the three observations, as described in Section \ref{cont}. Fluxes are in units of $10^{-12}$ erg s$^{-1}$ cm$^{-2}$.
\end{table}

\begin{table*}[!h]
\begin{center}
\caption{Fits performed above 3 keV with one set of Fe lines}
\begin{tabular}{|ll|ccc|ccc|}
\hline
\multicolumn{2}{|l|}{Parameters} & \multicolumn{3}{c|}{Obs. CN} & \multicolumn{3}{c|}{Obs. XN} \\
\hline
\hline
\multicolumn{2}{|l|}{Line} & $E_{rest}$ (keV) &  EW (eV) or $\tau$ & $\Delta$C / $ \Delta$dof & $E_{rest}$ (keV) &  EW (eV) or $\tau$ & $\Delta\chi^{2}$ / $\Delta$dof \\
\hline
\multirow{2}{*}{\fexxvi\ Ly$\alpha$} & em. & $7.04$ &	 $233_ { - 46}^{+46}$ & 28.28 / 3 & $7.13$& 	 $91_{ - 16}^{+16}$ & 96.50 / 3\\ 
& abs. & $9.16$ & 	 $259_{- 71}^{+71}$ & 15.55 / 3 & $9.15$& 	 $212_{-23}^{+23}$ & 134.18 / 3 \\ 
\multicolumn{2}{|l|}{\fexxvi\ Ly$\beta$ abs.} & $10.84$& 	 $142_{- 88}^{+88}$ &3.09 / 1 &$10.83$ & $132_{-29}^{+29}$  & 23.58 / 1 \\ 
\multicolumn{2}{|l|}{\fexxvi\ K-edge} & $12.20$& 	 $0.10_{-0.05}^{+0.05}$ & 4.31 / 1 & $12.18$& $0.06_{-0.03}^{+0.03}$  & 5.06 / 1\\ 
\hline
\multirow{2}{*}{$v_{out}/c$} & em. &     \multicolumn{3}{c|}{$-0.010_{-0}^{+0.02}$} & \multicolumn{3}{c|}{$-0.022_{-0.012}^{+0.012}$} \\
& abs. & \multicolumn{3}{c|}{$-0.267_{-0.023}^{+0.0312}$} & \multicolumn{3}{c|}{$-0.266_{ -0.0052}^{+0.0051}$} \\
\hline
\multirow{2}{*}{$\sigma$ (keV)} & em. &      \multicolumn{3}{c|}{$0.600^{ +0}_{ -0.0791}$} & \multicolumn{3}{c|}{$0.442^{ +0.0874}_{ -0.0769}$} \\
& abs. &  \multicolumn{3}{c|}{$0.457^{+ 0.1432}_{- 0.1568 }$} & \multicolumn{3}{c|}{$0.319^{+0.0423}_{-0.0390 }$}\\
\hline
\multicolumn{2}{|l|}{C or $\chi^{2}$ / dof} &   \multicolumn{3}{c|}{379.09 / 358 =1.059} & \multicolumn{3}{c|}{1150.37 / 1179 = 0.9757}\\  
\hline
\end{tabular}
 \label{Fe1} 
\end{center}
Energies are in keV, velocities in units of $c$, and line widths $\sigma$ are in keV. The equivalent width (EW, in eV) and the depth $\tau$ give information about the strength of the gaussian line (in the first case) and of the edge (in the second case). The significance of each line is given by $\Delta$C or $\Delta\chi^{2}$.
\end{table*}

\begin{table*}[!h]
\begin{center}
\caption{Fits performed above 3 keV with two sets of Fe lines}
\begin{tabular}{|ll|ccc|ccc|}
\hline
 \multicolumn{2}{|c|}{Parameters} & \multicolumn{3}{c|}{Obs. CN} & \multicolumn{3}{c|}{Obs. XN} \\
\hline
\hline
\multicolumn{2}{|l|}{Line} & $E_{rest}$ (keV) &  EW (eV) or $\tau$ & $\Delta$C / $\Delta$dof & $E_{rest}$ (keV) &  EW (eV) or $\tau$ & $\Delta\chi^{2}$ / $\Delta$dof \\
\hline
UFO 1$\&$2 & \fexxvi\ Ly$\alpha$ em. & $7.11$ &  $233_{-46}^{+46}$  & 27.19 / 3 & $7.13$ &	 $97_{-15}^{+15}$ & 104.29 / 3\\ 
\hline	
\multirow{3}{*}{UFO 1} & \fexxvi\ Ly$\alpha$ abs. & $9.05$ & 	 $231_{-59}^{+59}$ & 16.29 / 3& $9.13$ & 	 $159_{-18}^{+18}$ & 126.36 / 3\\ 
&\fexxvi\ Ly$\beta$ abs. & $10.71$ & $46_{-0}^{+0}$
 &-0.58 / 1&$10.80$ & 	 $432_{-11}^{+11}$ & 1.89 / 1 \\ 
&\fexxvi\ K-edge & $12.04$ & 	 $0.052_{-0.080}^{+0.042}$
 & 1.36 / 1 & $12.15$ & 	 $0_{-0}^{+0.012}$  & $-$\\ 
\hline
\multirow{3}{*}{UFO 2} & \fexxvi\ Ly$\alpha$ abs. & $11.50$ & 	 $161_{-76}^{+76}$ & 9.61 / 3& $11.59$ & 	 $161_{-57}^{+57}$
 & 28.68 / 3\\ 
&\fexxvi\ Ly$\beta$ abs. & $13.62$ & 	$161_{-129}^{+129}$
 &2.98 / 1&$13.72$ & 	$161_{-77}^{+77}$  & 5.18 / 1\\ 
&\fexxvi\ K-edge & $15.32$ & 	 $0.021_{-0.062}^{+0.011}$
 & 0.09  / 1& $15.43$ & 	 $0_{-0}^{ +0.005}$  & $-$ \\ 
\hline
UFO 1$\&$2 &$v_{out}/c$ em. &     \multicolumn{3}{c|}{$-0.020_{-0}^{+0.0229}$} & \multicolumn{3}{c|}{$-0.023_{-0.0033}^{+0.0114}$} \\
UFO 1 & $v_{out}/c$ abs. & \multicolumn{3}{c|}{$-0.255_{-0.019}^{+0.021}$} & \multicolumn{3}{c|}{$-0.263_{-0.0050}^{+0.0049}$} \\
UFO 2 & $v_{out}/c$ abs. & \multicolumn{3}{c|}{$-0.463_{-0.016}^{+0.006}$} & \multicolumn{3}{c|}{$-0.469_{-0.0279}^{+0.0321}$} \\
\hline
UFO 1$\&$2 &$\sigma$ (keV) em. &      \multicolumn{3}{c|}{$0.600^{+ 0}_{-0.071}$} & \multicolumn{3}{c|}{$0.449^{+0.0808}_{ -0.0710}$} \\
UFO 1 & $\sigma$ (keV) abs. &  \multicolumn{3}{c|}{$0.472^{+0.128}_{-0.131}$} & \multicolumn{3}{c|}{$0.282^{+0.0462}_{-0.0438}$}\\
UFO 2 & $\sigma$ (keV) abs. &  \multicolumn{3}{c|}{$0.100^{+0.060}_{-0} $} & \multicolumn{3}{c|}{$0.560^{+0.0400}_{-0.2465}$}\\
\hline
\multicolumn{2}{|l|}{C or $\chi^{2}$ / dof} &   \multicolumn{3}{c|}{372.95 / 353 = 1.057} & \multicolumn{3}{c|}{1137.82 / 1174 = 0.9692}\\ 
\hline
\multicolumn{2}{|l|}{$\Delta$C or $\Delta\chi^{2}$ / $\Delta$dof} &   \multicolumn{3}{c|}{6.14 / 5} & \multicolumn{3}{c|}{12.55 / 5}\\ 
\hline
\end{tabular}
 \label{Fe2} 
\end{center}
Parameters are defined as in Table \ref{Fe1}. The last line of the table shows the significance of the improvement of the fits when considering the second UFO in addition to the first one.
\end{table*}

\begin{figure*} [!h]
\begin{tabular}{cc}
\includegraphics[trim = 18mm 10mm 0mm 0mm, clip, scale=0.45,angle=0]{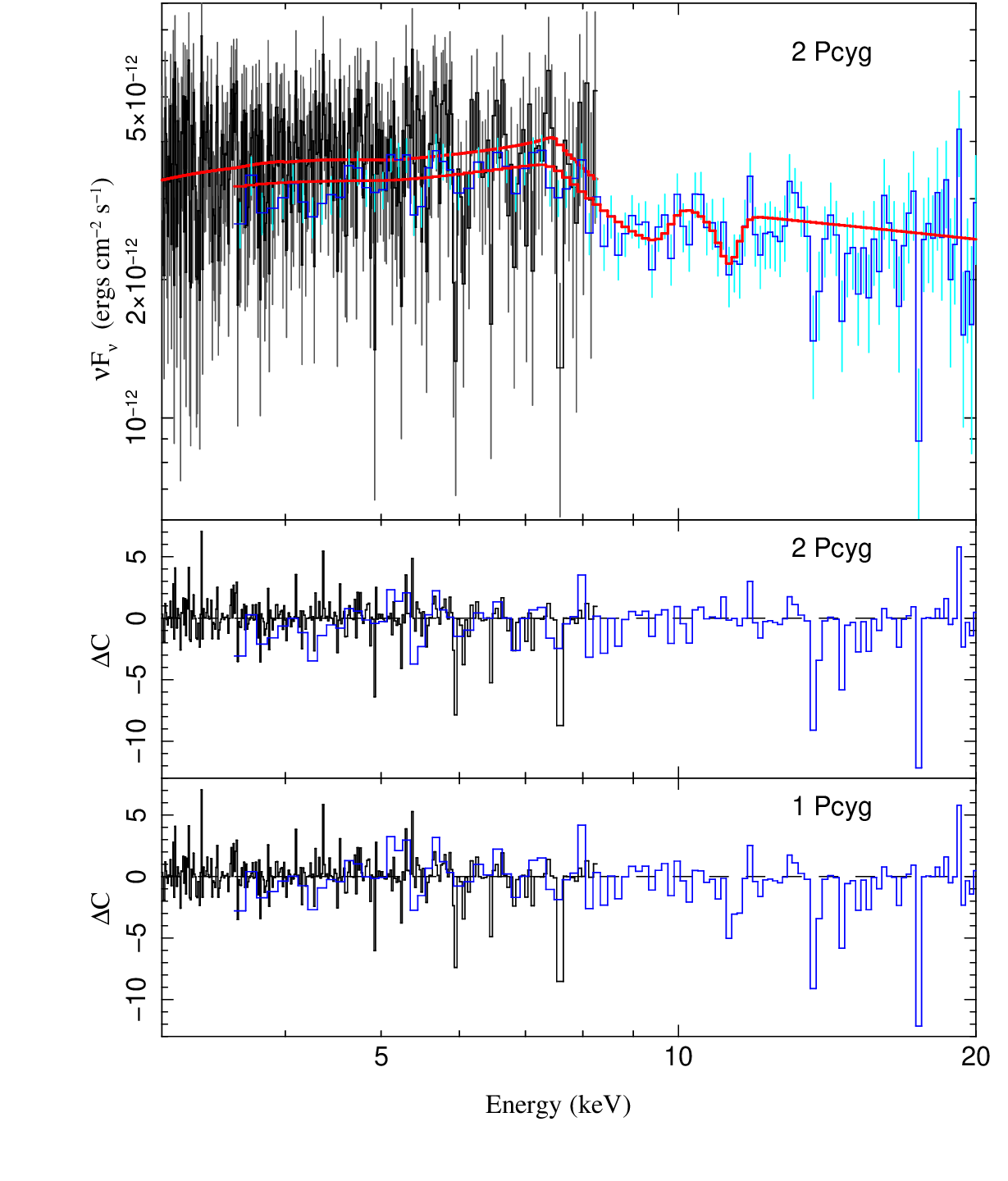} & \includegraphics[trim = 18mm 10mm 0mm 0mm, clip, scale=0.45, angle=0]{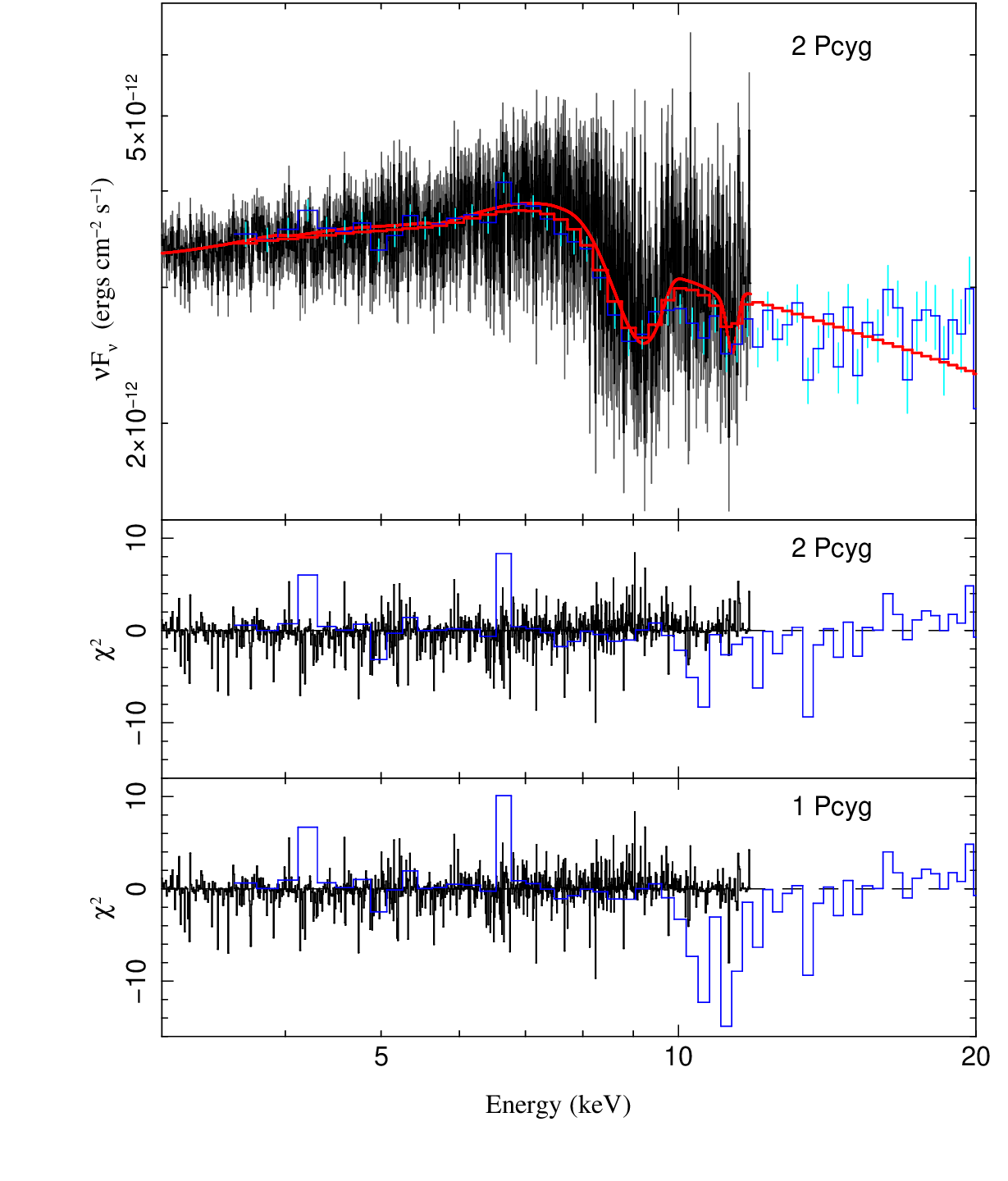} \\
\end{tabular}
\caption{Unfolded hard spectra and residuals of observations ``CN'' (left) and ``XN'' (right) fitted  above 3 keV with two (top and middle) or one (bottom) P Cygni profiles (corresponding to two UFOs or one UFO, as described in Section \ref{pcyg}). Especially note the improvement on the residuals above 10 keV, when taking into account the presence of the second UFO.}
\label{PcygFig}
\end{figure*}

\subsection{Fe K emission and absorption features}
\label{Fe}

After establishing the continuum model, we looked for absorption features above 7 keV. We fitted the spectra above 3 keV, adding to the base continuum model Gaussian and edge models, in order to constrain the emission and absorption profiles in the Fe-K band similarly to \cite{Nardini2015}. We first added \texttt{gabs} lines, one in emission (using a negative normalization) and two in absorption, and an edge, in order to account for the \fexxvi\ Ly$\alpha$ emission and absorption lines (at $E_{lab}=$6.97 keV), the \fexxvi\ Ly$\beta$ absorption line (at $E_{lab}=$8.25 keV) and the \fexxvi\ K-edge (at $E_{lab}=$9.28 keV). We tied the widths and shifts of the absorption lines together. The width and shift of the emission line was also free to vary, independently from the ones of the absorption features. This way we were able to measure the outflow velocities of the UFO in emission and in absorption. We did this fit for both ``CN'' and ``XN'' observations, because the \textit{NuSTAR} data helped to define the continuum so well. The significance of each line is given by $\Delta$C for observation ``CN'' and $\Delta\chi^{2}$ for observation ``XN'', adding first to the continuum the emission line (with three parameters of interest), then the Ly$\alpha$ absorption line (three free parameters), then the Ly$\beta$ absorption line (linked in shift and width, so only one free parameter), and finally the edge (linked in shift, leaving one  parameter of interest). Parameters of this fit are shown in Table \ref{Fe1}. Our results show that the \fexxvi\ Ly$\alpha$ emission and absorption line are significantly detected, in addition to the \fexxvi\ Ly$\beta$ absorption line and the \fexxvi\ K-edge (at lower significance), with a slight emission blueshift of $-0.01$-$-0.02c$ and a large absorption blueshift of $-0.27c$.
Note that we assessed the statistical significance of the lines using $\Delta\chi^{2}$ and $\Delta$C, however this method only gives an approximated significance according to \cite{Protassov2002}.

In order to assess the presence of a second higher-velocity UFO as claimed by \cite{Reeves2018}, we added a second set of absorption lines (using two additional absorption Gaussian lines and another edge), with linked widths and shifts, these parameters being allowed to be different from those of the first UFO and of the emission component. Parameters of this fit are shown in Table \ref{Fe2}. The results show that the \fexxvi\ Ly$\alpha$ emission and absorption lines for both UFOs ($v_{out1}$=-0.26c and $v_{out2}$=-0.47c) are detected significantly, the \fexxvi\ Ly$\beta$ absorption line from the second UFO being detected at a lower confidence level. However, the \fexxvi\ Ly$\beta$ absorption line of the first UFO and the \fexxvi\ K-edges of both winds are not detected significantly. Furthermore, the width of the lines from the second UFO is hardly constrained in the observation ``CN'', since it can only be established from \textit{NuSTAR} data that have a smaller spectral resolution compared to data from \textit{XMM-Newton}. 
Considering the improvement of the fit when taking into account the second UFO, looking at the statistics (last line of Table \ref{Fe2}), and at the residuals (in Figure \ref{FeFig}), we assess that a second faster UFO is required by the data, significantly for observation ``XN'' (at about 97\% confidence level) but only marginally for observation ``CN'' (slightly above 1$\sigma$ confidence level).


\subsection{P Cygni-like profiles}
\label{pcyg}

The combination of the broad emission and the broad and blueshifted absorption \fexxvi\ line has a P Cygni-like profile associated with the expansion of spherically symmetric stellar winds. Similarly to \cite{Nardini2015}, we applied a model from \cite{Done2007}, based on the Sobolev approximation with exact integration (SEI; \citealt{Lamers1987}) to reproduce Fe-K absorption features in this object. 
The velocity field $w=v/v_{\infty}$, i.e. the ratio between the wind velocity $v$ and the terminal velocity $v_{\infty}$, is defined in the model by $w=w_{0}+(1-w_{0})(1-1/x)^{\gamma}$, with $w_{0}$ the velocity at the photosphere and $x=r/R_{0}$ the radial distance $r$ normalized to the photospheric radius $R_{0}$. The optical depth of the line is described by $\tau(w)\approx\tau_{tot}w^{\alpha_{1}}(1-w)^{\alpha_{2}}$, with $\alpha_{1}$ and $\alpha_{2}$ characterizing the sharpness of the P Cygni profile.
We first applied a P Cygni model to replace the \fexxvi\ Ly$\alpha$ emission and absorption lines. We fixed the parameters $\gamma$ and $w_{0}$ to 2 and 0.001 respectively (note that these parameters do not have any strong influence on the shape of the profile; by trying different values, we found that those are suitable for the fits on both ``CN'' and ``XN'' observations). We thus let five parameters free to vary during the fit: the characteristic energy of the profile $E_{0}$ (corresponding to the beginning of the absorption feature), the terminal velocity $v_{\infty}$, and the parameters  $\tau_{tot}$, $\alpha_{1}$ and $\alpha_{2}$. We then added another P Cygni model in order to fit the second UFO. We chose the same free parameters, however we tied the energy $E_{0}$ of the second P Cygni profile to that of the first P Cygni profile, because of the assumption of a single \fexxvi\ Ly$\alpha$ emission line for both UFOs (as also supposed in section \ref{Fe} and in the analysis from \cite{Reeves2018b,Reeves2018}. This fit with the second UFO was thus done with nine free parameters. We performed the fits on observations ``CN'' and ``XN'', showing the results in Table \ref{PcygTable}.

The statistics of the fits (last line of Table \ref{PcygTable}) as well as the residuals (plotted on Figure \ref{PcygFig}) show that the use of a second P Cygni profile significantly improves the fit in both observations ``CN'' (at about 99\% confidence level) and ``XN'' (at more than 99\% confidence level), supporting the hypothesis of the presence of a second UFO having a higher velocity than the first one in both datasets. 

\begin{table}[!h]
\begin{center}
\caption{Fits performed above 3 keV with two P Cygni profiles}
\begin{tabular}{|ll|c|c|}
\hline
 \multicolumn{2}{|l|}{Parameters} & Obs. CN &Obs. XN \\
\hline
P Cygni 1$\&$2 & $E_{0}$ (keV) & $6.32_{-0.05}^{+0.05}$ & $6.35_{-0.04}^{0.04}$ \\
\hline
\multirow{4}{*}{P Cygni 1} & $v_{\infty}/c$ & -0.32$_{-0.027}^{+0.012}$& -0.32$_{-0.012}^{+0.049}$\\
& $\tau_{tot}$ & $0.17_{-0.03}^{+0.02}$ & $0.13_{-0.016}^{+0.001}$\\
& $\alpha_{1}$ & $0.93_{-0.84}^{+1.33}$ & $3.60_{-0.37}^{+1.38}$ \\
& $\alpha_{2}$ & $0.17_{-0.42}^{+0.77}$ & $1.58_{-0.91}^{+0.91}$\\
& C or $\chi^{2}$ / dof &   389.6/361=1.08 & 1187.5/1182=1.005\\ 
\hline
\multirow{4}{*}{P Cygni 2}  & $v_{\infty}/c$ & -0.52 $_{-0.015}^{+0.024}$ & -0.53 $_{-0.022}^{+0.034}$\\
& $\tau_{tot}$ & $0.04_{-0.004}^{+0.004}$ & $0.02_{-0.005}^{+0.003}$\\
& $\alpha_{1}$ & $6.17_{-7.18}^{+3.28}$ & $4.13_{-3.22}^{+1.87}$ \\
& $\alpha_{2}$ & $-0.42_{-0.63}^{+0.68}$ & $-0.68_{-0.17}^{+0.22}$\\
& C or $\chi^{2}$ / dof &   375.4/357=1.05 & 1170.3/1178=0.99\\ 
\hline
\multicolumn{2}{|l|}{$\Delta$C or $\Delta\chi^{2}$ / $\Delta$dof} & 14.2 / 4 & 17.2 / 4 \\
\hline
\end{tabular}
 \label{PcygTable} 
\end{center}
\end{table}

\subsection{Photoionization modeling}
\label{warmabs}

After our line identification at high energy (above 3 keV), we then added to our analysis the soft parts of the data (down to 0.4 keV) and studied the broad energy band spectra. We initially performed a blind line search and identified individual lines as described in the Appendix. Together with previous sections \ref{Fe} and \ref{pcyg} focusing on high energy individual features, this blind line search aimed at detecting individual lines as signatures of the outflows, particularly at lower energy. Our analysis strategy is to study both individual lines (model-independent analysis) and global models (model-dependent analysis), in order to check the consistency of our results using a dual-approach. For this purpose, we implemented a self-consistent photoionization modeling to reproduce the emission and absorption features seen in our broadband spectra of the three datasets ``CN'', ``XN'' (from 0.4 to 30 keV) and ``C'' (from 0.4 to 8 keV). To do so, we used the XSTAR photoionization code (version 2.39, \citealt{Kallman1996,Kallman2001,Kallman2004,Kallman2009}). However, instead of using XSTAR tabulated grids for our fits (as it has been done in previous works, e.g. \citealt{Nardini2015,Matzeu2017b,Reeves2018b,Reeves2018}), we employed the analytic XSTAR models \texttt{warmabs} for reproducing the absorption lines, and \texttt{photemis} for generating the emission lines. Using such analytic functions requires a larger amount of computation time in comparison to the calculation of XSTAR tabulated grids. However, it allows to explore the entire range of values for the column density $N_{H}$, and the ionization parameter $\xi=L_{ion}/(nr^{2})$ (with $L_{ion}$ the ionizing luminosity, $n$ the hydrogen density and $r$ the distance to the ionizing source), instead of discrete values that depend on the refinement of the grid. Furthermore, multiple parameters can freely vary during the fitting procedure, including the turbulent velocity.

Similarly to previous works \citep{Matzeu2016b,Reeves2018}, we adopted two SEDs that are identical in the UV band. This assumption of an unchanged UV spectrum is supported by recent results from \cite{Reeves2018b}. For observation ``XN'', we used the SED reported in \cite{Nardini2015}, derived from \textit{XMM-Newton} EPIC-\textit{pn} and \textit{OM}, and from \textit{NuSTAR}. For observations ``CN'' and ``C'', we used a SED that slightly differs from the former one only for the X-ray slope ($\Gamma=2.3$ instead of $2.4$, as found when fitting the continuum for ``CN'' data).


For our photoionization modeling, we added different wind emission and absorption components to the continuum of the three datasets, following the sequence described below and the order shown in the Table \ref{warmabsTable} that summarizes the results.
We first replaced the partial covering wind \texttt{zpcfabs} with a \texttt{warmabs} model convolved with a \texttt{partcov} model (to account for the partial covering absorption required in all observations). To find the best fit, this partial covering absorber was variable in ionization, column density, covering factor and velocity. Note that the outflowing velocities of the winds modeled with \texttt{warmabs} are derived from the redshift parameters which were variable in the fitting process. We then included a photoemission component in this baseline model, having variable ionization, normalization and velocity parameters, since emission lines such as slightly blueshifted \fexxvi\ Ly$\alpha$ have been detected, as described in the previous sections. We used free Doppler shift parameters for both these models, allowing the emission and absorption features to be blueshifted independently. As seen in the statistics of the best-fit wind emission ($\Delta$C and $\Delta\chi^{2}$ in Table \ref{warmabsTable}), the wind emission is significantly required by ``XN'' data, and marginally needed by ``CN'' and ``C'' data. A low-ionization warm absorber at the systematic Doppler shift is also required by ``XN'' data only, as shown by the significant improvement of the fit when including this new component in the previous model (see $\Delta\chi^{2}$ value for the warm absorption in Table \ref{warmabsTable}). We then applied a totally covering absorption to the model described above (including the new continuum, the wind emission, and the warm absorber for the ``XN'' observation only), with a variable column density, linking the ionization parameter to that of the photoemission, and linking the blueshift to the partial covering absorption, so we assumed that the partial covering absorber and the fast wind are outflowing at the same velocity but are independent in ionization parameters and column densities. Adding this first UFO strongly improves the fit for all observations, with $\Delta$C=21.4 for ``CN'',  $\Delta\chi^{2}$=406.0 for ``XN'', and $\Delta$C=19.4 for ``C'', for one degree of freedom, demonstrating that the slowest UFO is significantly required in all our datasets at a confidence level larger than 99.9\%. We then included a second total covering absorption in the model with the first UFO, with a varying column density and an independent blueshift (so an independent velocity), and an ionization fixed to the value of the first UFO and the emission component. We see that the inclusion of this larger velocity UFO in the model improves the fit for the observations ``CN'' and ``XN'' (see Table \ref{warmabsTable}), with $\Delta$C=8.0 for ``CN'' and  $\Delta\chi^{2}$=49.4 for ``XN'', for two degrees of freedom, showing that the fastest UFO is significantly required at confidence levels of >95\% and >99.9\% respectively.

For these fits, we allowed the continuum parameters (powerlaw, soft-excess and cross-calibration factor) to vary, since the application of \texttt{warmabs} models slightly changed the overall shape of the broadband spectra. Initially, we varied the turbulent velocity of each XSTAR component. We found that the turbulent velocity reached 20000 km/s for all components in all observations, so we fixed this parameter to this value.
Such a large turbulent velocity is consistent with the large widths of the lines described in the previous sections, and with the values used in previous works with XSTAR grids. All abundances have been fixed to solar values. 

\begin{table*}[!h]
\begin{center}
\caption{Photoionization models, applied to our data above 0.4 keV, and corresponding fit parameters for observations ``CN'', ``XN'' and ``C''}
\begin{tabular}{|c|c|}
\hline
Observation & Model\\
\hline
``CN'' & \texttt{cst*tbabs*warmabs[ufo1]*warmabs[ufo2]*(photemis+partcov*warmabs[pc]*powerlaw+zgauss)}\\

``XN'' & \texttt{cst*tbabs*warmabs[wa]*warmabs[ufo1]*warmabs[ufo2]*(photemis+partcov*warmabs[pc]*powerlaw+zgauss)}\\

``C'' & \texttt{cst*tbabs*warmabs[ufo1]*(photemis+partcov*warmabs[pc]*powerlaw+zgauss)}\\
\hline
\end{tabular}
\linebreak
\linebreak
\begin{tabular}{|ll|ccc|}
\hline
 \multicolumn{2}{|c|}{Parameters} & ``CN'' & ``XN'' & ``C'' 			\\
 \hline
 Cross-calibration & $C$& $0.90_{-0}^{+0.014}$ &$0.98_{-0.007}^{+0.006}$& - \\
 Galactic absorption & $N_{H}^{gal}$ ($10^{22}$ cm$^{-2}$) & 0.24 (f) & (f) & (f) \\
Powerlaw & $\Gamma$ & $2.21_{-0.008}^{+0.012}$ & $2.37_{-0.003}^{+0.003}$& $2.5_{-0.004}^{+0.001}$\\
& $F_{0.4-30\text{ keV}}$ ($10^{-12}$ erg s$^{-1}$ cm$^{-2}$)&$11.30_{-0.10}^{+0.10}$ & $15.33_{-0.018}^{+0.018}$& $17.10_{-0.17}^{+0.17}$\\
Soft-excess & E (keV)& $0.40_{-0.02}^{+0.07}$ & $0.55_{-0.007}^{+0.008}$& $0.087_{-0.010}^{+0.041}$\\
& $\sigma$ (keV)& $0.05_{-0.014}^{+0.008}$ & $0.24_{-0.003}^{+0.003}$& $0.057_{-0.0002}^{+0.0007}$\\
& $F_{0.4-30\text{ keV}}$ ($10^{-12}$ erg s$^{-1}$ cm$^{-2}$) & $62.9_{-15.8}^{+13.8}$&$2.27_{-0.011}^{+0.011}$ & $9.14_{-3.60}^{+2.98}$\\ 
Partial covering & log($N_{H}/10^{22}$ cm$^{-2}$) & $0.52_{-0.13}^{+0.12}$ & $1.22_{-0.04}^{+0.01}$& $0.83_{-0.08}^{+0.13}$\\
& log($\xi$/erg cm s$^{-1}$) & $3.20_{-0.58}^{+0.18}$ & $3.04_{-0.21}^{+0.02}$ & $2.89_{-0.68}^{+0.08}$\\
& $c_{f}$ & $0.33_{-0.054}^{+0.046}$ &$0.40_{-0.003}^{+0.001}$&$0.77_{-0.04}^{+0.03}$\\
& $v_{turb}$ (km s$^{-1}$)& 20000 (f) & (f) & (f) \\
& $v_{out}$ (c) & $-0.289_{-0.030}^{+0.025}$ & $-0.268_{-0.004}^{+0.007}$ & $-0.236_{-0.064}^{+0.025}$\\
\hline
\multirow{5}{*}{Wind emission} & log($\xi$/erg cm s$^{-1}$) & $6.03_{-0.28}^{+0.49}$ & $6.97_{-0.51}^{+0.20}$& $5.19_{-0.09}^{+0.36}$ \\
& norm ($\times 10^{-2}$)& $8.5_{-4.2}^{+5.2}$ & $5.8_{-2.2}^{+0.1}$ &  $5.3_{-0.028}^{+0.092}$\\
& $v_{turb}$ (km s$^{-1}$)& 20000 (f) & (f) & (f) \\
& $v_{out}$ (c) & $-0.056_{-0.042}^{+0.006}$ & $-0.089_{-0.001}^{+0.020}$& $-0.026_{-0.021}^{+0.002}$\\
&$\Delta$C or $\Delta\chi^{2}$ / $\Delta$dof & 6.1 / 3 &  68.1 / 3 & 4.8 / 3 \\
\hline
\multirow{5}{*}{Warm absorption } & log($N_{H}/10^{22}$ cm$^{-2}$) & - & $-0.99_{-0.002}^{+0.009}$& - \\
& log($\xi$/erg cm s$^{-1}$) & - & $0.73_{-0.21}^{+0.05}$& - \\
& $v_{turb}$ (km s$^{-1}$)& - & 100 (f) & - \\
& $v_{out}$ (c) & - & 0 (f)&- \\
&$\Delta\chi^{2}$ / $\Delta$dof & - & 74.5 / 2 &- \\
\hline
\multirow{6}{*}{Wind absorption 1 } & log($N_{H}/10^{22}$ cm$^{-2}$) & $1.32_{-0.22}^{+0.39}$& $1.90_{-0.46}^{+0.05}$& $0.99_{-0.24}^{+0.12}$\\
& log($\xi$/erg cm s$^{-1}$) & $6.03_{-0.28}^{+0.49}$ (t) &$6.97_{-0.51}^{+0.20}$(t) & $5.19_{-0.09}^{+0.36}$ (t)\\
& $v_{turb}$ (km s$^{-1}$)& 20000 (f) & (f) & (f) \\
& $v_{out}$ (c) & $-0.289_{-0.030}^{+0.025}$ (t) & $-0.268_{-0.004}^{+0.007}$ (t)& $-0.236_{-0.064}^{+0.025}$ (t)\\
& C or $\chi^{2}$ / dof & 1660.2 / 1603 =1.04& 1860.7 / 1696 =1.10& 1384.2 / 1325 =1.04\\
&$\Delta$C or $\Delta\chi^{2}$ / $\Delta$dof & 21.4 / 1 & 406.0 / 1 & 19.4 / 1\\
\hline
\multirow{6}{*}{Wind absorption 2 } & log($N_{H}/10^{22}$ cm$^{-2}$) & $1.13_{-1.31}^{+0.21}$& $1.71_{-0.38}^{+0.09}$& - \\
& log($\xi$/erg cm s$^{-1}$) & $6.03_{-0.28}^{+0.49}$ (t) & $6.97_{-0.51}^{+0.20}$(t)& - \\
& $v_{turb}$ (km s$^{-1}$)& 20000 (f) & (f) & - \\
& $v_{out}$ (c) & $-0.478_{-0.094}^{+0.031}$ & $-0.483_{-0.033}^{+0.001}$& - \\
& C or $\chi^{2}$ / dof &1652.2 / 1601 =1.03 & 1811.3 / 1694 =1.07& - \\
&$\Delta$C or $\Delta\chi^{2}$ / $\Delta$dof & 8.0 / 2 & 49.4 / 2 & - \\
\hline
\end{tabular}
 \label{warmabsTable} 
\end{center}
(f) refers to parameters that have been frozen during the fits. (t) refers to parameters that have been tied to other parameters during the fits (like the ionization parameters, and the velocities of the partial covering absorber and the slowest UFO). Each $\Delta$C or $\Delta\chi^{2}$ value refers to the improvement of the fit when adding the new component to the model including the components listed above in the table.
\end{table*}

The fits have been performed in ISIS using Cash statistics and the \texttt{powell} method for observations ``CN'' and ``C'', and $\chi^{2}$ statistics and the \texttt{mpfit} method for observation ``XN''. Using the XSTARDB\footnote{http://space.mit.edu/cxc/analysis/xstardb/index.html} library of ISIS S-lang scripts, we listed the strongest features predicted by our photoionization models with the fitted parameters given in Table \ref{warmabsTable}, for the three observations and for all components. For the UFOs, in addition to the expected \fexxv\ He$\alpha$, He$\beta$, \fexxvi\ Ly$\alpha$ and Ly$\beta$ absorption lines, the XSTAR model predicts \sixiv\ and \sxvi\ Ly$\alpha$ absorption lines at lower energy, consistent with the lines identified in the blind line search section (see appendix). Many emission features are also predicted by the \texttt{photemis} model, from \fexxv\ He$\alpha$ and \fexxvi\ Ly$\alpha$, but also from other ions (\cvi\ , \oviii\ , \nex\ , \mgxii\ , \fexxiv\ , \sixiv\ and \sxvi\ ). The lower-ionization partial covering absorber is expected to show absorption features from \ovii\ , \oviii\ , \fexvii\ , \fexviii\ , \neix\ , \nex\ , \mgxi\ and \sixiii\ ions. Some of these lines have also been identified in the blind line search section. Non-blueshifted signatures from \ovii\ and \nevi\ ions are predicted for the warm absorber found in the ``XN'' observation. We also used a Markov chain Monte Carlo (MCMC) approach to get the probability distribution of each parameter via Bayesian data analysis. For this purpose, we used the \texttt{isis\_emcee\_hammer} function\footnote{https://www.sternwarte.uni-erlangen.de/wiki/index.php/Emcee} developed in ISIS S-lang by MN. Best fit values and error bars are reported in Table \ref{warmabsTable}. Figure \ref{contours} shows the contour plots characterizing the absorbers in the three observations.

\section{Discussion}
\label{discussion}

\subsection{Modeling the highly-ionized features}
We initially modeled the absorption features at high energy (above 3 keV) with blueshifted and highly-ionized iron lines to characterize the ultra fast outflows in PDS 456. We found that the \fexxvi\ Ly$\alpha$ emission and absorption lines, the \fexxvi\ Ly$\beta$ absorption line and the Fe K-edge corresponding to the fit with only one UFO are significantly or marginally detected in both observations ``CN'' and ``XN''. Velocities in emission and in absorption derived from this fit are consistent with results from \cite{Nardini2015} (see Table \ref{Fe1}). Applying a second set of lines to account for the second possible UFO marginally improved the fit, but not all the lines were detected significantly (see Table \ref{Fe2}). This can be explained by the complexity of the model  fitted to data whose resolution is not sufficient. Indeed, signatures of the second UFO are located in the energy band covered predominantly by \textit{NuSTAR}.
However, the second UFO is required, more convincingly in the observation ``XN'' than in the observation ``CN'', according to the residuals in Figure \ref{FeFig}.

We replaced the sets of gaussian lines with P Cygni profiles, using the model from \cite{Done2007}, as proposed by \cite{Hagino2015} and \cite{Nardini2015}, describing a spherically symmetric outflowing wind. Applying two P Cygni models instead of a single one significantly improved the fits for both observation ``CN'' and ``XN'' (see the statistics in Table \ref{PcygTable} and the residuals on Figure \ref{PcygFig}). 
Table \ref{PcygTable} shows that the velocities implied by the P Cygni model are slightly higher (of about 15-20$\%$) than the one measured from individual Gaussian lines. The terminal velocity of the wind, $v_{\infty}$, is a characteristic parameter of the P Cygni model for a quasi-spherical fully-covering outflow. \textit{This value is the actual speed of the outflowing gas. }
However, $v_{\infty}$ strongly depends on the P Cygni model used to fit our data, so we instead considered the line-of-sight bulk velocity, that is found to have consistent values when fitting with gaussian lines and photoionization models, for our analysis and for energetic estimations.
We found an optical depth for the first P Cygni profile that is consistent with results from \cite{Nardini2015}, the one of the second P Cygni profile being smaller. Parameters $\alpha_{1}$ and $\alpha_{2}$ are different from values found by \cite{Nardini2015}, possibly because they were free to vary independently in our analysis. Their values are also very different between the observations, as well as between the two P Cygni profiles.

For both observations ``CN'' and ``XN'', the outflow velocities of both UFOs measured by one method or the other one are consistent between the two different epochs, and consistent also with previous results (e.g. \citealt{Nardini2015,Matzeu2017b,Reeves2018}).

\subsection{Photoionization results}

We applied self-consistent XSTAR photoionization models to our three different epochs of observations (using the entire detectors band, i.e. 0.4-30 keV when possible), in order to characterize the physical properties of the absorbers reproducing different absorption lines detected in our blind line search (see Appendix) and in high energy spectra (see sections \ref{Fe} and \ref{pcyg}). The statistics of the fits ($\Delta$C and $\Delta \chi^{2}$ rows in Table \ref{warmabsTable}) demonstrate that including the two UFO components in our modeling significantly improves the fits (at >99.9\% confidence level for the slowest UFO, and at >95->99.9\% confidence level for the fastest UFO). The high values of column densities $N_{H}$ and ionization parameters $\xi$ obtained from our modeling are typical of ultra-fast outflows (e.g. \citealt{Tombesi2013}) and consistent with previous studies of PDS 456.


\begin{sidewaysfigure*}[ht]
\vspace{250pt}
\begin{tabular}{ccc}
\includegraphics[trim = 18mm 12mm 0mm 0mm, clip, angle=0,scale=0.44]{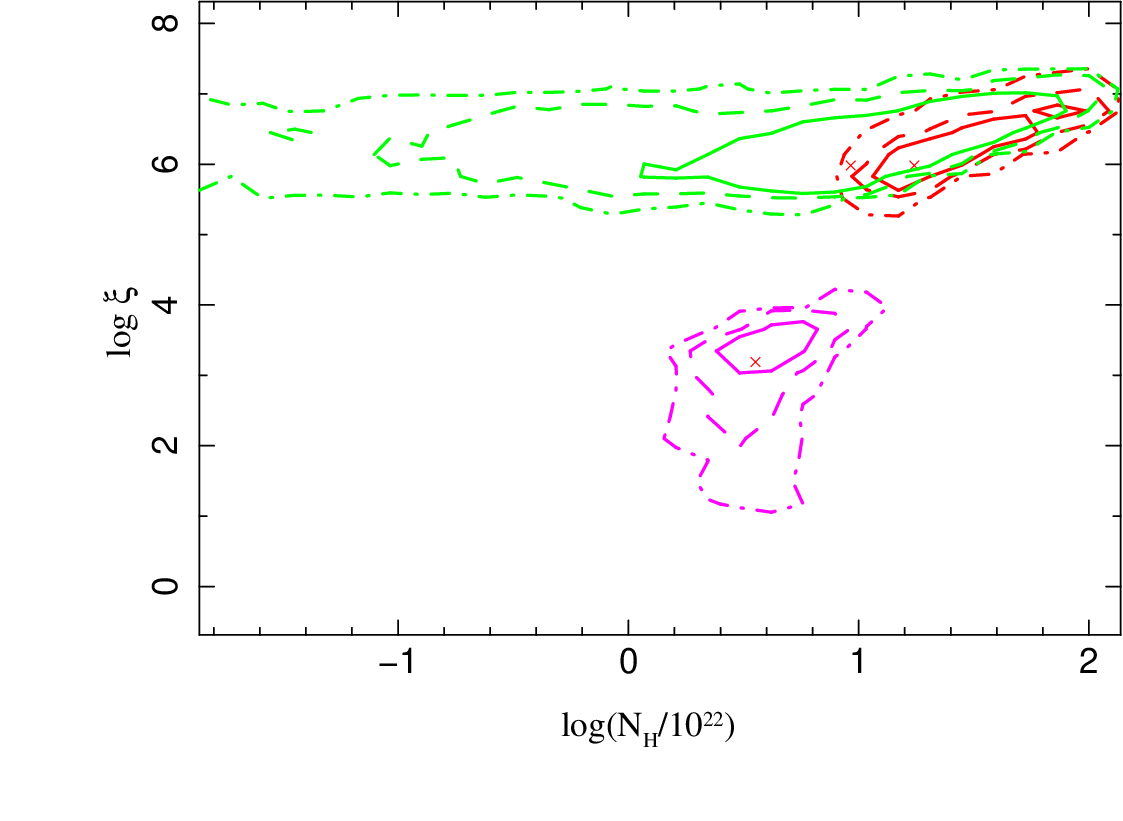} & 
\includegraphics[trim = 18mm 12mm 0mm 0mm, clip, angle=0,scale=0.44]{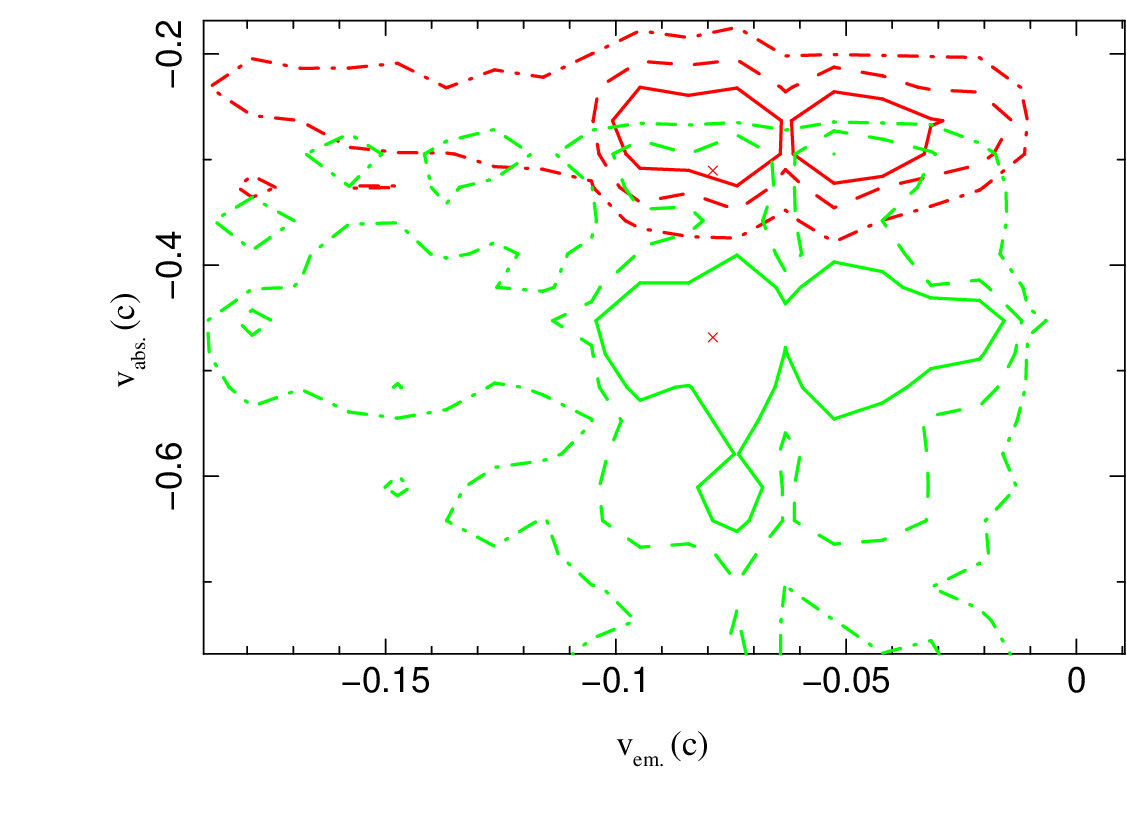} & 
\includegraphics[trim = 17mm 12mm 0mm 0mm, clip, angle=0,scale=0.44]{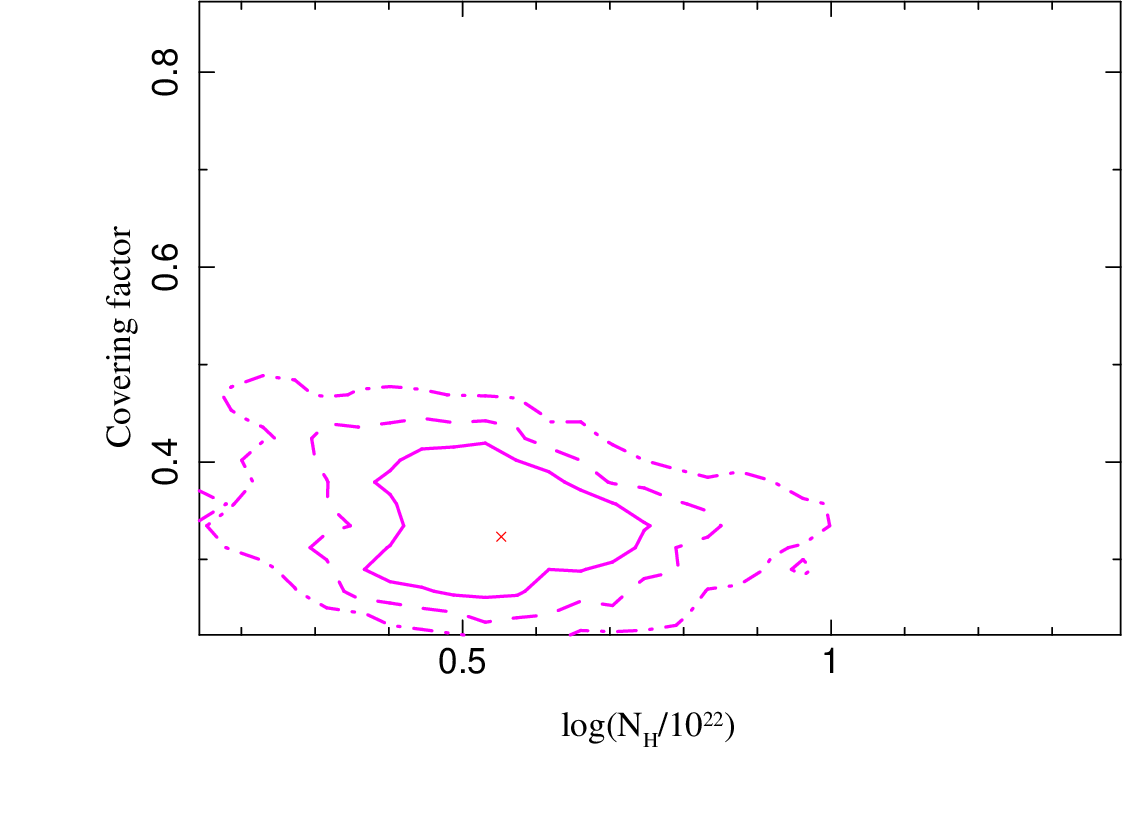} \\
\includegraphics[trim = 18mm 12mm 0mm 0mm, clip, angle=0,scale=0.44]{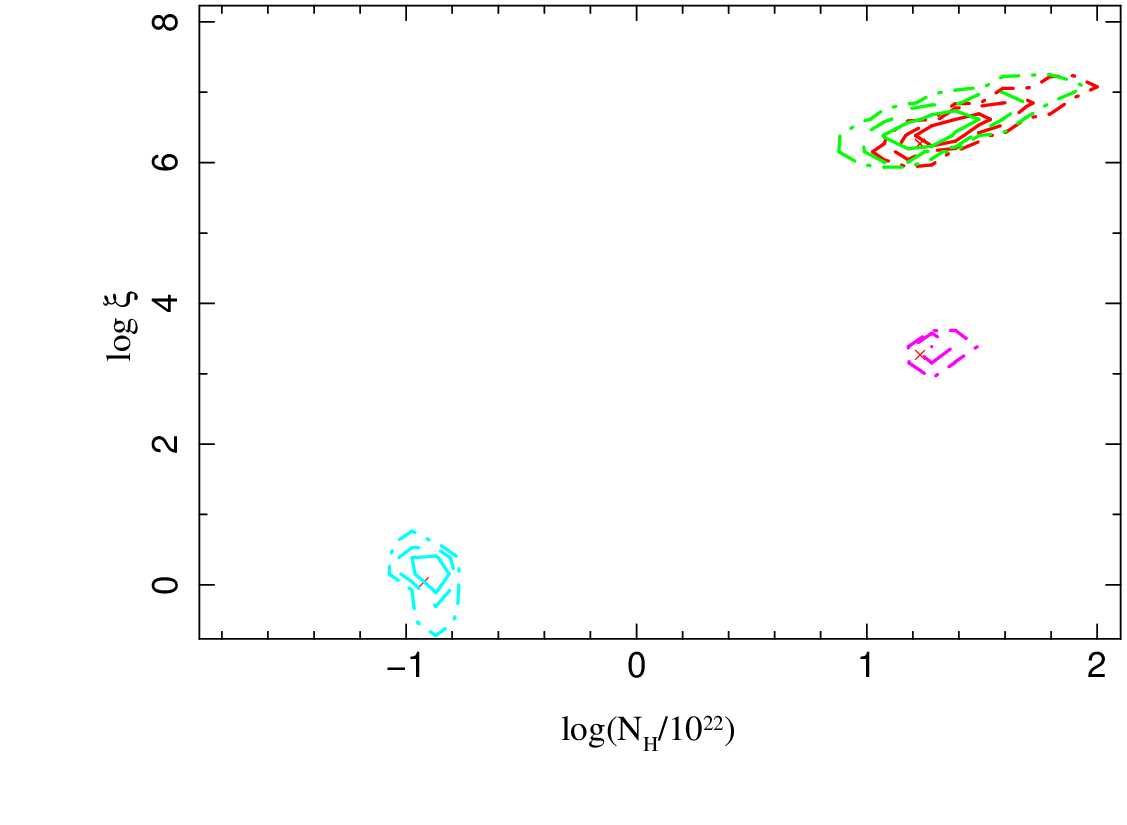} &
\includegraphics[trim = 18mm 12mm 0mm 0mm, clip, angle=0,scale=0.44]{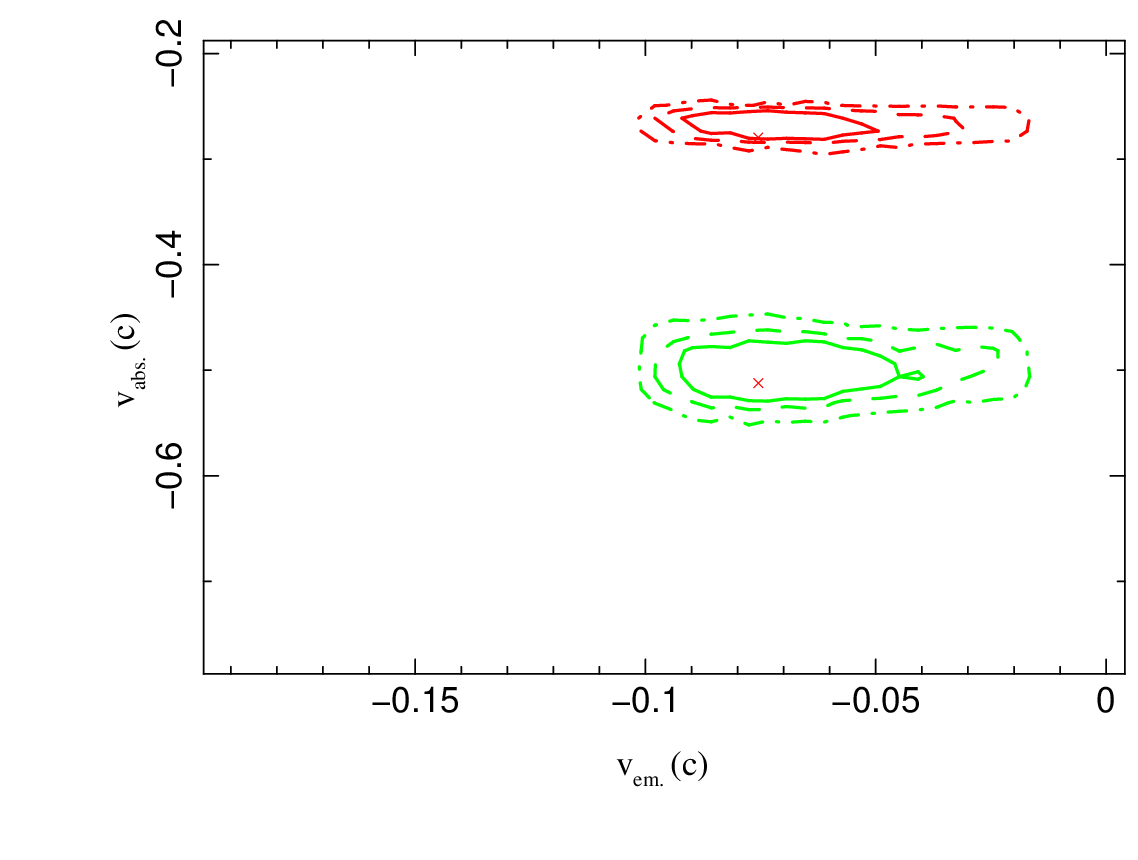} & 
\includegraphics[trim = 17mm 12mm 0mm 0mm, clip, angle=0,scale=0.44]{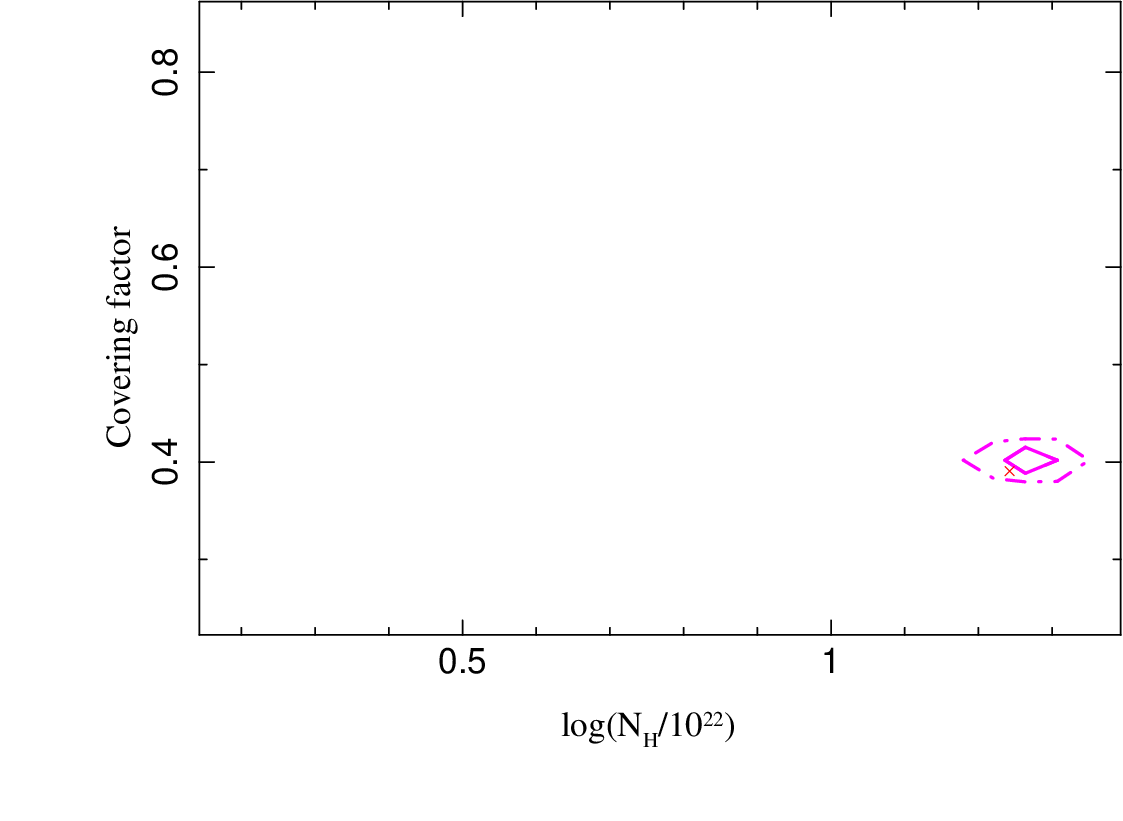} \\
\includegraphics[trim = 18mm 12mm 0mm 0mm, clip, angle=0,scale=0.44]{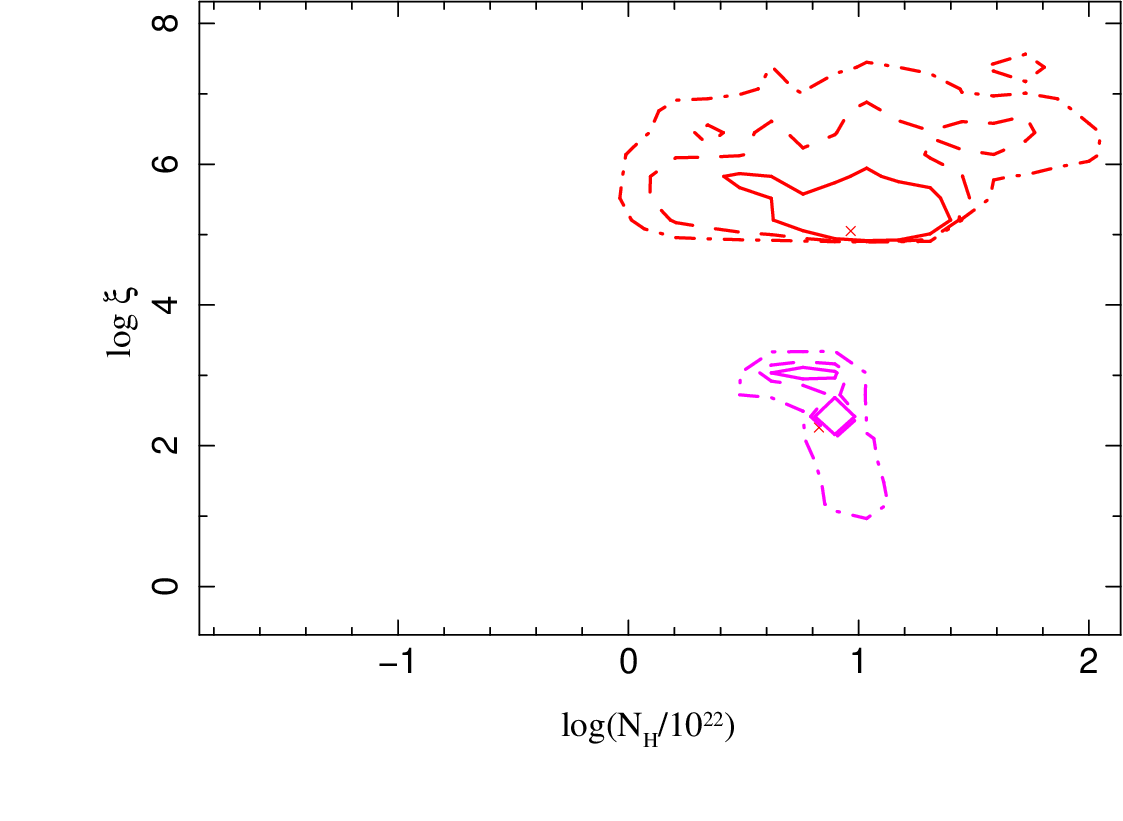} &
\includegraphics[trim = 18mm 12mm 0mm 0mm, clip, angle=0,scale=0.44]{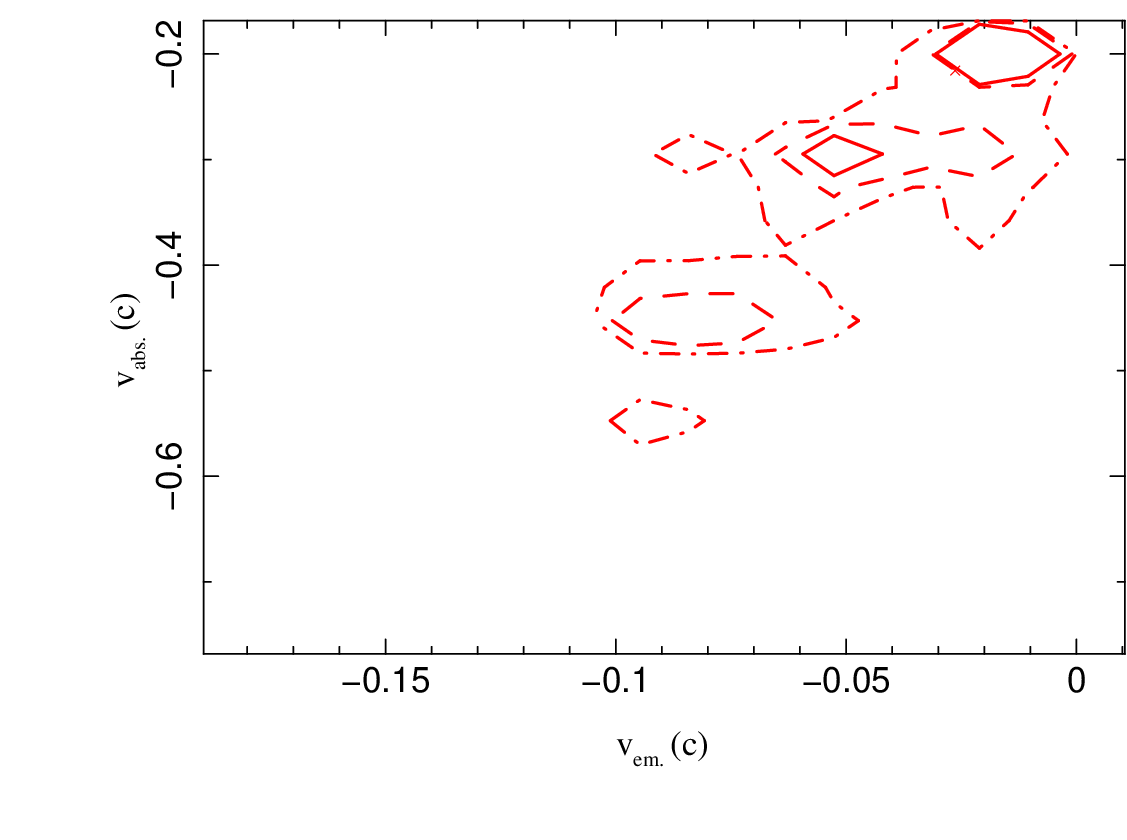} & 
\includegraphics[trim = 17mm 12mm 0mm 0mm, clip, angle=0,scale=0.44]{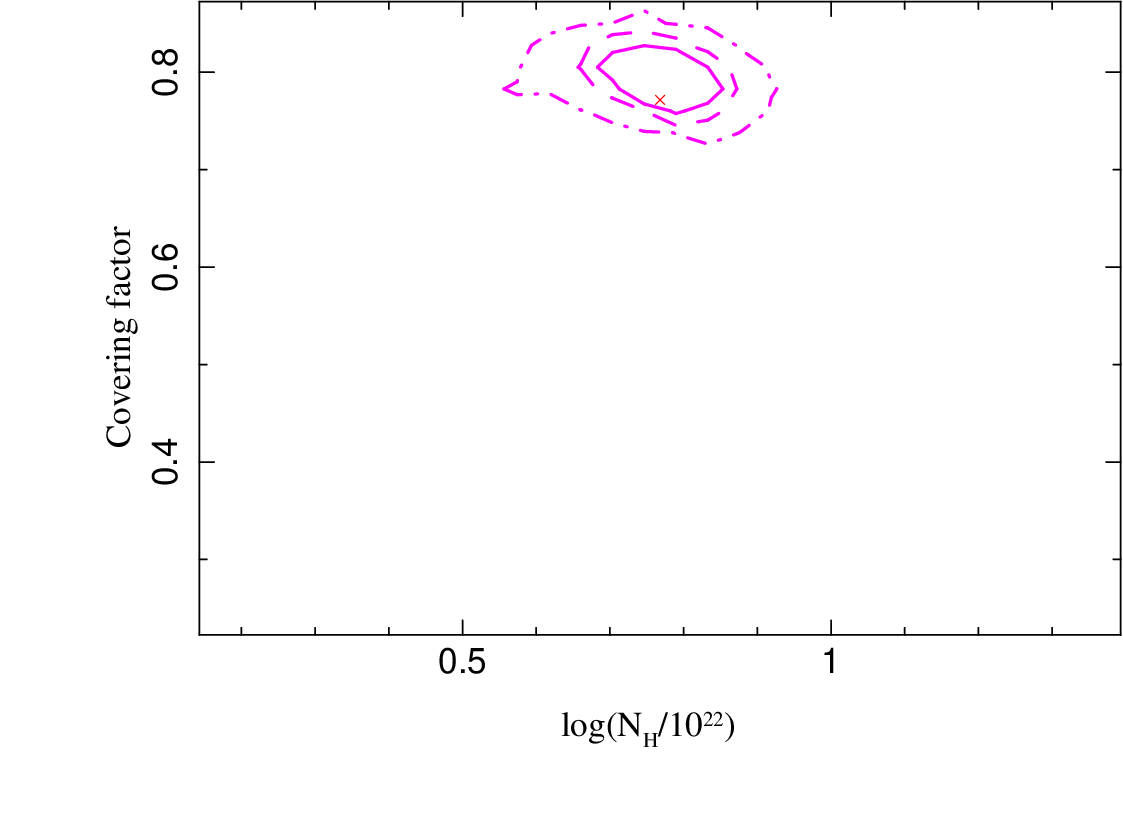} \\
\end{tabular}
\vspace{-8pt}
\label{contours}
\caption{Contours of parameters from the absorbers in the three observations (results of the photoionization modeling performed above 0.4 keV). Left: ionization parameter vs column density; Middle: velocity in absorption vs velocity in emission. Right: covering factor vs column density of the partial covering absorber. Top: observation ``CN''; Middle: observation ``XN''; Bottom: observation ``C''. Three confidence levels are represented: solid line for 68\%, dashed line for 90\%, and dotted-dashed line for 99\%. Like in previous figures, pink contours represent parameters from the partial covering absorber, red contours are for the slowest UFO, green contours are for the fastest UFO, and light blue contours are from the low-ionization warm absorber detected in observation ``XN''.}
\end{sidewaysfigure*}

As described in Section \ref{warmabs}, we adopted a fixed value for the turbulent velocity, that helped to constrain the other parameters better, as the model is complex with a lot of free parameters. The turbulent velocity of 20000 km/s  is consistent with the large widths of individual lines (e.g. see Table \ref{LinesRozenn}), and the values reported in previous studies (e.g. \citealt{Nardini2015}).

The ionization parameter of both UFOs has been tied to that of the emission associated with the disk wind to facilitate the fits, as implemented in \cite{Nardini2015} and \cite{Reeves2018}, so the ionization parameters of both UFO are the same, but different from the one of the partial covering absorber (even if the partial covering absorber has the same velocity as the slowest UFO). The high ionization parameters derived from our modeling are consistent with the detection of highly ionized \fexxvi\ lines and similar to previous results (e.g. \citealt{Nardini2015,Reeves2018b,Reeves2018}). It is also possible that two different sets of emission and absorption from different stratification layers of a single UFO are associated with the different ionizations. However, the available data would not be able to distinguish the different kinematic components, the overall model being in this case even more complex.

The velocity of the slowest UFO was tied to the velocity of the partial covering absorber. This scenario was adopted to obtain a better constraint. However, we saw in appendix that when we identified lines detected by blind search, the transitions had a slightly different blueshift for lines from the partial covering absorber than from the slowest UFO, in particular in the observation ``CN''. It may imply that the actual velocities of both components are slightly different. The large UFO velocities derived from photoionization modeling (about $v_{out}=-0.24$ - $-0.29c$ for the slowest UFO, and $v_{out}=-0.48c$ for the fastest UFO) are consistent with previous studies, as well as with results from modeling of high energy features (see sections \ref{Fe} and \ref{pcyg}) and results from the blind line search (see Appendix). An additional method for the characterization of the kinematic components of the UFOs is also described and applied to PDS 456 data in Appendix, and the results are also consistent with values found through photoionization modeling.

MCMC approach gave us the probability distribution of each parameter of the photoionization model, giving contours shown on Figure \ref{contours}. These contour plots show the behavior of the ionization parameter as a function of the column density, as well as the different velocities in emission and in absorption, for all absorbers detected in the three observations, and the variation of the covering factor as a function of the column density for the partial covering absorber along the three observations. 
We find that the velocities are consistent between the observations ``CN'' and ``XN'', with a better constraint for the observation ``XN'' compared to the observation ``CN'' because of the better signal-to-noise. Despite the poor constraint on the velocity of the UFO detected in the observation ``C'' (corresponding to the slowest one), we can see hints of a consistency with the other observations, with the best-fit value slightly higher than $v_{out}=-0.2c$ and contours coherent with an outflowing velocity of $v_{out}=-0.2$ - $-0.3c$. 
Contours of the UFO in the observation ``C'' are obviously improved when \textit{NuSTAR} data are combined with \textit{Chandra}/HETGS data as in the observation ``CN''. However, despite the lack of information at high energy in the observation ``C'', the photoionization modeling allowed us to constrain the parameters of the winds at a lower significance level.
The contours of the all parameters, and in particular parameters of the fastest UFO, are more precise for the observation ``XN'' compared to the observation ``CN'' (regarding velocities, ionization parameters, column densities and covering factors), as expected because the signal-to-noise in the \textit{NuSTAR} energy band brought by combining all the observations ``XN1'' to ``XN5'' is higher than that from the observation ``CN''. The ionization parameters and column densities of the UFOs are relatively stable between the different epochs.
The ionization of the partial covering absorber is lower than those of the UFOs, as expected from the detected absorption lines in the soft band, and this parameter is stable between observations, with a modest value ($log(\xi)\sim 3$) consistent with previous studies \citep{Nardini2015,Reeves2018b}.
Variability of the covering factor ($c_{f}\sim 0.3-0.8$) and of the column density of this absorber ($N_{H}=3.3-16.6\times 10^{22}\text{ cm}^{-2}$), as seen in the right panel of Figure \ref{contours}, could be responsible for the continuum shape variability observed in all observations of PDS 456, similar to NGC 5548 \citep{Kaastra2014,Nardini2015,Reeves2018b}.

An alternative scenario proposed to explain blueshifted absorption features is that they could be the signatures of reflection on an optically thick plasma that may cover the accretion disk whose inner parts may rotate at extremely high velocities, producing a relativistic blueshifted Fe K-shell feature (see example of PG 1211+143, \citealt{Gallo2013}, but counter-argument from \citealt{Lobban2016}). \cite{Nardini2015} found that such a reflection-dominated scenario under-predicts the strength of the absorption feature around 9 keV. \cite{Behar2009} found a persistent but small contribution of a reflection component in most observations before the \textit{XMM-Newton} observation in 2007. A \textit{Suzaku} observation from 2007 shows a marginally significant hard X-ray excess that could be modeled either by a strong reflection component or by a Compton-thick partial covering absorber \citep{Reeves2009}. Reprocessed and scattered X-ray emission off the surface of an accretion disk wind could explain the low flux and hard X-ray spectrum of the \textit{Suzaku} observation from 2011, but this scenario is only applicable for this particular state \citep{Reeves2014}. There is no sign of dominating reflection in the spectra of observations ``CN'' and ``XN''. As adding a reflection component to the continuum did not significantly improve the fit (see section \ref{cont}), we tried to consider neutral reflection in our photoionization modeling (using \texttt{pexmon}), because the ionized absorbers provide a better constraint on the complicated spectral shape around 9 keV compared to the continuum model. However, adding this \texttt{pexmon} reflection component to the photoionization model with the two UFOs did not improve the fits ($\Delta$C=1.66 for observation ``CN'', and $\Delta \chi^{2}$=-12.3 for observation ``XN'', for three parameters of interest). Furthermore, the parameters of this reflection component, i.e. the reflection factor, the abundance and the inclination, were poorly constrained during the fits for both observations, even through the MCMC approach.

\subsection{Thermal stability}

\begin{figure} [!h]
\resizebox{\hsize}{!}{\includegraphics[trim = 18mm 10mm 0mm 0mm, clip, angle=0]{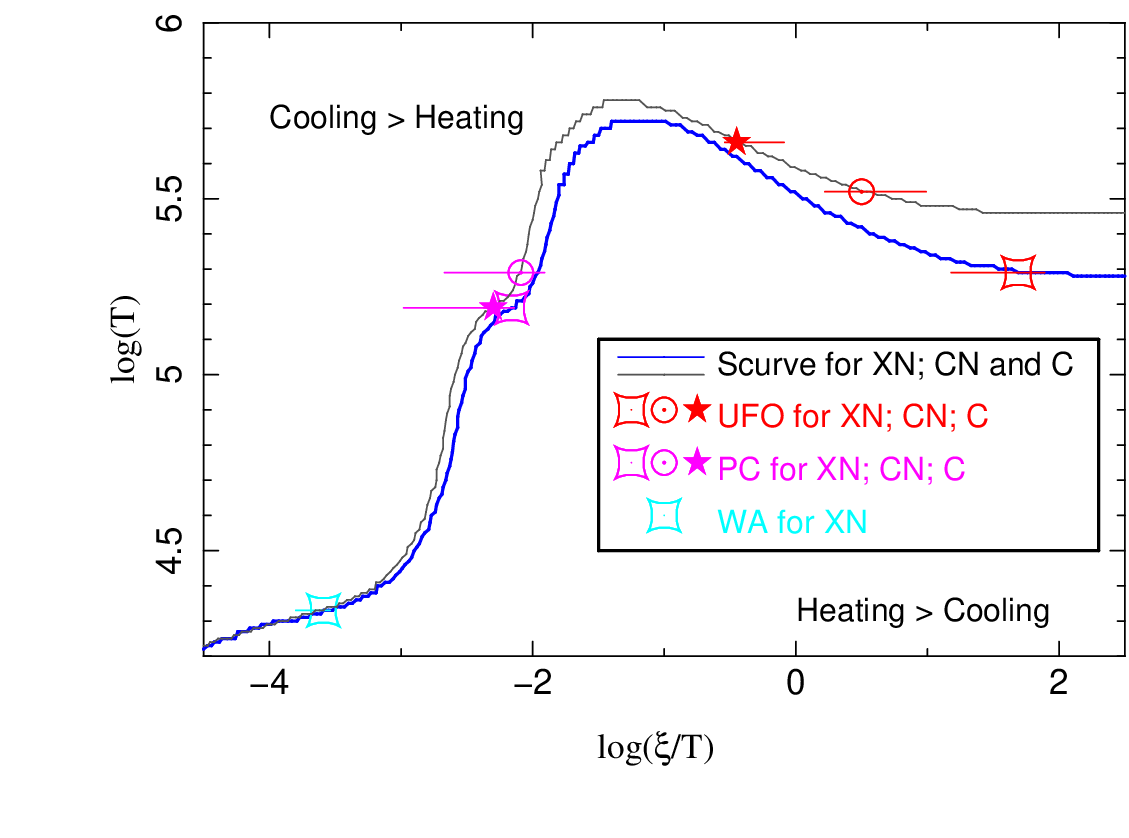}} 
\caption{Thermal stability curve for the photoionized gas in PDS 456, calculated for the SED described by \cite{Nardini2015} for observation ``XN'' (dark blue line) and for the slightly modified SED for observations ``CN'' and ``C'' (grey line). It shows the distribution of equilibrium temperature $log(T)$ as a function of $log(\xi/T)$. The position of the absorbers detected in the three observations are overplotted, according to their $\xi$ values and error bars at the 68\% confidence level resulting from the MCMC routine. Red points represent the ionization of the photoemission and the UFOs, pink color is used for the ionization of the partial covering (PC) absorber, and the light blue represents the ionization of the warm absorber (WA) of observation ``XN''.
}
\label{Scurve}
\end{figure}

To investigate the effects of the continuum on the ionization balance and thermal stability of the photoionized gas, we produced thermal stability curves, plotting the temperature of the plasma $log(T)$ as a function of $log(\xi/T)$ \citep{Krolik1981,Reynolds1995,Krolik2001,Chakravorty2009,Lee2013}.
These curves represent the thermal equilibrium of the gas. On one side of the curve, cooling dominates, while on the other side, at high ionization and low temperature, heating dominates. Generally, a positive gradient of the curve indicates thermal stability for the gas (indeed, a small increase of temperature will increase the cooling, while a perturbation decreasing the
temperature will increase the heating). Conversely, a negative gradient is present in regions of instability. 
We used the two SEDs described in Section \ref{warmabs} to produce the thermal stability curves shown in Figure \ref{Scurve}. We remind here that these two SEDs are identical in the UV band, and slightly differ in X-rays in order to match our three datasets. Note that the stability curve depends on the input assumptions provided to the photoionization code. Its shape is influenced by the ionizing SED (e.g. \citealt{Lee2013,Mehdipour2015}), and by the density and chemical composition of the absorber (e.g. \citealt{Chakravorty2009}).
The unusual decrease of temperature at high ionization can be explained by the fact that the SED we used for the calculation of the XSTAR models has a strong soft component (as shown in Figure 5 from \citealt{Matzeu2016b}), and as the Compton temperature ($T_{IC}=<E>/4k$)  depends on the mean energy photon, the value of the Compton temperature is low. When the gas is fully ionized, it reaches this Compton temperature \citep{Kallman2001}. However, when the gas is partially ionized, it can reach hotter temperatures, explaining the peak shown in Figure \ref{Scurve}. Indeed, in addition to the heating of the gas from the energetic electrons produced by photoionization, these electrons can heat the gas to a higher temperature through secondary collisional ionization of neutral atoms.

We overplotted points representing the absorbers, whose ionization parameters have been characterized by the photoionization models in the three observations, on top of their respective stability curves. We can see that the partial covering absorber is in a stable state in all observations (pink points), on a portion of the curve with a positive slope. This is also the case for the non-blueshifted warm absorber (light blue rectangle) detected in observation ``XN''. However, because the UFO components of the observations ``XN'', ``CN'' and ``C'' are on a negative gradient branch (red points), we might expect that such gas should be thermally unstable. 
However, the cooling time for the UFO gas is much shorter than the outflow time ($R/v_{out}$), so the gas should be stable in order for the UFO to be persistent and observable. In fact, the thermal stability of the gas can be determine by the slope of the branches, as described above, only in the case of ``S'' shapes. However, for our case, the thermal stability curve, with its unusual negative slope at high ionization, does not follow this rule. Indeed, as shown on Figure \ref{Scurve}, at high ionization the cooling dominates above the curve and the heating dominates below the curve. So on this high ionization branch, a small perturbation increasing the temperature will bring the gas to a region where the cooling dominates, and a small decrease of temperature will increase the heating. The UFOs located on this branch at high ionization are thus stable after all.

\begin{figure} [!h]
\resizebox{\hsize}{!}{\includegraphics[trim = 15mm 10mm 0mm 0mm, clip, angle=0]{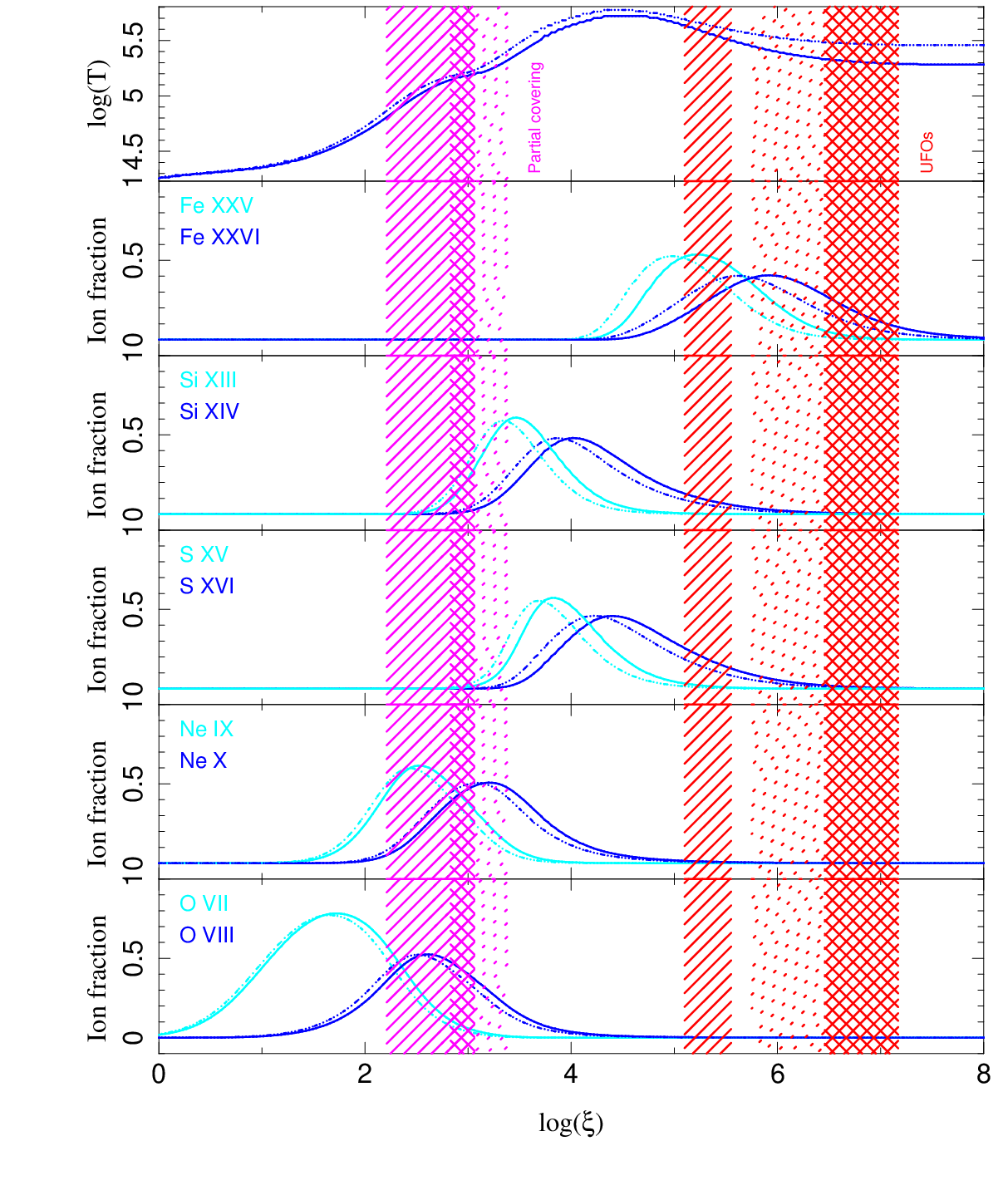}} 
\caption{Distribution of the temperature and of the ion fractions of elements that have been identified in our three datasets, as a function of the ionization parameter. Solid lines corresponds to the SED used by XSTAR for ``XN'' observation, while dashed lines represent the SED used for ``CN'' and ``C'' observations. The partial covering absorbers are identified in pink, while UFOs are represented in red (``CN'': dotted zone, ``XN'': checkered zone, ``C'': hatched zone).
}
\label{IonicCurve}
\end{figure}

Figure \ref{IonicCurve} shows the distribution of the temperature as a function of the ionization parameter, as well as the distribution of ion fractions of the H-like and He-like ions of the elements identified in the three datasets, as described in previous sections (Fe, Si, S, Ne and O). This distribution is dependent on the SED used by the photoionization code XSTAR, as we can see the difference between solid and dashed lines (see legend of the figure for more details). We overplotted the values of the ionization parameters obtained by photoionization modeling, for both partial covering absorbers (in pink) and UFOs (in red), in the three observations (``CN'': dotted zone, ``XN'': checkered zone, ``C'': hatched zone).  
We can see that the ions of Ne and O detected in the three observations can coexist in a single ionization zone at the same velocity, originating from the partial covering absorbers (pink zone). Iron ions are also detected in the three datasets, and can coexist with S and Si ions in a single ionization zone corresponding to the UFOs (red zone). This co-existence is more significant in observation ``C'' than in observations ``XN'' and ``CN''. S and Si ions could either be produced by the UFOs or by the partial covering absorbers. This uncertain origin is induced by the fact that, to simplify the model applied to our data, we linked together the velocities of the partial covering absorber and of the slowest UFO, as well as the ionization parameters between both UFOs. Higher quality data are required to allow the modeling with untied parameters and hence the precise determination of the origin of the detected S and Si ions. In the present study, despite their dependence on the assumptions made for the fitting, the distributions of ion fractions of different elements show that their co-existence is possible, giving indications of their origins, and thus support the results from the photoionization modeling, as well as the identification of the lines detected by the blind line search.

\subsection{Compare results from different epochs}
\label{comparison}

\begin{figure} [!h]
\resizebox{\hsize}{!}{\includegraphics[trim = 0mm 0mm 0mm 0mm, clip, angle=0]{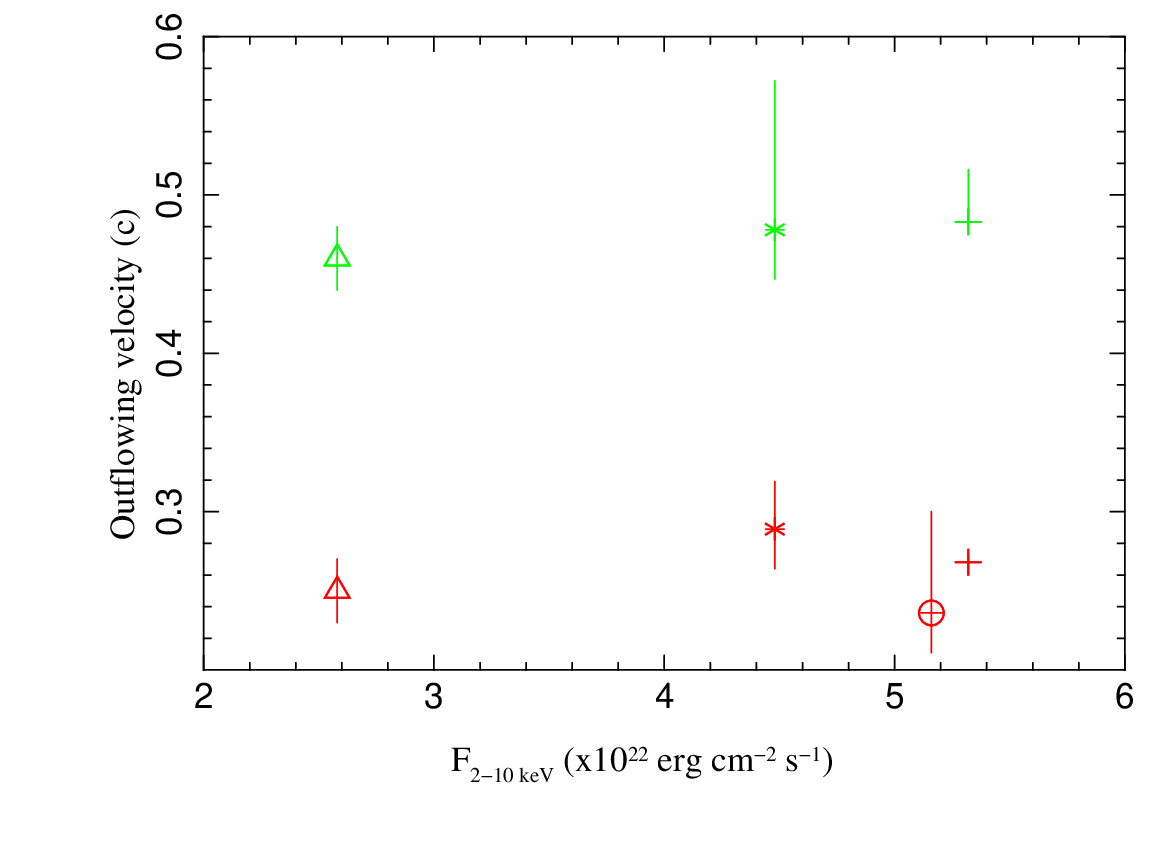}} 
\resizebox{\hsize}{!}{\includegraphics[trim = 0mm 0mm 0mm 0mm, clip, angle=0]{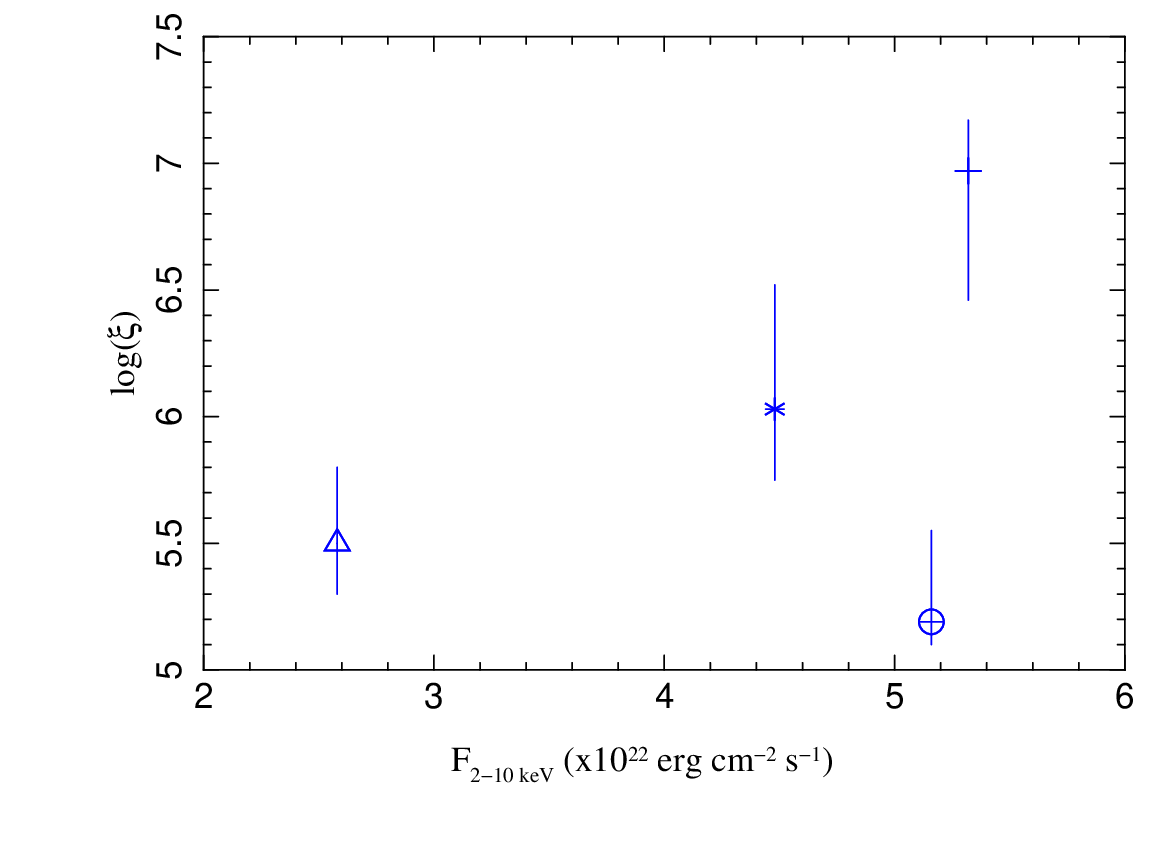}} 
\caption{Top: Outflow velocity as a function of the intrinsic flux between 2 and 10 keV. Red points represent the values for the slowest UFO, green points are for the fastest UFO. Triangles represent results of \cite{Reeves2018}, ``x'' symbol is used for our ``CN'' observation, ``+'' symbol for observation ``XN'' and circles for observation ``C''. Bottom: Ionization parameter of the UFOs resulting from photoionization modeling as a function of the intrinsic flux between 2 and 10 keV. The symbols used in bottom figure are the same as in top figure.}
\label{corr}
\end{figure}
 
We found that two UFOs are significantly detected in our observations ``CN'' and ``XN'', the detection being even more convincing in observation ``XN''. The velocity of the slowest UFO obtained by photoionization modeling on observation ``XN'' ($v_{out1}=-0.268_{-0.004}^{+0.007}c$) is consistent with the average value measured on the five individual ``XN1'' to ``XN5'' observations by \cite{Nardini2015} ($v_{out1, average}=-0.25_{-0.01}^{+0.01}c$). 

The second UFO was not detected in the individual ``XN'' observations as claimed by \cite{Reeves2018b,Reeves2018}. Indeed, after finding a second, faster absorber in recent \textit{XMM-Newton} and \textit{NuSTAR} data from 2017, the authors re-analyzed the ``XN1'' to ``XN5'' observations, but did not find any signature of a second UFO above 10 keV. 
For our work, we chose to combine all the ``XN'' observations together, as explained in Section \ref{bin},
as we wanted to get optimum signal-to-noise using all available data. 
We performed a simple check of the detection of the fastest UFO in the individual ``XN'' observations, by applying the fitting procedure described in Section \ref{Fe} on these five datasets, using two sets of Fe lines to look for signatures of both UFOs. 
Regarding the improvement of the fits statistics when considering two UFOs instead of one, and according to the significant detection of the \fexxvi\ Ly$\alpha$, Ly$\beta$ and K-edge features from the fastest UFO, we found that the second UFO is required by the data ``XN5'' ($\Delta\chi^{2} / \Delta \text{dof}$=19.03/5, >99.9\% confidence level), ``XN2'' ($\Delta\chi^{2} / \Delta \text{dof}$=15.82/5, about 99.5\% confidence level) and ``XN4'' ($\Delta\chi^{2} / \Delta \text{dof}$=10.16/5, >90\% confidence level). However, we did not detect the second UFO in observations ``XN3'' and ``XN1''. Looking at their recent observations from 2017, \cite{Reeves2018} detected the second UFO with a significance >99.9\% (the addition of the Ly$\alpha$ from the fastest UFO improving the fit by $\Delta \chi^{2}$=39.3 for two free parameters). The detection of the lines from the second UFO is less significant in ``XN5'', ``XN2'' and ``XN4'' data than in the more recent data from 2017; this is probably why \cite{Reeves2018b,Reeves2018} considered that this fastest component was not visible in the individual ``XN'' observations. According to the count rate of each observation shown in Table \ref{Obs}, and to the lower flux of the recent data from 2017, there seems to be a hint of an anti-correlation between the statistical confidence level for the detection of the fastest UFO and the flux intensity of PDS 456. Such a trend is consistent with the proposition from \cite{Reeves2018b} who suggested that PDS 456 has to be in a low state to allow the detection of the second UFO (see further discussion in section \ref{energetics}). 

In our observations ``CN'' and ``XN'', the column density of the fastest UFO is smaller than the one of the slowest UFO (see Table \ref{warmabsTable}), consistent with results from \cite{Reeves2018}.

Figure \ref{corr} shows the outflow velocities (top figure) and the ionization parameters (bottom figure) of the UFOs detected in our analysis and in the \textit{XMM-Newton/NuSTAR} data from 2017 \citep{Reeves2018}, as a function of the intrinsic flux of PDS 456 between 2 and 10 keV. We can see that the velocities are similar between the different observations. In the study of twelve previous X-ray observations of PDS 456, \cite{Matzeu2017b} depicted a strong correlation between the outflow velocity and the X-ray luminosity, which supports the hypothesis of a radiatively driven wind in PDS 456, boosted by line driving \citep{Hagino2016}. The recent detection of the \civ\ BAL at the velocity of 0.3c in \textit{HST} UV observations may also boost the opacities in UFOs for radiative driving \citep{Hamann2018}. This hypothesis of radiative driving is also strongly supported by the fact that the supermassive black hole of PDS 456 is accreting at a regime near the Eddington limit \citep{King2003,Gofford2014,Matzeu2017b}.
We checked the correlation between velocity and luminosity of the data shown in Figure \ref{corr}. We only found a possible small correlation between intrinsic flux and velocity of the fastest UFO (Pearson correlation coefficient r=0.995, p-value=0.06), and no correlation between the flux and the velocity of the slowest UFO (r=0.11, p-value=0.89). The outflowing velocities seem rather to have stable values. We only found a non-significant correlation for the expected positive relation between the ionizing flux and the ionization parameter (r=0.40, p-value=0.60), probably because of the less precise constraint on the slowest UFO with \textit{Chandra} data from the observation ``C'' compared to the parametric constraints derived from additional high energy data available in the observations ``CN'' and ``XN''.

\subsection{Winds mass outflow rates and energetics}
\label{energetics}

The mass outflow rate is:
\begin{equation}
\dot{M}_{out}\sim\Omega N_{H} m_{p} v_{out} R_{in}
\end{equation}
where $\Omega$ is the solid angle, $N_{H}$ is the column density, $m_{p}$ is the mass of the proton, $v_{out}$ is the outflowing velocity, and $R_{in}$ is the starting point of the wind (see details for example in \citealt{Nardini2015}). We used an approximated value of $\Omega=2\pi$ for the solid angle, as justified by \cite{Nardini2015} (see their discussion in supplementary material for further details, as well as the argumentation from \citealt{Reeves2018b}). The column density is directly derived from the photoionization modeling, as well as the outflow velocity (see Table \ref{warmabsTable}). In the case of a radiatively accelerated wind (e.g. \citealt{Matzeu2017b}), $R_{in}$ can be approximated as: 
\begin{equation}
R_{in}\sim 2(\alpha \frac{L}{L_{Edd}} -1)(\frac{v_{\infty}}{c})^{-2}
\end{equation}
where $v_{\infty}$ is the wind terminal velocity and $\alpha$ is a force multiplier factor \citep{Reeves2018}. PDS 456 is accreting at about the Eddington limit, so we could approximate $L/L_{Edd}=1$. Using a factor of $\alpha=2$ (as done by \citealt{Reeves2018}) and the outflow velocities derived from photoionization modeling for each of our three observations, we find that the launching radius is $\sim5\times10^{15}$ cm (or 30$R_{g}$ for $M_{BH}=10^{9}M_{\odot}$) for the slowest UFO, and $\sim1\times10^{15}$ cm (or 9$R_{g}$) for the fastest UFO. These values  are close to the escape radii $R_{escape}=\frac{2GM}{v_{out}^{2}}$, i.e. the minimum radii from which winds of a given outflow velocity can be launched. 
We are probably observing a stratified ultra-fast outflow in PDS 456, with several components from multiple stratification layers having different velocities and being launched from the accretion flow close to the supermassive black hole (e.g. \citealt{Tombesi2013,Reeves2018}). 
\cite{Reeves2018} proposed that PDS 456 has to be in a low state for the source to not be extremely luminous (and the iron fully ionized), to allow the detection of the second UFO at such small distances. This trend seems to be verified considering the high confidence level of the detection of the second UFO in individual ``XN'' observations as a function of the flux, as explained in section \ref{comparison}.
An alternative explanation could be that a partially covering dense gas (as the one already observed in previous studies) is shielding the innermost wind, preventing it from getting too highly ionized \citep{Matzeu2016a,Reeves2018}.

The maximum radial distance of the absorbers can be estimated by considering that $\Delta R/R<1$, i.e. their thickness cannot exceed their distance from the ionizing source (e.g. \citealt{Reeves2003,Reeves2018}). Using the definition of the ionization parameter given in section \ref{warmabs}, $R_{max} < L_{ion}/N_{H}\xi$. This gives maximum radii of $7\times10^{17}$ cm (or 4700$R_{g}$) for observation ``CN'', $2\times10^{16}$ cm (or 130$R_{g}$) for observation ``XN'', and $6\times10^{18}$ cm (or 40000$R_{g}$) for observation ``C''. Thus, the outflows in PDS 456 may extend to the Broad Line Region. 

For the slowest UFO, we found a mass outflow rate  of $1.2M_{\odot}/yr=0.05 \dot{M}_{Edd}$ for observation ``CN'', $4.4M_{\odot}/yr=0.2 \dot{M}_{Edd}$ for observation ``XN'', and $0.6M_{\odot}/yr=0.03 \dot{M}_{Edd}$ for observation ``C''. For the fastest UFO, we found $0.4M_{\odot}/yr=0.02 \dot{M}_{Edd}$ for observation ``CN'', and $1.6M_{\odot}/yr=0.07 \dot{M}_{Edd}$ for observation ``XN''. We found a different value for the mass outflow rate of the slowest UFO for observation ``XN'' compared to previous studies (with for example estimated kinetic power and mass outflow rate of $\sim$ 15\% and $\sim$ 50\% of Eddington values
respectively for \citealt{Nardini2015}, 5\% and 40\% for \citealt{Gofford2014}) because we obtained slightly different parameters resulting from the photoionization fit and because we used a smaller value for the inner radius, calculated in the case of radiatively driven winds, while others used a timing approach to estimate it at a few hundreds of gravitational radii \citep{Reeves2009,Nardini2015,Matzeu2016a,Matzeu2016b}. We see that the mass outflow rate of the fastest UFO is smaller than the one from the slowest UFO. We calculated the kinetic power of the winds ($P_{kin}=0.5\dot{M} v_{out}^{2}$) and found kinetic powers of $0.02L_{Edd}$ for observation ``CN'', $0.07-0.08L_{Edd}$ for observation ``XN'' and $0.008L_{Edd}$ for observation ``C'' (again a bit smaller than in previous works, for the same reason as above). Both winds are found to have a similar kinetic energy that contradicts the prediction of a larger power for the faster UFO, estimated by \cite{Reeves2018} using the approximation that $P_{kin}\propto v_{out}^{3}$. The high velocities and high column density characterizing the UFOs detected in our three observations result in a large amount of kinetic power of about 0.8-8\% of the bolometric luminosity, that is sufficient to induce significant AGN feedback according to models of black hole and host galaxy co-evolution \citep{King2003,DiMatteo2005,Hopkins2010}.

\section{Conclusion}
\label{conclusion}
We presented the analysis of simultaneous \textit{Chandra}/HETGS and \textit{NuSTAR} observations of PDS 456 from 2015 (``CN''), simultaneous \textit{XMM-Newton} and \textit{NuSTAR} data from 2013-2014 (``XN''), and \textit{Chandra}/HETGS data from 2003 (``C''). We performed a dual-approach study of these selected observations of the quasar, analyzing data from the three different epochs in a consistent way, using both model-independent and model-dependent techniques.

We confirmed the presence of the persistent ultra-fast outflow at velocity of $v_{out}=-0.24$-$-0.29c$, that was observed in previous studies (e.g. \citealt{Reeves2009,Nardini2015}). We also detected a faster UFO ($v_{out}=-0.48c$) in the ``CN'' and ``XN'' observations, that was reported previously only in very recent observations \citep{Reeves2018b,Reeves2018}. In the model-independent approach, we observed their signatures via deep absorption troughs at about 9 and 11 keV, corresponding to blueshifted, highly ionized iron K-shell transitions, that form P Cygni profiles when considering the associated blueshifted emission. We also identified other lines in the HETGS spectra, blueshifted at the same extreme velocities at lower energies, e.g. \sixiv\ Ly$\alpha$, \sxvi\ Ly$\alpha$, \oviii\ Ly$\alpha$, \nex\ Ly$\alpha$, and possible contribution from nickel at high energy. 

In the model-dependent approach, we performed photoionization modeling to characterize both UFOs ($log(\xi)\sim 6-7 \text{ erg cm s}^{-1}$, $N_{H}\sim 1-8 \times 10^{23} \text{ cm}^{-2}$) as well as the partial covering absorber ($log(\xi)\sim 3 \text{ erg cm s}^{-1}$, $N_{H}\sim 3 \times 10^{22} - 2 \times 10^{23} \text{ cm}^{-2}$, $c_{f}\sim 0.3-0.8$). We found that all the winds detected in the three datasets are thermally stable and can coexist.

The outflow of PDS 456 is probably composed of several components
from multiple layers having different velocities and ionizations, launched from the accretion flow close to the supermassive
black hole, and certainly radiatively driven. Both relativistic components of the outflow are powerful enough to play a role in the evolution of the host galaxy, with mass outflow rates of 2-20\% and kinetic powers of 0.8-8\% of the Eddington values. 

We performed an analysis using different methods that led to consistent results. However, we made some assumptions in order to be able to constrain the winds characteristics, because of the complexity of the models applied to our data. Further simultaneous and high signal data are required in order to test our assumptions.
Future high resolution instruments such as \textit{ARCUS} and \textit{Athena} will be useful to determine more precisely the structure of the high velocity winds in PDS 456.
 

\begin{acknowledgements}
We thank very much the anonymous referee for useful comments and corrections which helped to improve this paper. We gratefully acknowledge Claude Canizares for the \textit{Chandra}/HETGS GTO time to observe PDS 456. 
We sincerely thank Tim Kallman for our private communication about stability curves for photoionized gas. RBM acknowledge Emanuele Nardini and Chris Done for providing the P Cygni profile model used in this paper. We thank Fiona Harrison, PI of {\em NuSTAR}, for DDT time. Support for this work was provided in part by the National Aeronautics and
Space Administration (NASA) through the Smithsonian Astrophysical Observatory (SAO)
contract SV3-73016 to MIT for support of the {\em Chandra} X-Ray Center (CXC),
which is operated by SAO for and on behalf of NASA under contract NAS8-03060. 
\end{acknowledgements}


\newpage
\appendix

\section*{Blind line detection}
\label{blindApp}

We performed a blind line search in the broad energy band for the observations ``CN'' and ``C'', and the hard energy band for the observation ``XN''. To perform this blind line search, we added 50 gaussian lines on top of our continuum (described in section \ref{cont}) that we allowed to be either in emission or in absorption. We used \texttt{gabs} models in order to take into account eventual saturation of the lines. We first fixed the continuum, added one gaussian model (with free energy, width and strength, i.e. three parameters of interest), fitted the data, varied the continuum parameters and fitted again the data, and then fixed everything before running again this sequence for 50 iterations. We kept only statistically significant lines for which we got a $\Delta$C or $\Delta\chi^{2}$ larger than 6.25, corresponding to 90\% confidence level for three degrees of freedom. Note that these estimations of the significance of the lines might be slightly overestimated according to \cite{Protassov2002}. The results of this blind line search and line identification are presented in Table \ref{LinesRozenn} and Figure \ref{RozennIDlines} for observation ``CN'', Table \ref{LinesNardini} and Figure \ref{NardiniIDlines} for observation ``XN'', and Table \ref{LinesChandraOld} and Figure \ref{ChandraOldIDlines} for observation ``C''. The tables give the rest energy of each line, together with its width, equivalent width and significance. We tentatively identified some of the detected emission and absorption lines (reported in the tables), considering the transitions of the strongest lines resulting from the photoionization modeling described in section \ref{warmabs}. 
We also considered weaker contributions from higher Z elements, such as the H-like transition of Nickel. Even if Ni is 20 times less abundant than Fe \citep{Grevesse1998}, and thus negligible in XSTAR photoionization modeling, such a transition could possibly exist in gas with such high column densities and high ionizations as found in Table \ref{warmabsTable}.

For the observation ``CN'', some lines detected blindly have been tentatively identified as blueshifted lines with three different ranges of values (see Figure \ref{RozennIDlines}), as shown by the $z_{out}$ column in Table \ref{LinesRozenn}. Taking into account errors on the energy of the line and its width, and taking also into account that the actual shape of the line may be different from a Gaussian curve (like for radiative recombination continuum), we found a velocity in emission $v_{em}$ that is consistent with the value and error bars found when fitting the data with XSTAR models, i.e. $v_{em}=-0.098$ - $-0.050c$. This is also the case for the velocity in absorption of the second wind $v_{abs2}=-0.512$ - $-0.436c$, with absorption lines from \fexxv\ He$\alpha$, \fexxvi\ Ly$\alpha$, \fexxvi\ Ly$\beta$ and \nixxviii\ Ly$\alpha$, and the lower significance \sixiv\ Ly$\alpha$ line. The velocity of the first wind determined using photoionization models, $v_{abs1}=-0.319$ - $-0.264c$, is consistent with the identification of absorption lines \fexxv\ He$\alpha$, \fexxvi\ Ly$\alpha$, \sxvi\ Ly$\alpha$, \nex\ Ly$\alpha$ and \nixxviii\ Ly$\alpha$. However, the lines tentatively identified as \fexxvi\ Ly$\beta$, \fexvii\ 2p-3d, \oviii\ Ly$\alpha$ (detected only at lower significance), \oviii\ Ly$\beta$ and \neix\ He$\alpha$ show a slightly higher velocity of $-0.355$ - $-0.332c$. This is consistent with the fact that the \fexxvi\ Ly$\beta$ absorption line of the first wind was not detected significantly when fitting the data with sets of Fe K lines (see section \ref{Fe}). However, this \fexxvi\ Ly$\beta$ transition may contribute to the broad absorption feature around 11 keV. The \oviii\ and \neix\ He$\alpha$ absorption lines modeled by our broadband fitting with \texttt{warmabs} come mostly from the partial covering absorber. In our model, we chose to tie the velocity of this partial covering component to the one of the slowest  UFO, consistently with \cite{Nardini2015}, to improve its constraint when fitting the data. Furthermore, we also linked the turbulent velocities for the same reason. The photoionization modeling thus tended to make these \oviii\ and \neix\ lines broader than when detected during the blind line search, it can thus explain the slight difference of blueshift for these lines, and it suggests that the partial covering component may have a slightly different velocity than the slowest UFO.	
		
For the observation ``XN'', some lines have been tentatively identified in the spectrum above 5 keV, consistent with the identification for the observation ``CN'' (see Table \ref{LinesNardini} and Figure \ref{NardiniIDlines}). Due to the large widths of the detected lines, several transitions can be attributed to the same line, as they contribute to the absorption feature. Considering the uncertainties, the blueshift values resulting from this identification are close to the values resulting from the photoionization modeling (see Table \ref{warmabsTable}). The slight difference could be due to the fact that the blind line search has been done only above 5keV, while the photoionization model takes the broadband spectrum into account, determination of the velocities are thus more precise in the later case.

For the observation ``C'', only some emission lines and absorption lines from one UFO have been identified. The blueshifted \fexxv\ He$\alpha$ and \fexxvi\ Ly$\alpha$ absorption features detected in the other observations were not detected but some other lines (\sixiv\ , \sxvi\ ) coming from the slowest UFO (and from the partial covering wind that is set to have the same velocity in our photoionization model) have been identified, as shown in Table \ref{LinesChandraOld} and Figure \ref{ChandraOldIDlines}.  Some emission lines have also been tentatively identified. The velocities derived from these identifications are globally consistent with the velocity values resulting from photoionization modeling (see Table \ref{warmabsTable}).

\section*{Additional kinematic diagnostic}
\label{vshiftApp}
\begin{figure} [!h]
\begin{tabular}{cc}
\includegraphics[trim = 15mm 10mm 0mm 0mm, scale=0.5,clip, angle=0]{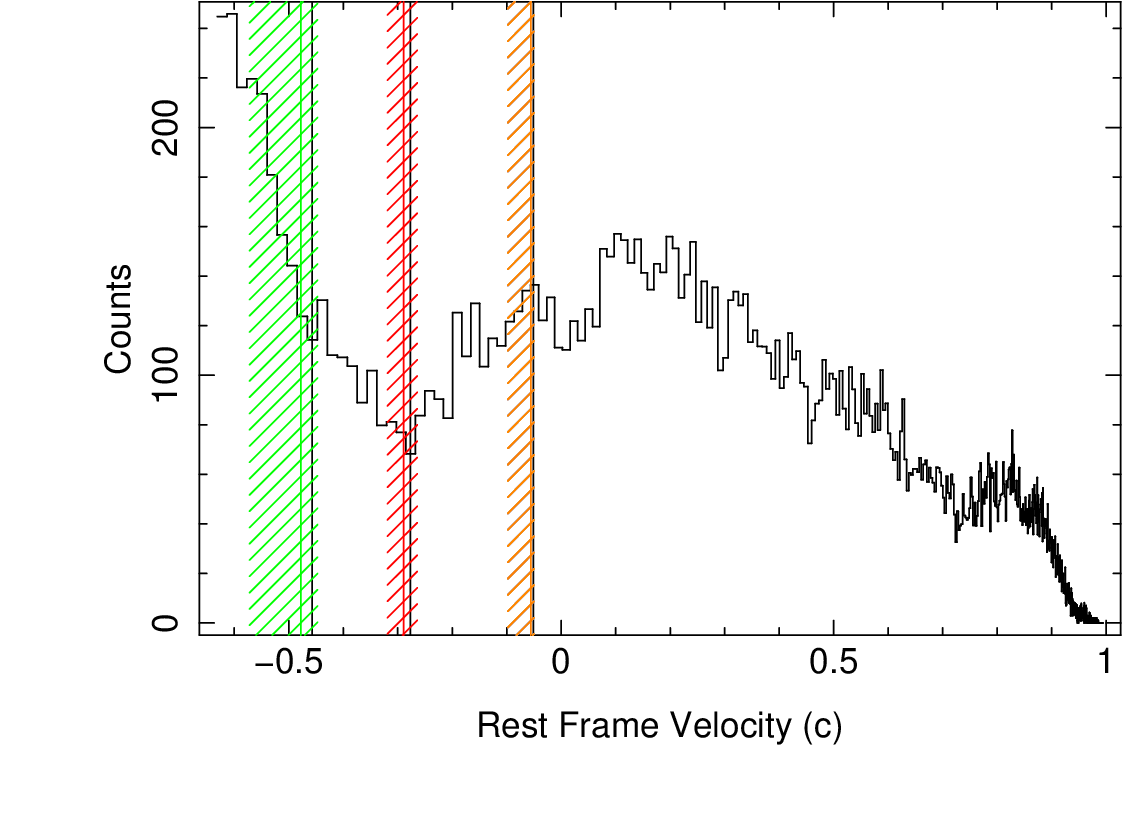} &
\includegraphics[trim = 15mm 10mm 0mm 0mm, scale=0.5, clip, angle=0]{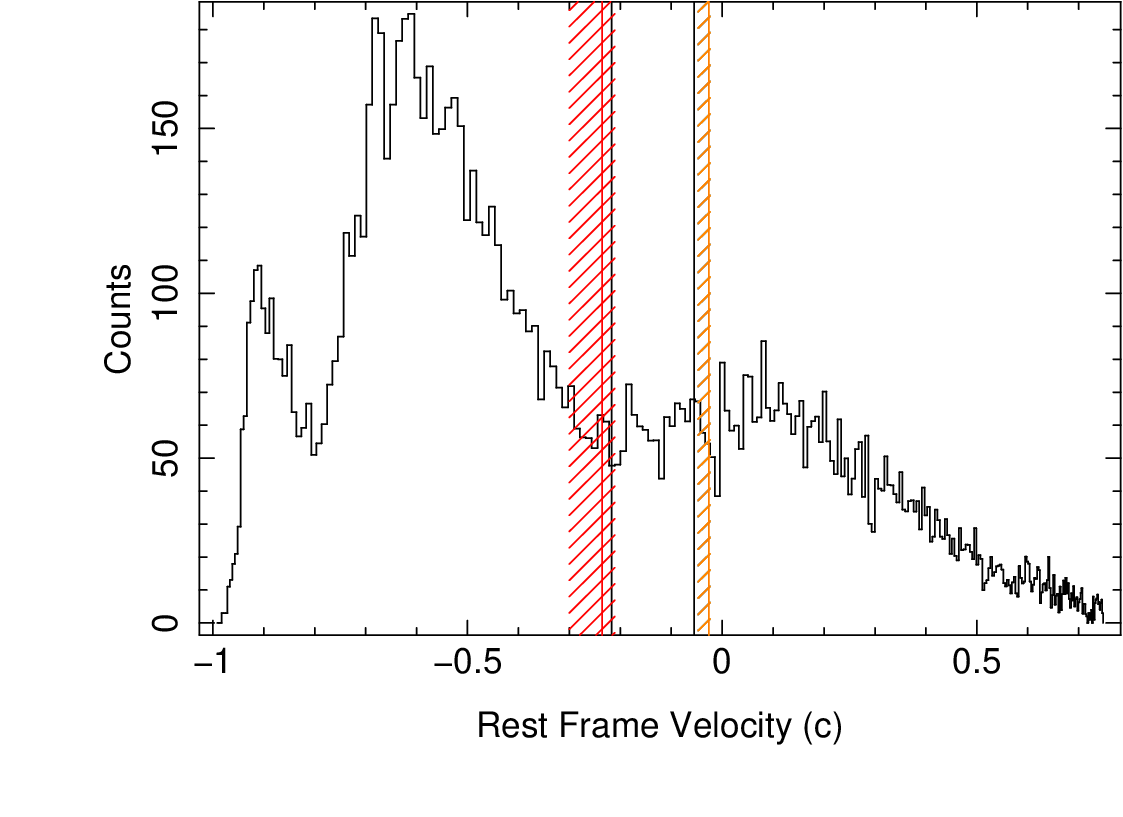} \\
\end{tabular}
\caption{Chandra/HETGS 1st-order detector counts transformed in velocity bins centered on the identified lines in Tables \ref{LinesRozenn} and \ref{LinesChandraOld}. Left: observation ``CN'', right: observation ``C''. Peaks of emission and absorption consistent with photoionization results are indicated with black vertical lines. Velocity values resulting from photoionization modeling are also plotted for comparison (orange zone: emission, red zone: absorption from UFO 1, green zone: absorption from UFO 2).}
\label{vshift}
\end{figure}

\begin{figure*} [!h]
\begin{tabular}{cc}
\includegraphics[trim = 18mm 10mm 0mm 0mm, clip, scale=0.5,angle=0]{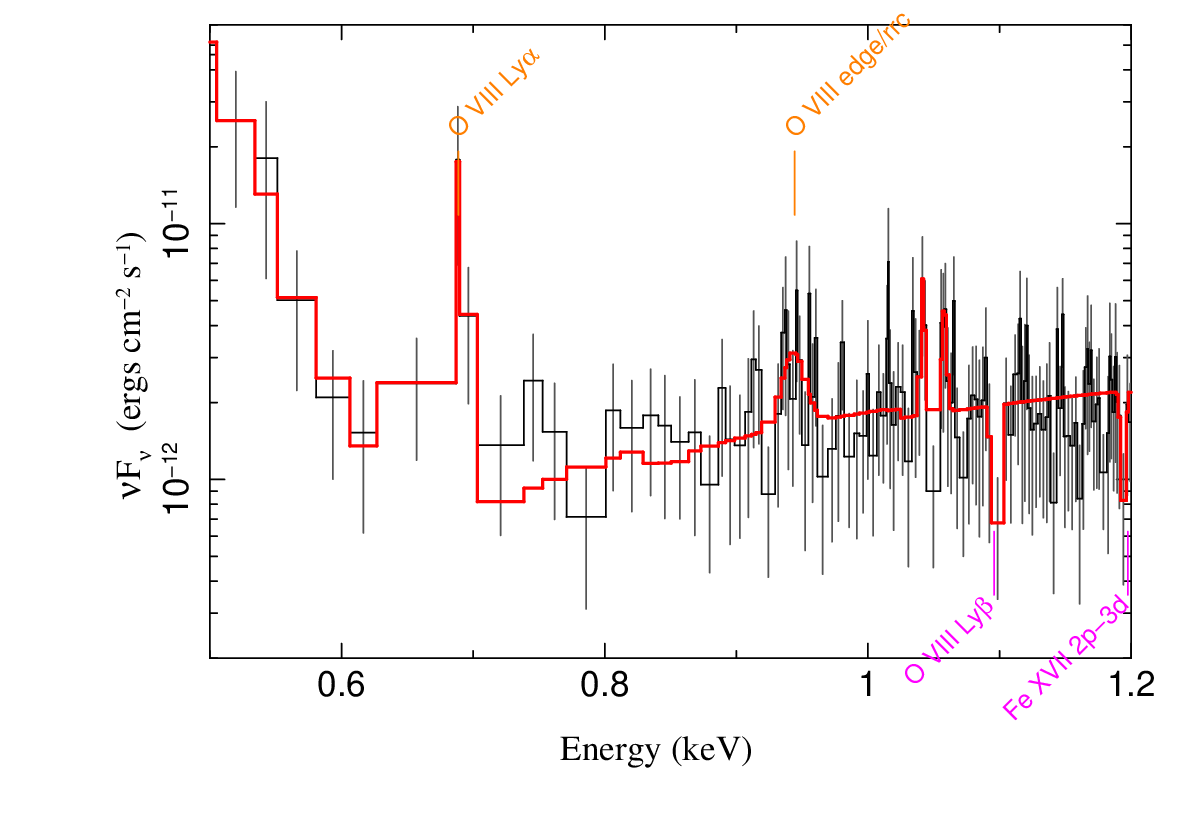} & \includegraphics[trim = 18mm 10mm 0mm 0mm, clip, scale=0.5,angle=0]{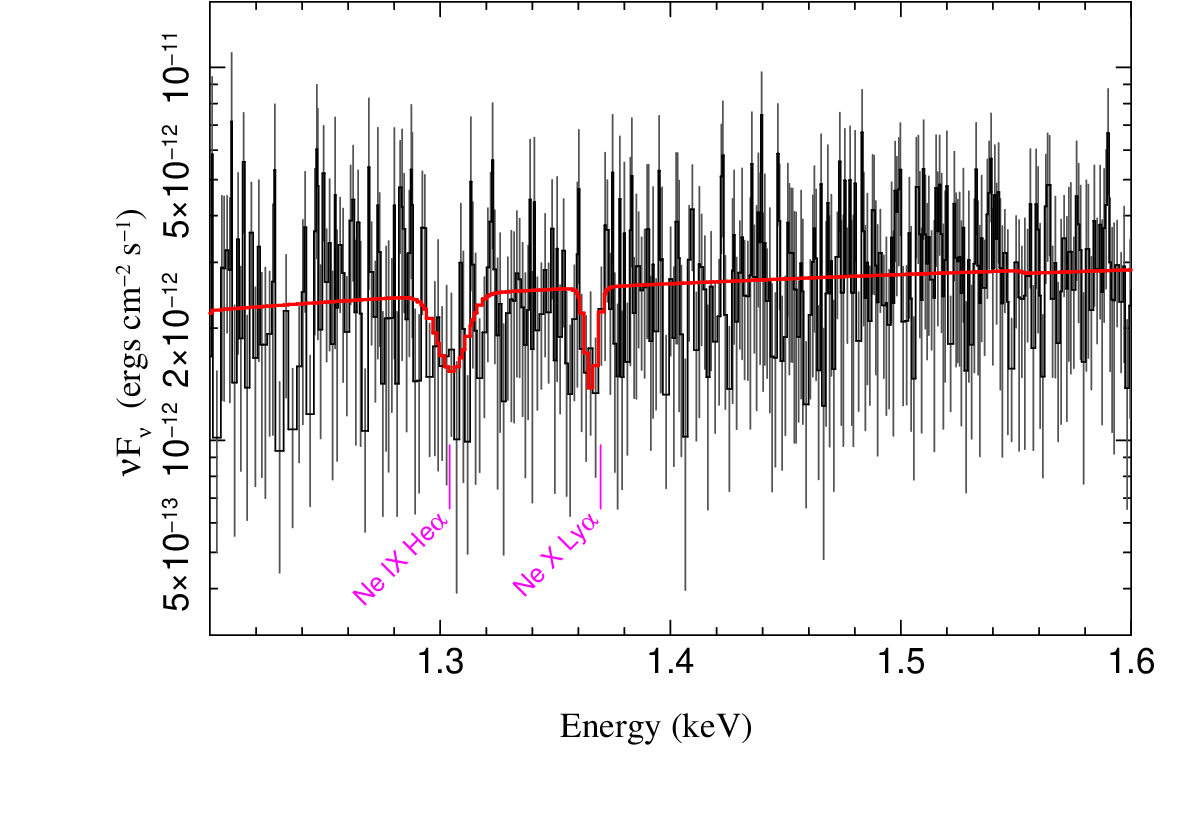}\\
\includegraphics[trim = 18mm 10mm 0mm 0mm, clip, scale=0.5,angle=0]{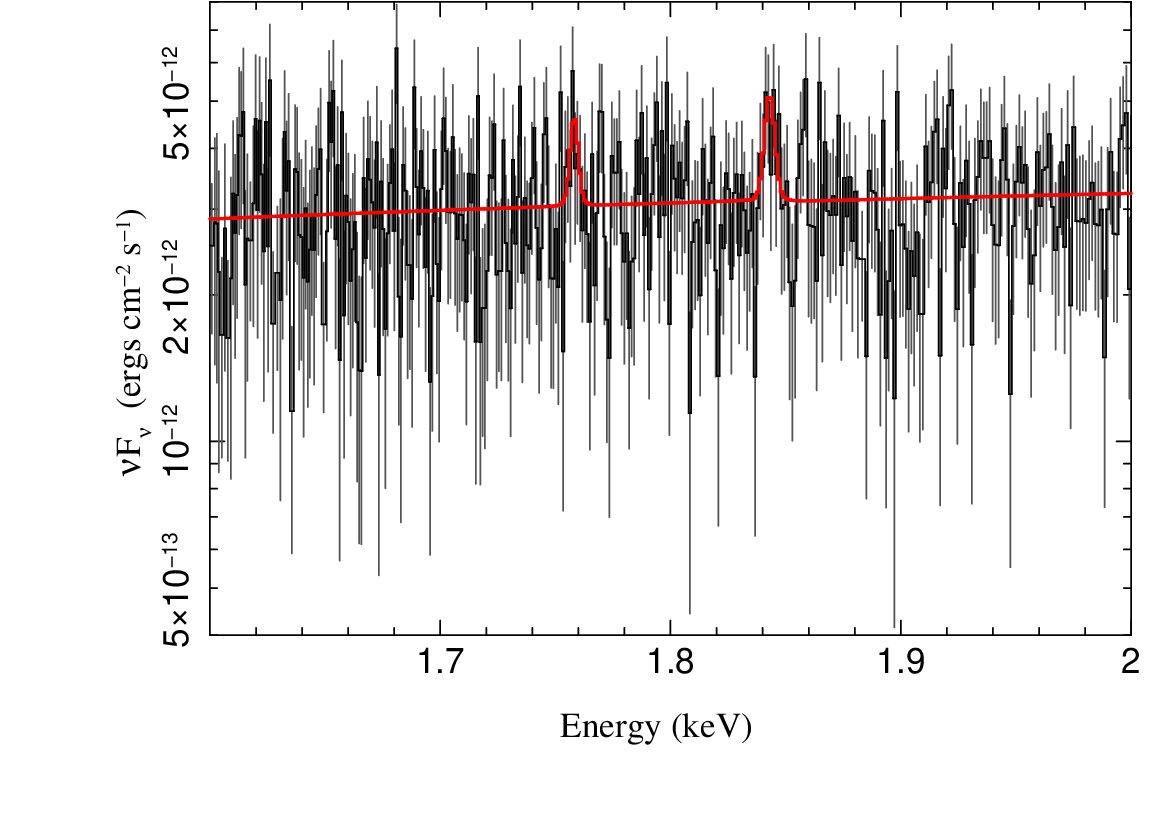} & \includegraphics[trim = 18mm 10mm 0mm 0mm, clip, scale=0.5,angle=0]{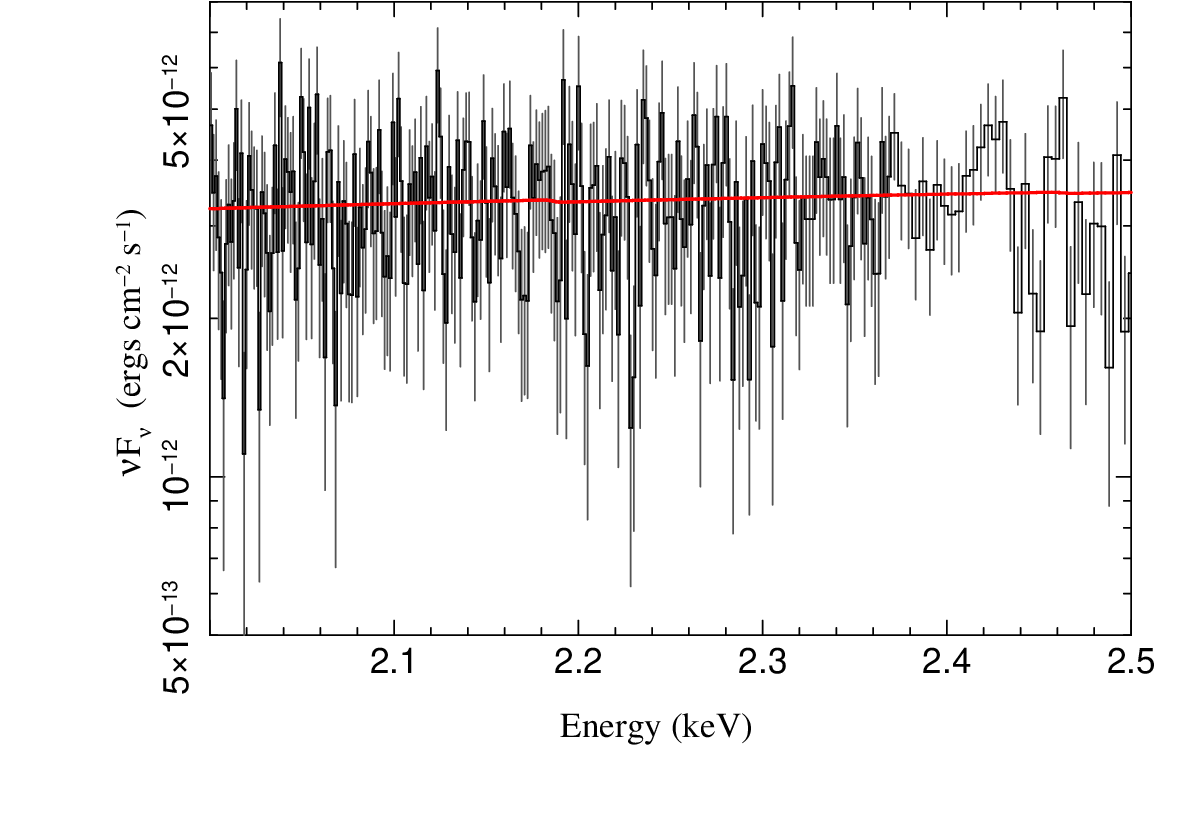}\\
\includegraphics[trim = 18mm 10mm 0mm 0mm, clip, scale=0.5,angle=0]{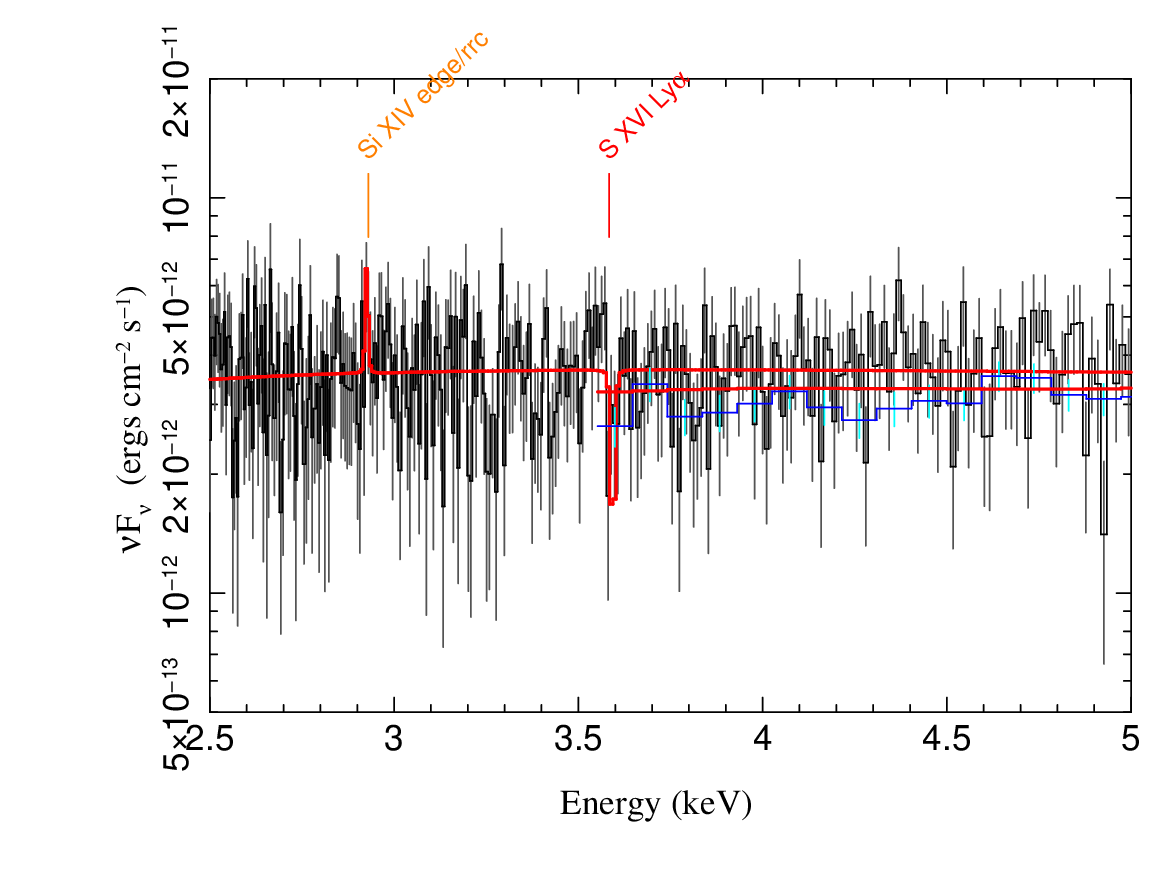} & \includegraphics[trim = 18mm 10mm 0mm 0mm, clip, scale=0.5,angle=0]{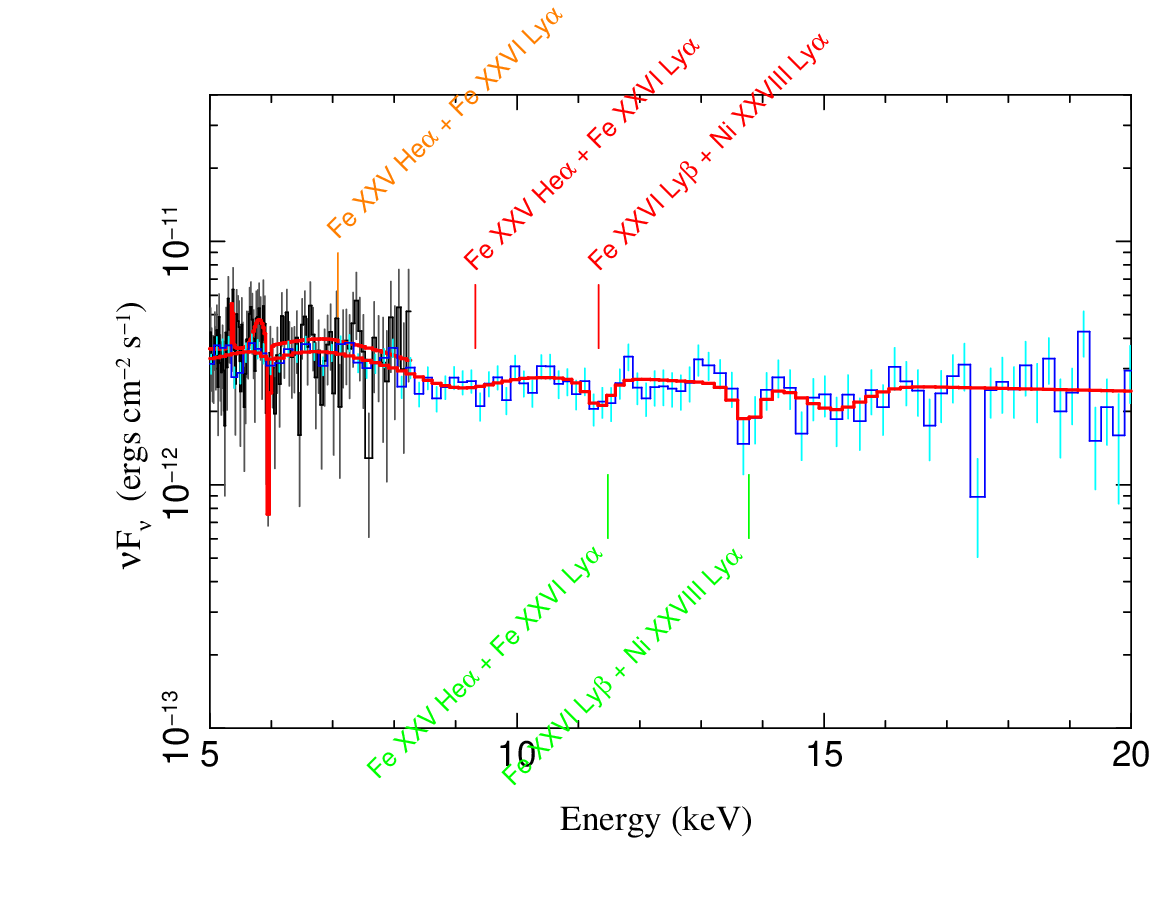}\\
\end{tabular}
\caption{Result of the blind line search described in Section \ref{blindApp}, for observation ``CN'' (entire spectrum). Some lines have been identified in concordance with results from the fitting with photoionization models described in Section \ref{warmabs}. Orange lines come from photoemission, pink absorption lines come mostly from the partial covering absorber, red lines come from the first UFO (with the smallest velocity), and green lines come from absorption by the faster UFO.}
\label{RozennIDlines}
\end{figure*}

\begin{figure*} [!h]
\includegraphics[trim = 18mm 10mm 0mm 0mm, clip, scale=0.5, angle=0]{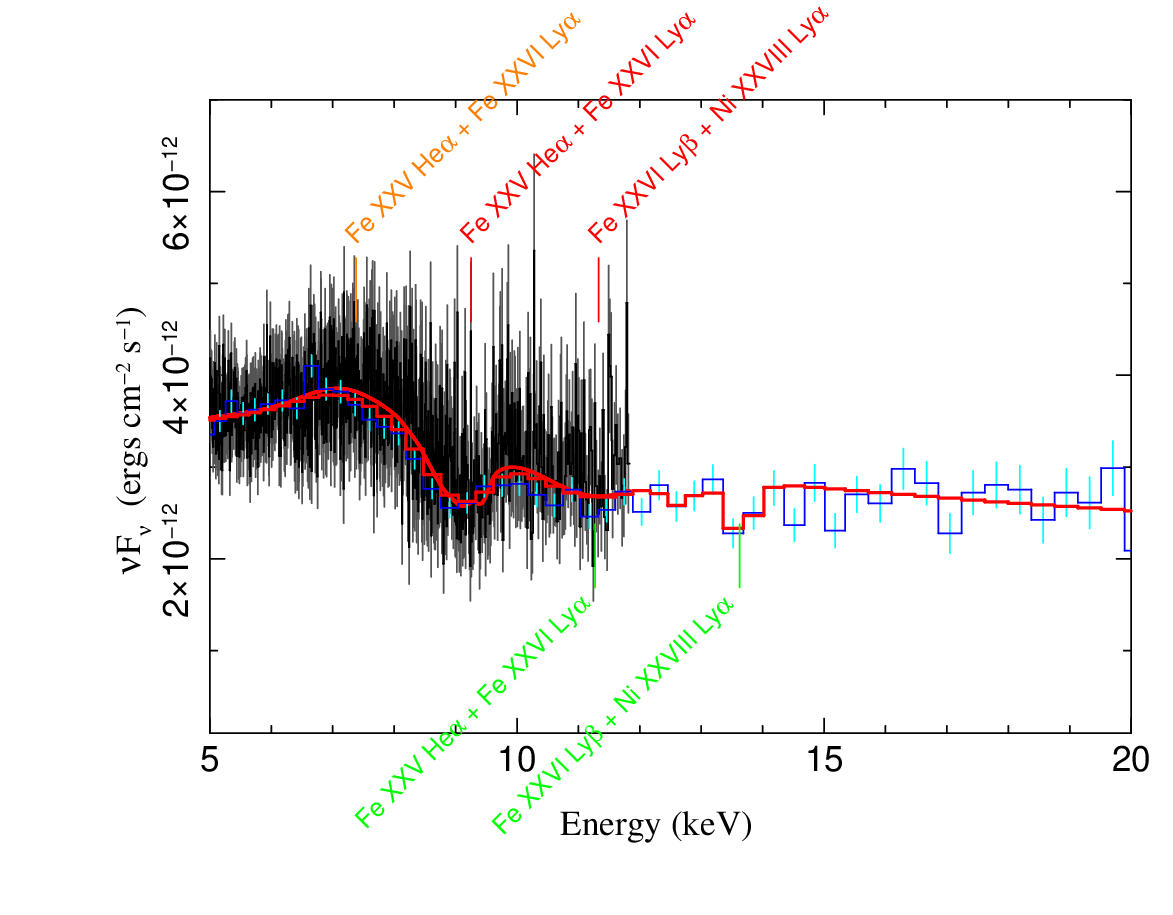}
\caption{Result of the blind line search described in Section \ref{blindApp}, for observation ``XN'' (hard spectrum), similarly to Fig. \ref{RozennIDlines}.}
\label{NardiniIDlines}
\end{figure*}

\begin{figure*} [!h]
\begin{tabular}{cc}
\includegraphics[trim = 18mm 10mm 0mm 0mm, clip, scale=0.5,angle=0]{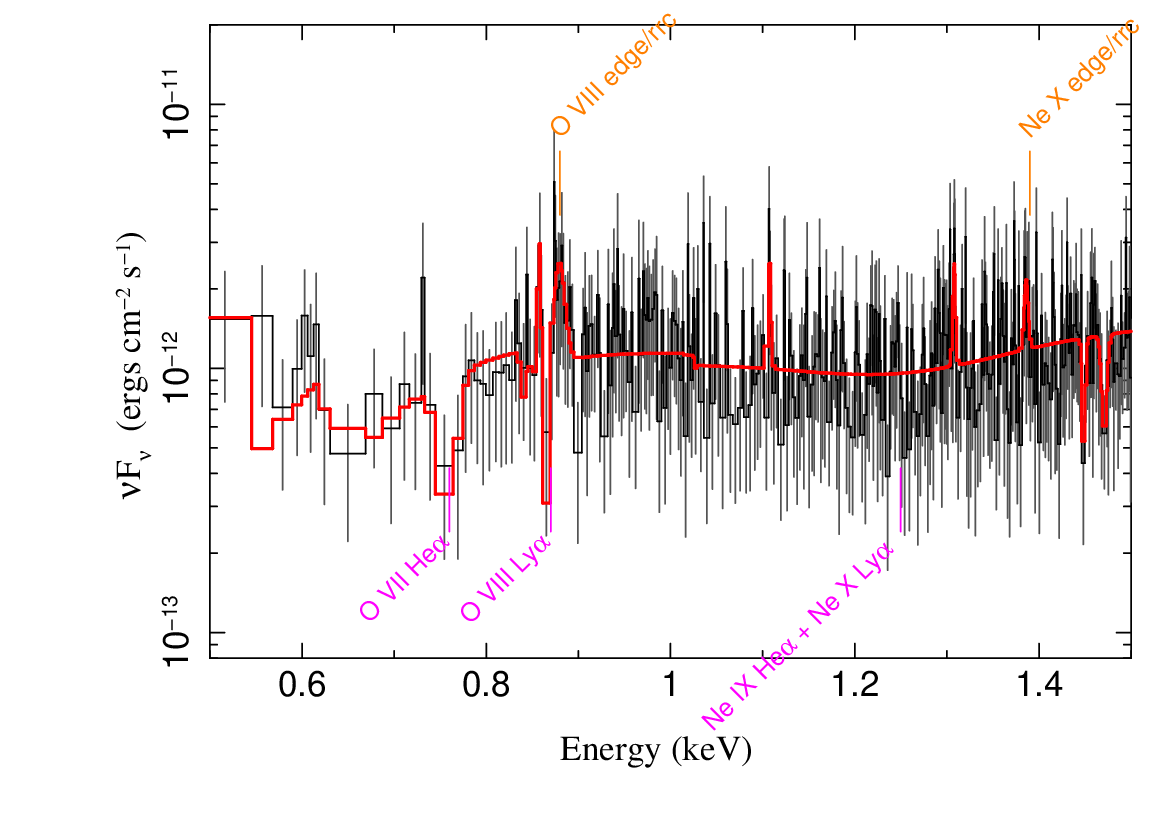} & \includegraphics[trim = 18mm 10mm 0mm 0mm, clip, scale=0.5,angle=0]{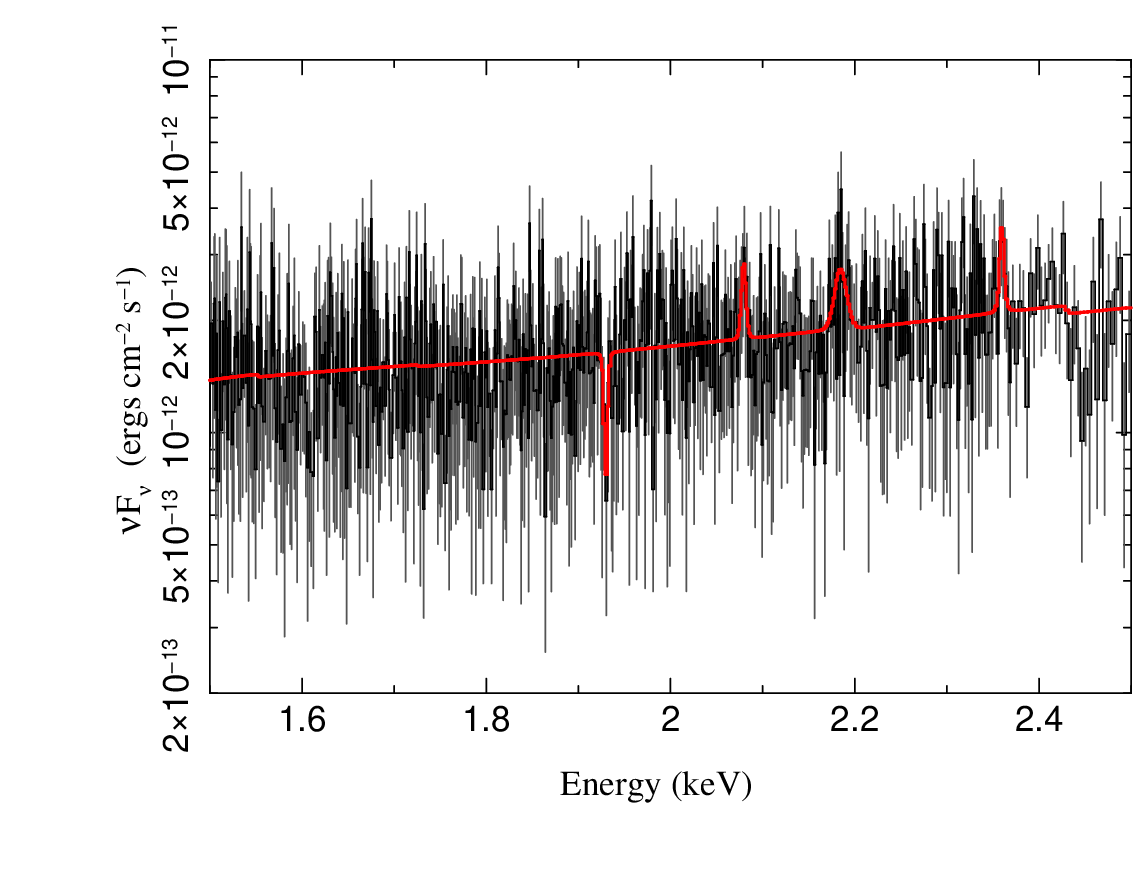}\\
\includegraphics[trim = 18mm 10mm 0mm 0mm, clip, scale=0.5,angle=0]{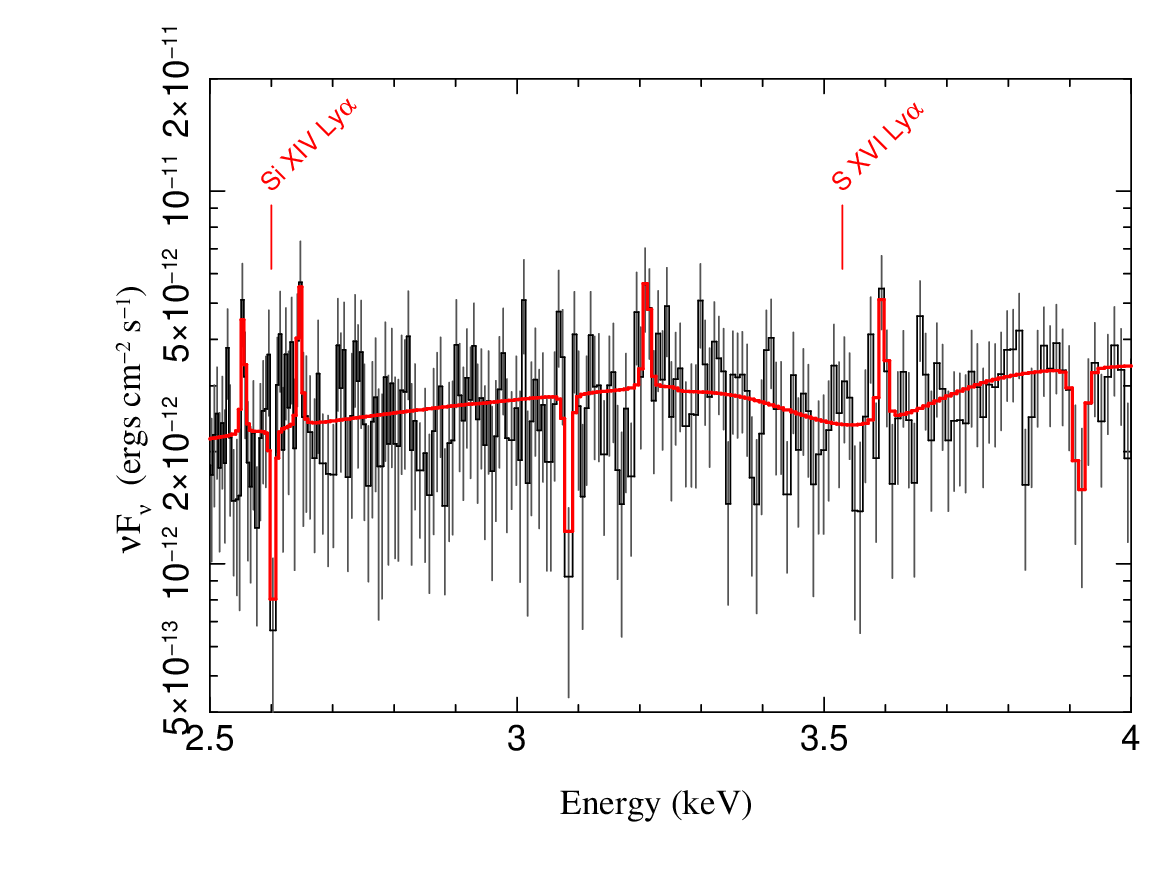} & \includegraphics[trim = 18mm 10mm 0mm 0mm, clip, scale=0.5,angle=0]{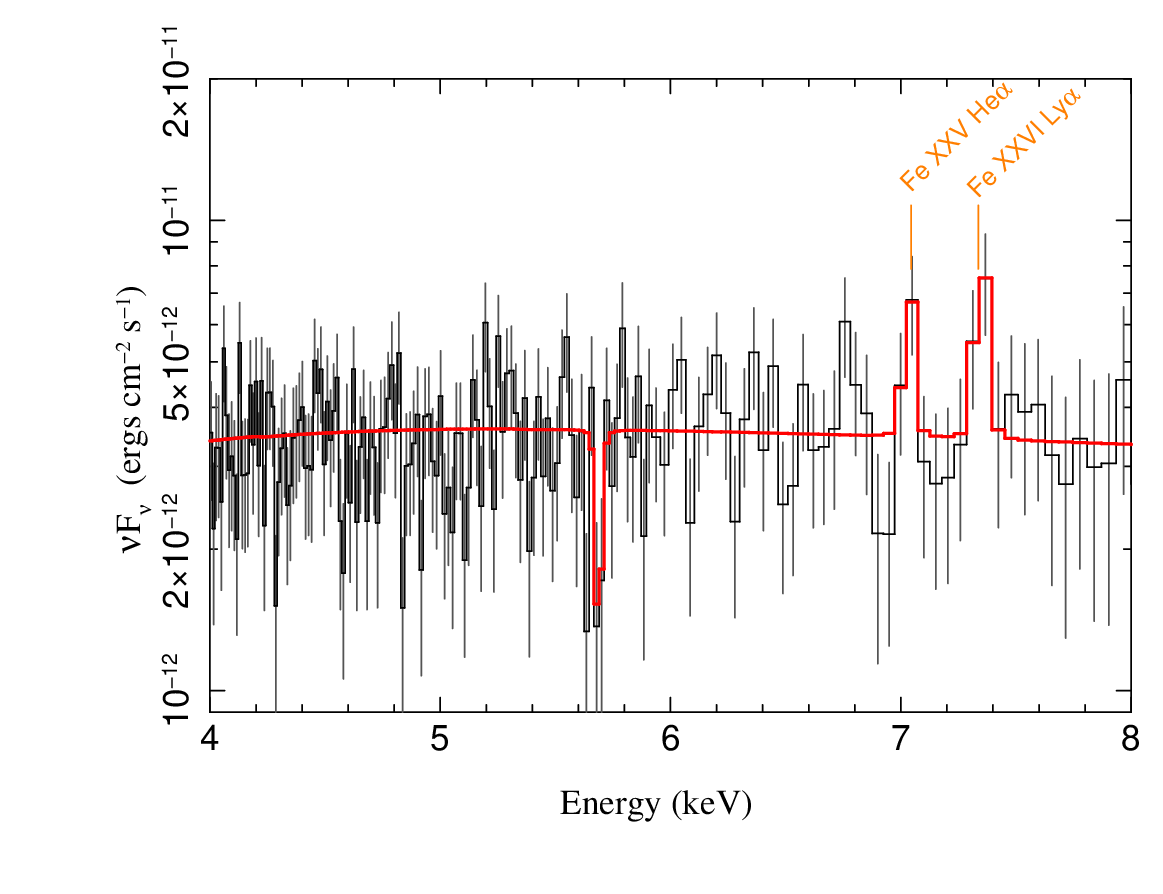}\\
\end{tabular}
\caption{Result of the blind line search described in Section \ref{blindApp}, for observation ``C'' (entire spectrum), similarly to Fig. \ref{RozennIDlines}.}
\label{ChandraOldIDlines}
\end{figure*}

An additional diagnostic for identifying unknown kinematics of the absorbers, as proposed by \cite{Ash2017} in PG 1211+143, is to take the 1st-order \textit{Chandra}/HETGS counts, and use the lines tentatively identified in Tables \ref{LinesRozenn} and \ref{LinesChandraOld} as a set of reference wavelengths, to transform the grid repeatedly to velocities, resulting in Figure \ref{vshift} for the observations ``CN'' and ``C''. We can see that velocity values resulting from photoionization modeling (orange zone: emission, red zone: absorption from the slowest UFO, green zone: absorption from the fastest UFO for the observation ``CN'' only) are consistent (considering uncertainties) with the peaks of emission and absorption at certain velocities (black vertical lines) for both observations ``CN'' and ``C''. This analysis provides an additional confirmation of the kinematic components of the UFO detected in PDS 456.

\newpage

\begin{table*}[!h]
\begin{center}
\caption{Blind line search, performed above 0.4 keV, on observation ``CN''.}
\begin{tabular}{|ccccccc|}
\hline
 $\text{E}_{\text{rest}}$ &	 $\sigma$ &	 EW (eV) &	 $\Delta$C 	&	ID	&	$\text{E}_{\text{lab}}$ 	&	$\text{z}_{\text{out}}$ 			\\
 \hline
$0.69	^{+	0.001	}_{-	0.002	}$&$	0.005	^{+	0.0059	}_{-	0.0019	}$&$	220.00	\pm	164.00	$&	30.39	&	\oviii\ Ly$\alpha$ ($v_{em}$)	&	0.65	&$	-0.058	_{	-0.005	}^{	+0.009	}$	\\
$0.94^{+	0.004	}_{-	0.004	}$&$	0.007	^{+	0.0043	}_{-	0.0032	}$&$	15.90	\pm	6.99	$&	9.43	&	\oviii\ edge/rrc ($v_{em}$)	&	0.87	&$	-0.074	_{	-0.007	}^{	+0.008	}$	\\
$1.04^{+	0	}_{-	0.089	}$&$	0.001	^{+	0.001	}_{-	0	}$&$	6.52	\pm	2.80	$&	8.10	&		&		&							\\
$1.06	^{+	0.001	}_{-	0	}$&$	0.001	^{+	0.0017	}_{-	0.0004	}$&$	5.57	\pm	2.78	$&	6.71	&		&		&							\\
$1.10	^{+	0.004	}_{-	0.001	}$&$	0.001	^{+	0.0006	}_{-	0	}$&$	-5.95	\pm	1.55	$&	9.01	&	\oviii\ Ly$\beta$ ($v_{abs1}$, pc)	&	0.77	&$	-0.300	_{	-0.001	}^{	+0.003	}$	\\
$1.19	^{+	0.001	}_{-	0.001	}$&$	0.001	^{+	0.0013	}_{-	0.0001	}$&$	-2.95	\pm	0.98	$&	7.08	&	\fexvii\ 2p-3d ($v_{abs1}$, pc)	&	0.83	&	$-0.303	_{	-0.001	}^{	+0.001	}$	\\
$1.31^{+	0.003	}_{-	0	}$&$	0.005	^{+	0.0021	}_{-	0.0021	}$&$	-5.34	\pm	1.09	$&	7.19	&	\neix\ He$\alpha$ ($v_{abs1}$, pc)	&	0.92	&$	-0.298	_{	-0.001	}^{	+0.003	}$	\\
$1.37^{+	0.001	}_{-	0.001	}$&$	0.002	^{+	0.0013	}_{-	0.001	}$&$	-2.78	\pm	0.90	$&	6.48	&	\nex\ Ly$\alpha$ ($v_{abs1}$, pc)	&	1.02	&	$-0.255	_{	-0.001	}^{	+0.001	}$	\\
$1.76	^{+	0.001	}_{-	0.001	}$&$	0.001	^{+	0.0008	}_{-	0	}$&$	1.93	\pm	0.77	$&	6.93	&		&		&							\\
$1.84	^{+	0.001	}_{-	0.001	}$&$	0.001	^{+	0.0009	}_{-	0.0004	}$&$	3.08	\pm	1.23	$&	12.18	&		&		&							\\
$2.93	^{+	0.003	}_{-	0.002	}$&$	0.001	^{+	0.0031	}_{-	0	}$&$	7.24	\pm	3.15	$&	6.70	&	\sixiv\ edge/rrc ($v_{em}$)	&	2.67	&$	-0.089	_{	-0.001	}^{	+0.002	}$	\\
$3.59	^{+	0.003	}_{-	0.005	}$&$	0.002	^{+	0.0014	}_{-	0.0006	}$&$	-9.77	\pm	3.15	$&	7.52	&	\sxvi\ Ly$\alpha$ ($v_{abs1}$)	&	2.62	&$	-0.270	_{	-0.001	}^{	+0.001	}$	\\
$5.37	^{+	0.008	}_{-	0.002	}$&$	0.001	^{+	0.0535	}_{-	0.0002	}$&$	17.50	\pm	5.75	$&	8.26	&		&		&							\\
$5.80	^{+	0.032	}_{-	0.056	}$&$	0.071	^{+	0.0297	}_{-	0.0426	}$&$	49.60	\pm	20.90	$&	10.71	&		&		&							\\
$5.96^{+	0.009	}_{-	0.014	}$&$	0.006	^{+	0.0087	}_{-	0.0025	}$&$	-42.10	\pm	0.11	$&	11.52	&		&		&							\\
\multirow{2}{*}{$6.91	^{+	0.003	}_{-	0.337	}$}&\multirow{2}{*}{$	0.600	^{+	0	}_{-	0.0528	}$}&\multirow{2}{*}{$	238.00	\pm	22.40	$}&	\multirow{2}{*}{92.71}	&	\fexxv\ He$\alpha$ ($v_{em}$)	&	6.70	&$	-0.030	_{	-0.058	}^{	+0.000	}$	\\
&&&	&	+\fexxvi\ Ly$\alpha$ ($v_{em}$)	&	6.97	&$	0.009	_{	-0.060	}^{	+0.000	}$	\\
\multirow{2}{*}{$9.11	^{+	0.238	}_{-	0.425	}$}&\multirow{2}{*}{$	0.434	^{+	0.1607	}_{-	0.426	}$}&\multirow{2}{*}{$	-169.00	\pm	0.00	$}&	\multirow{2}{*}{7.74}	&	\fexxv\ He$\alpha$ ($v_{abs1}$)	&	6.70	&$	-0.265	_{	-0.076	}^{	+0.031	}$	\\
&&&	&	+\fexxvi\ Ly$\alpha$ ($v_{abs1}$)	&	6.97	&$	-0.235	_{	-0.079	}^{	+0.032	}$	\\
\multirow{4}{*}{$11.34	^{+	0.075	}_{-	0.082	}$}&\multirow{4}{*}{$	0.016	^{+	0.146	}_{-	0.0065	}$}&\multirow{4}{*}{$	-112.00	\pm	0.00	$}&	\multirow{4}{*}{8.94}	&	\fexxv\ He$\alpha$ ($v_{abs2}$)	&	6.70	&$	-0.409	_{	-0.005	}^{	+0.011	}$	\\
&&&&	+\fexxvi\ Ly$\alpha$ ($v_{abs2}$)	&	6.97	&$	-0.385	_{	-0.005	}^{	+0.012	}$	\\
&&&&	+\fexxvi\ Ly$\beta$ ($v_{abs1}$)	&	7.88	&$	-0.305	_{	-0.005	}^{	+0.013	}$	\\
&&&&	+\nixxviii\ Ly$\alpha$ ($v_{abs1}$)	&	8.11	&$	-0.285	_{	-0.006	}^{	+0.014	}$	\\
\multirow{2}{*}{$13.76	^{+	0.061	}_{-	0.1	}$}&\multirow{2}{*}{$	0.024	^{+	0.0783	}_{-	0.0158	}$}&\multirow{2}{*}{$	-161.00	\pm	0.62	$}&	\multirow{2}{*}{6.88}	&	\fexxvi\ Ly$\beta$ ($v_{abs2}$)	&	7.88	&$	-0.427	_{	-0.005	}^{	+0.006	}$	\\
&&&&	+\nixxviii\ Ly$\alpha$ ($v_{abs2}$)	&	8.11	&$	-0.411	_{	-0.005	}^{	+0.006	}$	\\
$15.18^{+	0.127	}_{-	0.633	}$&$	0.349	^{+	0.2506	}_{-	0.0781	}$&$	-28.30	\pm	16.10	$&	7.18	&		&		&							\\
			 \hline
\end{tabular}
 \label{LinesRozenn} 
\end{center}
\end{table*}			

\newpage

\newpage
\begin{table*}[!h]
\begin{center}
\caption{Blind line search, performed above 5 keV,  on observation ``XN''.}
\begin{tabular}{|ccccccc|}
\hline
 $\text{E}_{\text{rest}}$ 	&	 $\sigma$ &	 EW (eV) &	 $\Delta\chi^{2}$ 	&	ID	&	$\text{E}_{\text{lab}}$ 	&	$\text{z}_{\text{out}}$ 			\\
 \hline
\multirow{2}{*}{$7.19	_{-	0.068	}^{+	0.069	}$}&\multirow{2}{*}{$	0.549	_{-	0.055	}^{+	0.050	}$}&	\multirow{2}{*}{124.64 \pm 10.84}	&	\multirow{2}{*}{124.90}	&	\fexxv\ He$\alpha$ ($v_{em}$)	&	6.70	&$	-0.068	_{	-0.016	}^{	+0.015	}$	\\
						&						&		&		&	+ \fexxvi\ Ly$\alpha$ ($v_{em}$)	&	6.97	&$	-0.031	_{	-0.017	}^{	+0.016	}$	\\
\multirow{2}{*}{$9.22	_{-	0.082	}^{+	0.066	}$}&\multirow{2}{*}{$	0.317	_{-	0.023	}^{+	0.031	}$}&\multirow{2}{*}{	-174.48 \pm 7.62}	&	\multirow{2}{*}{208.38}	&	\fexxv\ He$\alpha$ ($v_{abs1}$)	&	6.70	&$	-0.273	_{	-0.008	}^{	+0.008	}$	\\
						&						&		&		&	+\fexxvi\ Ly$\alpha$ ($v_{abs1}$)	&	6.97	&$	-0.244	_{	-0.009	}^{	+0.008	}$	\\
\multirow{4}{*}{$11.24	_{-	0.131	}^{+	0.069	}$}&\multirow{4}{*}{$	0.593	_{-	0.091	}^{+	0.007	}$}&\multirow{4}{*}{	-173.50 \pm 25.12}	&	\multirow{4}{*}{58.42}	&	\fexxvi\ Ly$\beta$ ($v_{abs1}$)	&	7.88	&$	-0.299	_{	-0.014	}^{	+0.005	}$	\\
						&						&		&		&	+ \nixxviii\  Ly$\alpha$ ($v_{abs1}$)	&	8.11	&$	-0.278	_{	-0.015	}^{	+0.005	}$	\\
						&						&		&		&	+\fexxv\ He$\alpha$ ($v_{abs2}$)	&	6.70	&$	-0.404	_{	-0.012	}^{	+0.004	}$	\\
						&						&		&		&	+\fexxvi\ Ly$\alpha$ ($v_{abs2}$)	&	6.97	&$	-0.449	_{	-0.012	}^{	+0.004	}$	\\
$12.64_{-	0.154	}^{+	0.062	}$&$	0.005	_{-	0.003	}^{+	0.003	}$&	-45.17 \pm 23.88	&	6.79	&		&		&							\\
\multirow{2}{*}{$13.62	_{-	0.065	}^{+	0.063	}$}&\multirow{2}{*}{$	0.009	_{-	0.002	}^{+	0.002	}$}&\multirow{2}{*}{	-106.03 \pm 25.86}	&	\multirow{2}{*}{18.15}	&	\fexxvi\ Ly$\beta$ ($v_{abs2}$)	&	7.88	&$	-0.422	_{	-0.003	}^{	+0.003	}$	\\
							&						&		&		&	+ \nixxviii\ Ly$\alpha$ ($v_{abs2}$)	&	8.11	&$	-0.405	_{	-0.003	}^{	+0.003	}$	\\
			 \hline
\end{tabular}
 \label{LinesNardini} 
\end{center}
\end{table*}

\begin{table*}[!h]
\begin{center}
\caption{Blind line search, performed above 0.4 keV,  on observation ``C''.}
\begin{tabular}{|llrlccc|}
\hline
 $\text{E}_{\text{rest}}$ 	 &	 $\sigma$ &	 EW (eV) &	 $\Delta$C 	&	ID	&	$\text{E}_{\text{lab}}$ 	&	$\text{z}_{\text{out}}$ 			\\
 \hline
$0.76	^{+	0.005	}_{-	0.005	}$&$	0.008	^{+	0.0074	}_{-	0.0069	}$&$	-16.8	\pm	37.7	$&	8.88	&	\ovii\ He$\alpha$ ($v_{abs1}$, pc)	&	0.57	&$	-0.250	_{	-0.012	}^{	+0.012	}$	\\							
$0.87		^{+	0.001	}_{-	0.003	}$&$	0.001	^{+	0.0005	}_{-	0	}$&$	-5.34	\pm	0.96	$&	6.98	&	\oviii\ Ly$\alpha$ ($v_{abs1}$, pc)	&	0.65	&$	-0.253	_{	-0.003	}^{	+0.001	}$	\\							
$0.88		^{+	0.002	}_{-	0.002	}$&$	0.004	^{+	0.0021	}_{-	0.0012	}$&$	9.97	\pm	4.09	$&	11.01	&	\oviii\ edge/rrc ($v_{em}$)	&	0.87	&$	-0.011	_{	-0.004	}^{	+0.005	}$	\\							
\multirow{2}{*}{$1.25	^{+	0.042	}_{-	0.022	}$}&\multirow{2}{*}{$	0.102	^{+	0.0302	}_{-	0.0258	}$}&\multirow{2}{*}{$	-23.90	\pm	0.00	$}&	\multirow{2}{*}{32.53}	&	\neix\ He$\alpha$ ($v_{abs1}$, pc)	&	0.92	&$	-0.264	_{	-0.029	}^{	+0.040	}$								\\
&&&&	+\nex\ Ly$\alpha$ ($v_{abs1}$, pc)	&	1.02	&$								-0.184	_{	-0.032	}^{	+0.045	}$	\\
$1.31^{+	0	}_{-	0.001	}$&$	0.001	^{+	0	}_{-	0	}$&$	3.80	\pm	0.02	$&	8.54	&		&		&							\\							
$1.39	^{+	0.001	}_{-	0.001	}$&$	0.002	^{+	0.0009	}_{-	0.0006	}$&$	3.64	\pm	1.64	$&	7.68	&	\nex\ edge/rrc ($v_{em}$)	&	1.36	&$	-0.022	_{	-0.001	}^{	+0.001	}$	\\							
$1.47	^{+	0.001	}_{-	0.001	}$&$	0.002	^{+	0.0013	}_{-	0.001	}$&$	-3.66	\pm	1.28	$&	8.50	&		&		&							\\							
$1.93	^{+	0.001	}_{-	0.001	}$&$	0.001	^{+	0.0011	}_{-	0	}$&$	-2.67	\pm	0.82	$&	7.18	&		&		&							\\							
$2.22	^{+	0.215	}_{-	0.064	}$&$	0.204	^{+	0.086	}_{-	0.1517	}$&$	14.30	\pm	8.07	$&	9.55	&		&		&							\\							
$2.45	^{+	0.003	}_{-	0.003	}$&$	0.004	^{+	0.0036	}_{-	0.0029	}$&$	-5.98	\pm	6.81	$&	6.43	&		&		&							\\							
$2.55	^{+	0.001	}_{-	0.001	}$&$	0.001	^{+	0	}_{-	0	}$&$	7.03	\pm	2.76	$&	8.35	&		&		&							\\							
$2.60	^{+	0.002	}_{-	0.001	}$&$	0.001	^{+	0.0004	}_{-	0	}$&$	-6.46	\pm	1.91	$&	8.97	&	\sixiv\ Ly$\alpha$ ($v_{abs1}$)	&	2.01	&$	-0.227	_{	-0.0003	}^{	+0.0007	}$	\\							
$2.65	^{+	0.001	}_{-	0.001	}$&$	0.001	^{+	0.0023	}_{-	0	}$&$	10.30	\pm	4.00	$&	9.51	&		&		&							\\							
$3.08	^{+	0.002	}_{-	0.002	}$&$	0.001	^{+	0	}_{-	0	}$&$	-8.23	\pm	0.17	$&	6.64	&		&		&							\\							
$3.21	^{+	0.004	}_{-	0.004	}$&$	0.001	^{+	0.0112	}_{-	0	}$&$	11.30	\pm	4.63	$&	8.51	&		&		&							\\							
$3.47	^{+	0.192	}_{-	0.007	}$&$	0.033	^{+	0.1709	}_{-	0	}$&$	-22.60	\pm	32.80	$&	9.84	&	\sxvi\ Ly$\alpha$ ($v_{abs1}$)	&	2.62	&$	-0.245	_{	-0.002	}^{	+0.071	}$	\\							
$3.55	^{+	0.003	}_{-	0.328	}$&$	0.001	^{+	0.0004	}_{-	0.0003	}$&$	-9.47	\pm	1.06	$&	6.26	&		&		&							\\							
$3.92	^{+	0.004	}_{-	0.004	}$&$	0.005	^{+	0.0063	}_{-	0.0039	}$&$	-11.70	\pm	25.40	$&	10.76	&		&		&							\\							
$5.69	^{+	0.005	}_{-	0.005	}$&$	0.003	^{+	0.0035	}_{-	0.001	}$&$	-23.40	\pm	0.00	$&	7.91	&		&		&							\\							
$7.03	^{+	0	}_{-	0	}$&$	0.002	^{+	0.0225	}_{-	0.0007	}$&$	54.90	\pm	39.00	$&	7.14	&	\fexxv\ He$\alpha$ ($v_{em}$)	&	6.70	&$	-0.047	_{	-0.0001	}^{	+0.003	}$	\\							
$7.35	^{+	0.017	}_{-	0.01	}$&$	0.013	^{+	0.027	}_{-	0.0125	}$&$	90.30	\pm	80.00	$&	11.03	&	\fexxvi\ Ly$\alpha$ ($v_{em}$)	&	6.97	&$	-0.052	_{	-0.0029	}^{	+0.006	}$	\\							
			 \hline
\end{tabular}
 \label{LinesChandraOld} 
\end{center}
\end{table*}

\newpage

\newpage
\bibliography{pds456}

\end{document}